\documentclass[apj,twocolappendix]{emulateapj}

\usepackage{mathtools}
\usepackage{graphicx}
\usepackage[bottom]{footmisc}
\usepackage{mathptmx}
\usepackage{gensymb}
\usepackage{color}
\usepackage[breaklinks=true, colorlinks=true, linkcolor=red, urlcolor=blue, citecolor=black]{hyperref}
\definecolor{myblue}{RGB}{0,0,0}

\def\Mpc{{\rm Mpc}}
\def\kpc{{\rm kpc}}

\def\kms{{\rm km}\,{\rm s}^{-1}}
\def\yr{{\rm yr}}
\def\Msun{{\rm M}_\odot}
\def\Mstel{M_\ast}

\def\logM{\log\Mstel/\Msun}

\def\resp{respectively}
\def\bfr{\bf\color{red}}
\def\bfb{\color{myblue}}

\def\ssfr{{\rm sSFR}}
\def\sfr{{\rm SFR}}

\def\tobs{t_{\rm obs}}

\def\beq{\begin{equation}}
\def\eeq{\end{equation}}
\def\bitem{\begin{itemize}}
\def\eitem{\end{itemize}}
\def\benum{\begin{enumerate}}
\def\eenum{\end{enumerate}}

\mathchardef\mhyphen="2D

\def\cite{{\bfr CITE}}
\def\glass{GLASS}
\def\inn{{\sc inner}}
\def\mid{{\sc middle}}
\def\out{{\sc outer}}

\def\psb{PSB}
\def\pas{PAS}
\def\csf{CSF}
\def\ssf{SSF}

\def\ntot{23265} 
\def\nspec{3022} 
\def\nz{2176} 
\def\npot{76} 
\def\ngood{29} 

\def\facilities{{\it Facilities:}}
\def\software{{\it Software:}}

\shortauthors{Abramson et al.}
\shorttitle{GLASS Resolved SFHs at $z>1$}

\begin{document}

\title{The Grism Lens-Amplified Survey from Space (GLASS). XII.\\ 
Spatially Resolved Galaxy Star Formation Histories and True Evolutionary Paths at \lowercase{{\it z }}$>1^{\ast}$}

\slugcomment{Submitted to AAS Journals, 1 October 2017}

\author{
L.E.\ Abramson\altaffilmark{1,$\dagger$},
A.B.\ Newman\altaffilmark{2},
T.\ Treu\altaffilmark{1},
K.H. Huang\altaffilmark{3,4},
T.\ Morishita\altaffilmark{5},
X.\ Wang\altaffilmark{1},
A.\ Hoag\altaffilmark{3},
K.B.\ Schmidt\altaffilmark{6}, 
C.A.\ Mason\altaffilmark{1},
M.\ Brada\v{c}\altaffilmark{3},
G.B.\ Brammer\altaffilmark{5},
A.\ Dressler\altaffilmark{2},
B.M.\ Poggianti\altaffilmark{7},
M.\ Trenti\altaffilmark{8},
B.\ Vulcani\altaffilmark{7,8}
}


\begin{abstract}

Modern data empower observers to describe galaxies as the spatially and biographically complex objects they are. We illustrate this through case studies of four, $z\sim1.3$ systems based on deep, spatially resolved, 17-band + G102 + G141 {\it Hubble Space Telescope} grism spectrophotometry. Using full spectrum rest-UV/-optical continuum fitting, we characterize these galaxies' observed $\sim$kpc-scale structures and star formation rates (SFRs) and reconstruct their {\it history} over the age of the universe. The sample's diversity---passive to vigorously starforming; stellar masses $\logM=10.5$ to 11.2---enables us to draw spatio-temporal inferences relevant to key areas of parameter space (Milky Way- to super-Andromeda-mass progenitors). Specifically, we find signs that bulge mass-fractions ($B/T$) and SF history shapes/spatial uniformity are linked, such that higher $B/T$s correlate with ``inside-out growth'' and central specific SFRs that peaked above the global average for all starforming galaxies at that epoch. Conversely, the system with the lowest $B/T$ had a flat, spatially uniform SFH with normal peak activity. Both findings are consistent with models positing a feedback-driven connection between bulge formation and the switch from rising to falling SFRs (``quenching''). While sample size forces this conclusion to remain tentative, this work provides a proof-of-concept for future efforts to refine or refute it: {\it JWST}, {\it WFIRST}, and the 30\,m class telescopes will routinely produce data amenable to this and more sophisticated analyses. These samples---spanning representative mass, redshift, SFR, and environmental regimes---will be ripe for converting into thousands of sub-galactic-scale empirical windows on what individual systems actually looked like in the past, ushering in a new dialog between observation and theory.

\end{abstract}


\altaffiltext{1}{
	UCLA, 430 Portola Plaza, Los Angeles, CA 90095, USA
}
\altaffiltext{2}{
	The Carnegie Observatories, 813 Santa Barbara Street, Pasadena, CA 91101, USA
}
\altaffiltext{3}{
	UC Davis, 1 Shields Avenue, Davis, CA 95616, USA
}
\altaffiltext{4}{
	Translational \& Molecular Imaging Institute, Icahn School of Medicine at Mt.\ Sinai, 1470 Madison Avenue, New York, NY 10029, USA
}
\altaffiltext{5}{
	Space Telescope Science Institute, 3700 San Martin Drive, Baltimore, MD 21218, USA
}
\altaffiltext{6}{
	Leibniz-Institut f\"{u}r Astrophysik Potsdam, An der Sternwarte 16, 14482 Potsdam, Germany
}
\altaffiltext{7}{
	INAF-Osservatorio Astronomico di Padova, Vicolo Osservatorio 5, I-35122 Padova, Italy
}
\altaffiltext{8}{
	School of Physics, University of Melbourne, VIC 3010, Australia
}
\altaffiltext{$\ast$}{
	This work incorporates data obtained using the Keck 10\,m telescopes at Maunakea, Hawai`i.
}
\altaffiltext{$\dagger$}{
	\href{mailto:labramson@astro.ucla.edu}{labramson@astro.ucla.edu}
}


\section{Introduction}
\label{sec:intro}

{\bfb A core goal of studying} galaxy evolution is understanding why a galaxy has a given stellar mass ($\Mstel$), star formation rate ($\sfr$), size ($r_{e}$), color, structure, and chemical composition at a given time. Large amounts of data describing those properties in different galaxies at different times circumscribe this endeavor: Deep and wide ground- and space-based imaging yields censuses {\bfb to $z\gtrsim2$---over 75\% of cosmic history---complete to below the mass of the modern Milky Way} \citep[][]{Grogin11,Spitler12,Illingworth16}. Ground- and space-based spectroscopy adds details on stellar populations, and (gaseous) metallicities, kinematics, and hydrodynamics to those images \citep[][]{York00,Calvi11,Oemler13a,Kelson14a,Steidel14,Shapley14,Kriek15,Brammer12,Momcheva16}. Combined, the above form a rich database describing {\bfb many} aspects of the cosmic galaxy ensemble at different snapshots of its existence ready to be compared to the ever-advancing numerical and semi-analytic models that seek to explain them \citep[e.g.,][]{Benson12,Hopkins14,Vogelsburger14,McAlpine16}.

Indeed, at the {\it population} level, the mean and dispersion of many of the above {\bfb spatially unresolved} properties are probably well enough known to require little further empirical investigation: With samples of thousands of $\log\Mstel\gtrsim10$, $z\lesssim2$ systems, statistical uncertainties may no longer impede our understanding {\bfb as much as} systematic {\bfb interpretive} issues.\footnote[9]{This is not true with respect to environment, the circumgalactic medium, neutral gas content, or stellar chemical abundances ($Z$).} {\bfb That is,} knowing how {\bfb the galaxy population} evolves may {\bfb never} lead to a unique idea of how {\it a galaxy} develops---the objective stated above. 

This problem arises from the inescapable fact that data are cross-sectional (different systems at different times) while theories are longitudinal (the same systems at different times). This allows different physical models to describe the same (uncontested) observations. Projections of the size--$\sfr$--mass plane {\bfb provide rich grounds for such} debate \citep{PengLilly10, Pacifici13, Pacifici16, Gladders13b, Kelson14, Zolotov15, Barro16, Abramson16, Oemler16, Lilly16, Morishita17, Abramson17}: Whether, for example, red galaxies' small size compared to blue ones reflects a rapid quenching transformation or merely the universe's density evolution has strong implications for the interplay between {\bfb galactic} structure and star formation.

Progress requires {\bfb lifting} these descriptive degeneracies {\bfb using} different {\it kinds} of information, not just more of it \citep{Abramson16, Kelson16}. Principally, higher resolution observations and inferences may be key.

{\bfb Observationally,} despite the breadth and depth of current surveys, galaxies are still typically analyzed as monolithic entities with a {\bfb single} stellar mass, SFR, or metallicity. As such, they are condensed to points in parameter space through (or near) which more than one physical model can pass. 

The advent of large samples of spatially resolved integral field unit (IFU) spectroscopy is changing this norm. These data transform galaxies into many stellar populations corresponding to many mass surface densities/metallicities, {\bfb challenging models to reproduce distributions for individual objects,} not just integrated quantities \citep{Blanc09, ForsterSchreiber09, Cappellari11, Sanchez12, AllenSAMI15, Bundy15, Magdis16, Goddard17, Poggianti17, Mason17}.

{\bfb Inferentially,} empirical reconstructions of galaxy star formation histories (SFHs) are adding new {\it temporal} resolution, shedding light on a system's full growth curve, not just its integral ($\Mstel$) or current derivative ($\sfr(\tobs)$). This endeavor stretches back at least to \citet{Tinsley68} and has been used {\bfb to great effect to study the light-/mass-weighted ages of red galaxies (constraining quenching timescales and late-time star formation; e.g.,} \citealt{DresslerGunn83,Couch87,Poggianti99,Poggianti13a,Kelson01,Kauffmann03,Dressler04,Treu05,Fritz07,Fritz14,Kriek08,DominguezSanchez16}). However, {\bfb it also provides insight into} more abstract questions, such as the extended role of environment \citep{Thomas05,Trager00,Kelson06,Guglielmo15,McDermid15}, whether galaxy histories are self-similar \citep[][]{Pacifici16}, how their parameterization affects population-level physical inferences \citep{Tinsley68, Gallagher84, Oemler13b,Iyer17,Ciesla17}, and whether established ways of classifying galaxies or matching them to dark matter halos require rethinking (\citealt{Weisz14,Dressler16}; Dressler et al., in preparation).

{\bfb Expanding} the space in which we think about galaxies and ``longitudinalizing'' their cross-sectional snapshots {\bfb have great promise individually}, but {\it combining} them is optimal: {\bfb While} higher spatial resolution data allow for more realistic descriptions, {\bfb reconstructing} SFHs for {\bfb {\it parts} of galaxies} would enable full spatio-spectral {\bfb projections} of what individual {\bfb systems} actually looked like in the past (as opposed to, e.g., abundance matching-based inferences; \citealt{vanDokkum13,Morishita15}). Unfortunately, due to resolution and signal-to-noise {\bfb ratio} ($S/N$) requirements, {\bfb suitable} IFU surveys {\bfb reaching stellar continua} are limited to the local universe \citep{Ma14,McDermid15,Goddard17}, far from the peak of stellar mass production/galaxy growth \citep[$0.7\lesssim z\lesssim2$;][]{MadauDickinson14}. {\bfb Yet,} by brightening and extending sources by $\sim$2--10$\times$, gravitational lensing can {\bfb help} bypass this constraint. 

Combined with the native sensitivity and spatial resolution of the {\it Hubble Space Telescope} (HST), lensing---especially by foreground galaxy clusters---{\bfb yields} unprecedented views into galaxies at the peak of cosmic activity \citep[e.g.,][]{Wuyts12a,Wuyts12b}, resolving their star formation and gas/stellar properties in such detail as to meaningfully challenge theory \citep[][]{Wuyts14,Jones15,Newman15,Wang17,Mason17,Johnson17b,Johnson17a,Rigby17,Toft17}. The {\it Grism Lens-Amplified Survey from Space} \citep[GLASS;][]{Schmidt14,Treu15} builds on this foundation, turning HST's slitless {\it Wide Field Camera 3} (WFC3) IR grism spectrographs on the powerful lenses from the {\it Cluster Lensing And Supernova survey with Hubble} \citep[CLASH;][]{Postman12} and {\it Hubble Frontier Field} \citep[HFF;][]{Lotz17} campaigns. 

Exploiting space's low IR backgrounds{\bfb, lack of seeing limitations,} and lensing's {\bfb additional resolution} and $S/N$ boosts, GLASS provides rest-optical continuum spectra for SFR-unbiased, $z\gtrsim1$ galaxy samples resolved out to $\sim$2 half-light radii. These IFU{\bfb -like} data \citep{Vulcani15,Vulcani16,Vulcani17,Wang17} enable {\bfb individual systems'} size, mass, and SFR {\bfb evolution to be inferred over Gyr timescales}, providing new spatio-temporal {\it empirical} windows on how galaxies wove their way through the cosmic narrative.

{\bfb Here}, we use {\bfb these} data to reconstruct {\bfb the spatially resolved histories} of four $z>1$, $\log\Mstel\gtrsim10.5$ galaxies spanning a diversity of star formation states {\bfb from fully passive to strongly starforming}. {\bfb We then use these inferences to draw links between the systems' observed stellar structure and the shapes of their SF histories. While sample size limits the generality of our conclusions, our analysis} represents a first step towards building a new common ground for observational and theoretical studies, providing a proof of concept for methods that {\bfb could soon readily be} applied to {\bfb thousands of representative systems from {\it JWST}, {\it WFIRST}, and 30\,m class facilities.}

We {\bfb proceed} as follows: Section \ref{sec:data} describes the HST spectrophotometry and {\bfb lensing magnification estimates (\ref{sec:mag})}; Section \ref{sec:models} the SFH fitting process; Section \ref{sec:results} the integrated (\ref{sec:global}), spatially resolved (\ref{sec:resolved}), and spatio-temporally resolved (\ref{sec:longitudinal}) inferences derived therefrom; and Section \ref{sec:discussion} their astrophysical (\ref{sec:discussionPhysics}), and methodological/theoretical (\ref{sec:discussionTheory}) implications. Appendices \ref{sec:AA}--\ref{sec:AD} provide further details. Readers comfortable with modern spectral energy distribution (SED) fitting techniques can skip to Section \ref{sec:results}.

We take $(H_{0},\Omega_{m},\Omega_{\Lambda}) = (73\,\kms\,\Mpc^{-1},0.27,0.73)$ and quote AB magnitudes.


\section{Data}
\label{sec:data}

{\bfb We combine public GLASS spectroscopy\footnote[10]{\url{https://archive.stsci.edu/prepds/glass/}} with CLASH and HFF imaging\footnote{\url{https://archive.stsci.edu/prepds/clash/}}$^{,}$\footnote{\url{https://archive.stsci.edu/prepds/hff/}} to create integrated and spatially resolved SEDs in three radial bins at $|r|\lesssim2\,r_{e}$ for four, diverse, $z\sim1.3$ galaxies selected for their brightness and spectral quality. Fitting these SEDs provides our SFHs (Section \ref{sec:models}).} We defer to \citet{Schmidt14} and \citet{Treu15} for {\bfb GLASS'} details, but review some relevant aspects here. 

\subsection{GLASS Spectroscopy}
\label{sec:spectra}

\glass\ obtained {\bfb slitless} WFC3IR G102 and G141 grism spectra ($\lambda\sim8800$--16700\,\AA) covering {\bfb ten HFF/CLASH cluster-magnified sightlines}. To mitigate against {\bfb nearby object contamination}, {\bfb two spectra were taken per sightline at roughly orthogonal roll angles over 5\,(G102)\,+\,2\,(G141) orbits per orient, reaching similar depths in each grism. For the objects studied here, however, usable data come from only one orient} (Section \ref{sec:selection}).

A modified 3D-HST pipeline \citep{Brammer12,Momcheva16} associates these spectra with objects identified in {\bfb deeper} CLASH/HFF pre-imaging, and subtracts contaminating light from nearby traces. FITS cubes are produced for each source in each grism at each orient. We reference relevant layers of these in {\tt typewritten text} to aid reproducibility (Appendix \ref{sec:AA} provides a thorough discussion), but, at root, rely on each source's postage stamps, 2D spectra, and RMS and contamination maps. Section \ref{sec:extraction} details how these objects {\bfb guide} 1D spectral extraction.

The \glass\ database contains \ntot\ objects. Of these, \nspec\ have ${\rm F140W_{GLASS}}\leq24$ and {\bfb so were inspected} and assigned a fiducial spectroscopic redshift ($z_{\rm GLASS}$). All redshifts quoted here are rederived in the analysis (Section \ref{sec:models}), {\bfb but both estimates agree within uncertainties (Table \ref{tbl:sample}).}

\subsection{Sample Selection}
\label{sec:selection}

{\bfb We aim to spectrophotometrically reconstruct spatially resolved SFHs to draw more meaningful longitudinal inferences from cross-sectional data. In scheme, we follow \citet{Kelson14a} and \citet{Dressler16}, \citet{Newman14}, and \citet{DominguezSanchez16}, who demonstrated the potential of this approach using similar but spatially integrated data at $z\sim0.7$, $z\sim1.8$, and our $z=1$--1.5 range, \resp.}

The critical aspect of these data is that photometry samples the rest-UV/-IR where SFH information is low ({\it current} SFR and its integral, $\Mstel$, \resp). Meanwhile, spectroscopy covers the high information-density rest-optical, containing details on the intermediate age stellar populations that provide leverage on the path a galaxy took to reach those endpoints. ``High'' spectral resolution ($R_{\rm eff}\gtrsim70$) is thus present where it is needed. This said, we diverge from the above studies in that {\it Spitzer} IRAC's larger PSF precludes us from using those data for spatially resolved photometry. 

We stress that, {\bfb for the case studies possible at} this stage, data quality, not quantity, is key: While we will speculate on their broader implications (Sections \ref{sec:longitudinal}, \ref{sec:discussionPhysics}), we are not yet so interested in the representativeness of any one finding, but the potential of what can be learned from these data/methods. Given the critical role of spectroscopy in this exercise, the GLASS database determines sample selection. 

\begin{figure}[b!]
\centering
\includegraphics[width = \linewidth, trim = 0.5cm 0cm 0cm 0cm]{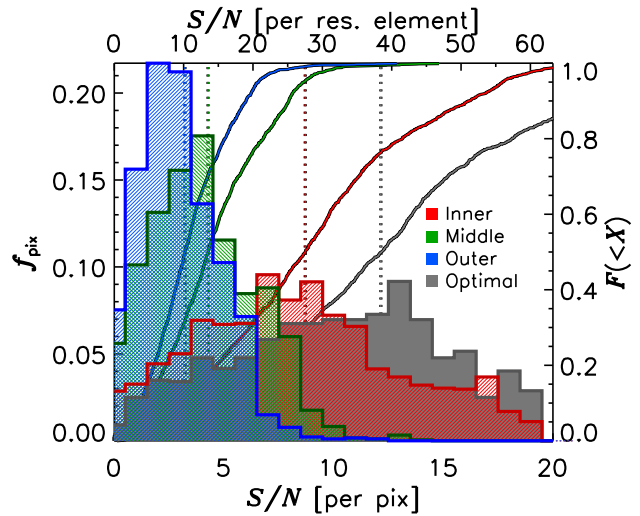}
\caption{$S/N$ histograms and cumulative distributions for the 16 sample spectra (four per galaxy). \inn\ (red) covers $|r/r_{e}|\lesssim0.4$; \mid\ (green) and \out\ (blue) reach $|r/r_{e}|\in (0.4,1.0]$, and $\in (1,2]$, \resp. Optimal extractions are plotted in grey. Dotted vertical lines show median $S/N$.}
\label{fig:snrs}
\end{figure}

From the \nz\ sources at $1.0\leq z\leq 1.8$---where $z$ is $z_{\rm GLASS}$ or a photometric redshift \citep[][]{Brammer08, Castellano16, Morishita17}---we draw \npot\ sources with ${\rm F140W_{GLASS}}\leq21.8$. This ensures important age-sensitive features lie in the grism bandpasses (the 4000\,\AA\ break, the G4300 to Mg$b$ continuum, as many Balmer lines as possible), avoids selecting against passive systems lacking a secure (emission line-based) $z_{\rm GLASS}$, and ensures $\langle S/N\rangle\approx10$ per $\sim$10\,pix resolution element in the (folded) \out\ extractions (Section \ref{sec:extraction}, Figure \ref{fig:snrs}). {\bfb We then select \ngood\ objects with at least one pair of grism spectra with $\leq$30\% of \out\ pix contaminated at $\leq$30\%, where ``contaminated'' is the ratio of the {\tt CONTAM} to {\tt MODEL} FITS layers (Appendix \ref{sec:AA})}. This quality control cut is arbitrary, but balances sample size with avoiding continuum biases from contamination subtraction residuals. (No source has $\leq$10\% of pix at $\leq$10\% contamination.)

Finally, we discard three obvious mergers, 18 poorly sampled sources (1\,pix wide extraction boxes, some of which are stars), one case of bad sky subtraction, two {\it post facto}-identified low-$z$ interlopers, and one source for which the lognormal SFH failed to converge (Section \ref{sec:models}). {\bfb Ongoing m}ergers may merit future study, but{\bfb, to the extent possible (Section \ref{sec:discussionTheory}),} we aim to infer {\it in situ} growth and so avoid spectral superpositions of objects with distinct SFHs.

This leaves a final sample of four sources. {\bfb Fortuitously, they constitute a meaningfully diverse cohort of a (1) continuously starforming (``\csf''), (2) {\it strongly} starforming (``\ssf''), (3) recently quenched/poststarburst (``\psb''), and (4) anciently passive galaxy (``\pas''). }

\begin{figure*}[t!]
\centering
\includegraphics[width = 0.8\linewidth, trim = 0cm 0cm 0cm 0cm]{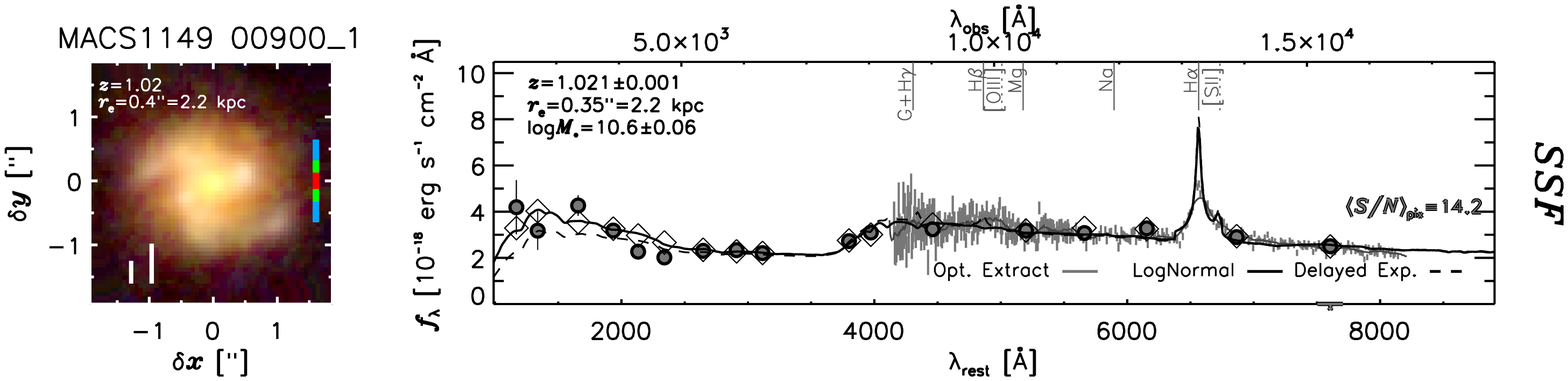}
\includegraphics[width = 0.8\linewidth, trim = 0cm 0cm 0cm 0cm]{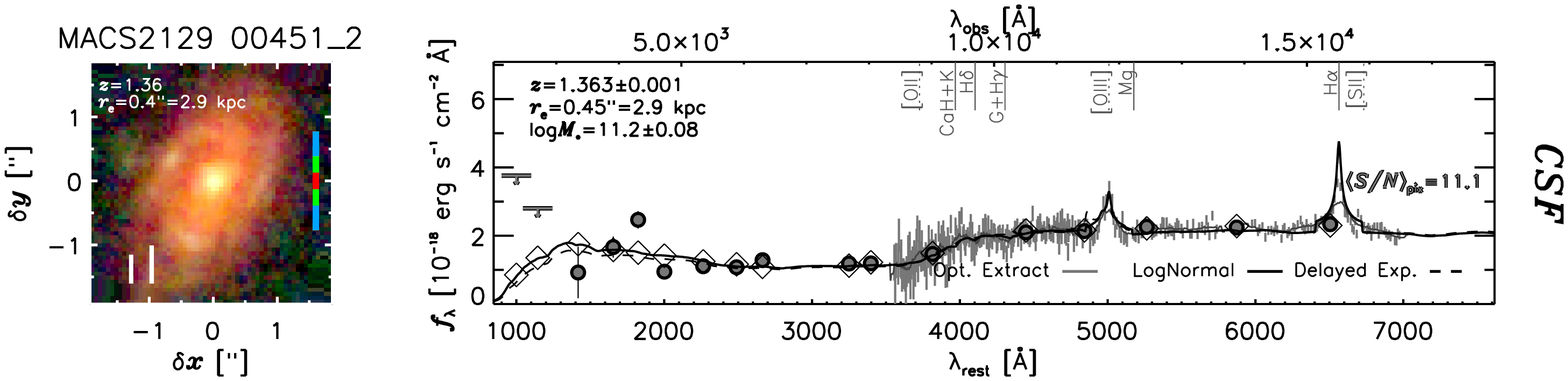}
\includegraphics[width = 0.8\linewidth, trim = 0cm 0cm 0cm 0cm]{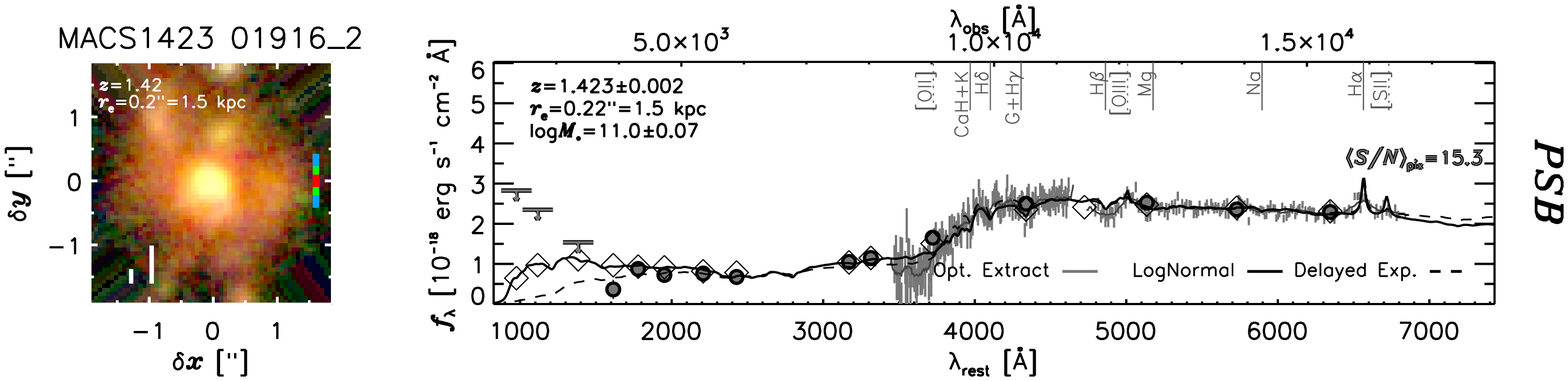}
\includegraphics[width = 0.8\linewidth, trim = 0cm 0cm 0cm 0cm]{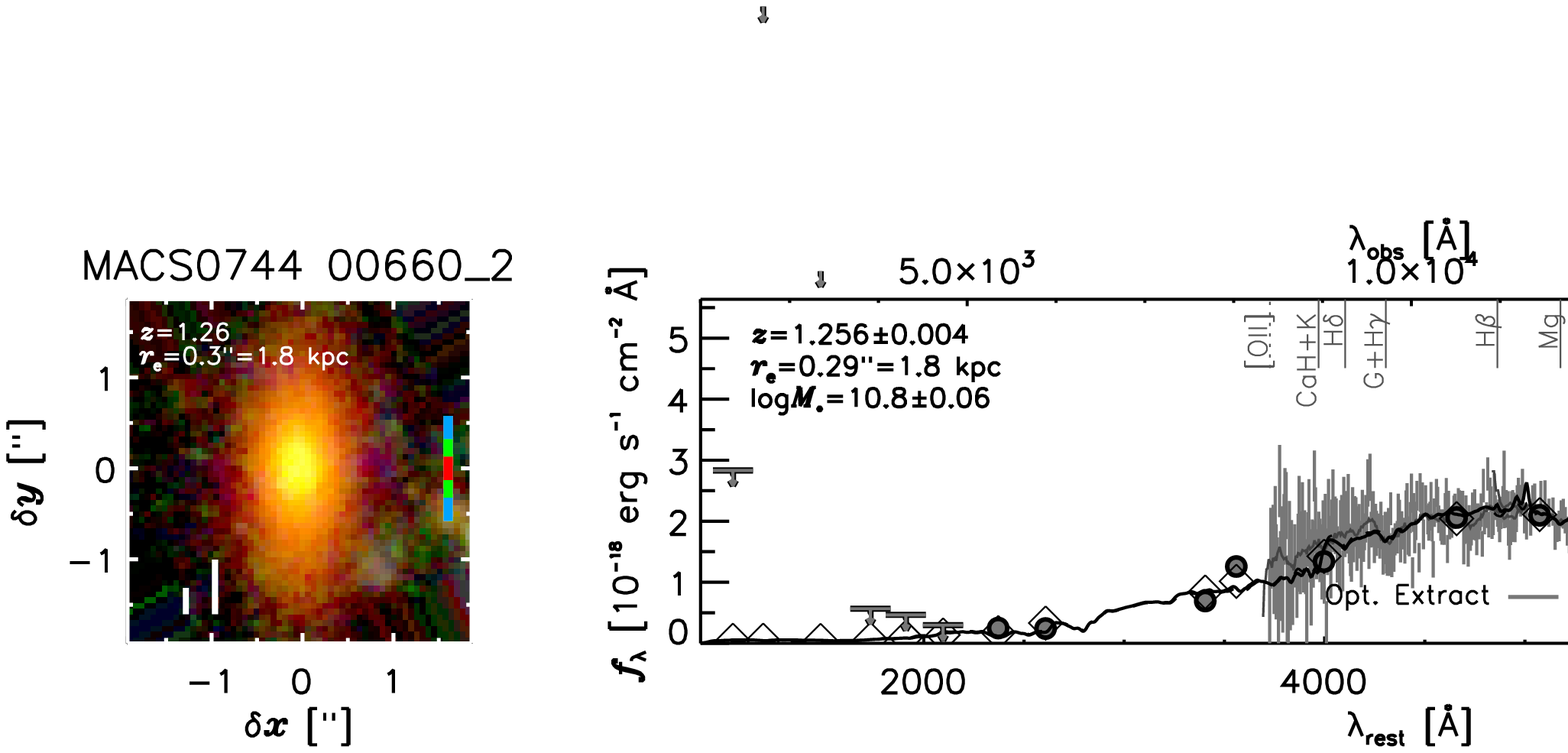}
\caption{Color CLASH/HFF stamps and optimal GLASS spectra of the four sample objects (Table \ref{tbl:sample}) ordered by decreasing $\ssfr$ from {\it top}. The smaller white scale-bar on each $\sim$4$''$$\times$4$''$ image shows $r_{e}$; the larger, 5\,kpc at the source redshift. Multicolored scale bars at right show the location/width of the radial spectral zones (Section \ref{sec:extraction}) {\bfb color-}coded as in Figure \ref{fig:mosaic} and all further plots. CLASH/HFF photometry/2\,$\sigma$ limits are denoted by circles and arrows, \resp, with open diamonds showing the {\tt PYSPECFIT} model values assuming lognormal (solid) or delayed exponential SFHs (dashed lines). Spectra are plotted as their 1\,$\sigma$ uncertainty interval at each $\lambda$. Numbers at the red ends of the spectra give mean per-pixel $S/N$. Other integrated physical parameters are listed at {\it top-left}.}
\label{fig:mosaic}
\end{figure*}

Figure \ref{fig:mosaic} shows these sources and their integrated spectrophotometry; Section \ref{sec:global}, Tables \ref{tbl:sample} and \ref{tbl:derived} describe them in detail. While modest ($\langle\mu\rangle\simeq1.7$), none of the sample would meet $S/N$ requirements without lensing magnification.

\begin{deluxetable*}{cccccccccc}
	\tabletypesize{\footnotesize}
	\tablewidth{0pt}
	\tablecolumns{10}
	\tablecaption{Basic Source Information}
	\tablehead{
	\colhead{Sightline} & 
	\colhead{GLASS ID$\_$PA\tablenotemark{a}} & 
	\colhead{Tag} &
	\colhead{RA [$\degree$]} & 
	\colhead{DEC [$\degree$]} & 
	\colhead{$z$\tablenotemark{b}} & 
	\colhead{$z_{\rm GLASS}$\tablenotemark{c}} & 	
	\colhead{Magnification ($\mu$)\tablenotemark{d}} & 
	\colhead{Pre-survey} & 
	\colhead{F140W$_{\rm GLASS}$\tablenotemark{e}} \\
	\colhead{} & 
	\colhead{} & 
	\colhead{} & 
	\colhead{(J2000)} & 
	\colhead{(J2000)} & 
	\colhead{} & 
	\colhead{} & 
	\colhead{} & 
	\colhead{} & 
	\colhead{(AB)} 
}
\startdata
	MACS0744 & 00660\_2 & \pas & 116.2281 & 39.46423 & 1.257$\pm$0.005 & 1.260$\pm$0.006 & 1.87$\pm$0.24 & CLASH & 21.42 \\
	MACS1149 & 00900\_1 & \ssf & 177.4089 & 22.40319 & 1.021$\pm$0.001 & 1.033$\pm$0.007 & 1.67$\pm$0.23 & CLASH, HFF & 21.22 \\
	MACS1423 & 01916\_2 & \psb & 215.9373 & 24.06136 & 1.423$\pm$0.002 & 1.435$\pm$0.006 & 1.52$\pm$0.24 & CLASH & 21.34 \\
	MACS2129 & 00451\_2 & \csf & 322.3719 & -7.68199 & 1.367$\pm$0.001 & 1.368$\pm$0.006 & 1.66$\pm$0.24 & CLASH & 21.72
\enddata
\tablecomments{$^{\rm a}$\,GLASS source number and orientation of the analyzed spectra; ``\_1'' corresponds to the GLASS position angle closest to zero. $^{\rm b}$\,Basis of this analysis; from full spectrophotometric fitting (Section \ref{sec:models}). $^{\rm c}$\,GLASS catalog redshift; error corresponds to a $\sim$10\,pix LSF FWHM at observed H$\alpha$; roughly $\sigma_{v}=2000\,{\rm km\, s^{-1}}$. $^{\rm d}$\,From the online calculator (HFF) or the maps of Zitrin et al.\ (CLASH; Section \ref{sec:mag}). $^{\rm e}$\,GLASS catalog SExtractor {\tt MAG\_AUTO} estimate.}
\label{tbl:sample}
\end{deluxetable*}

\begin{figure*}[t!]
\centering
\includegraphics[width = 0.77\linewidth, trim = 0cm 0cm 0cm 0cm]{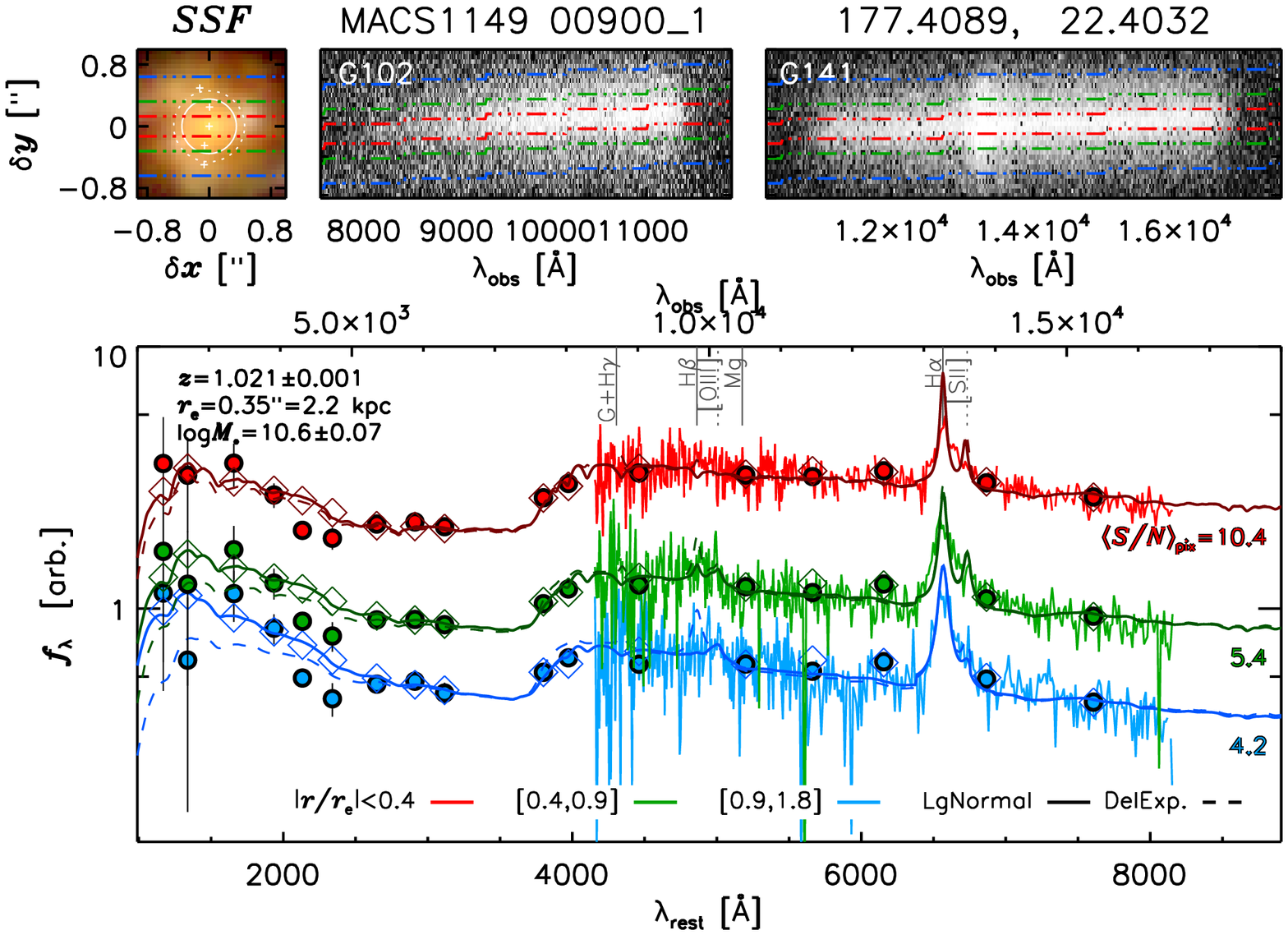}
\includegraphics[width = 0.77\linewidth, trim = 0cm 0cm 0cm 0cm]{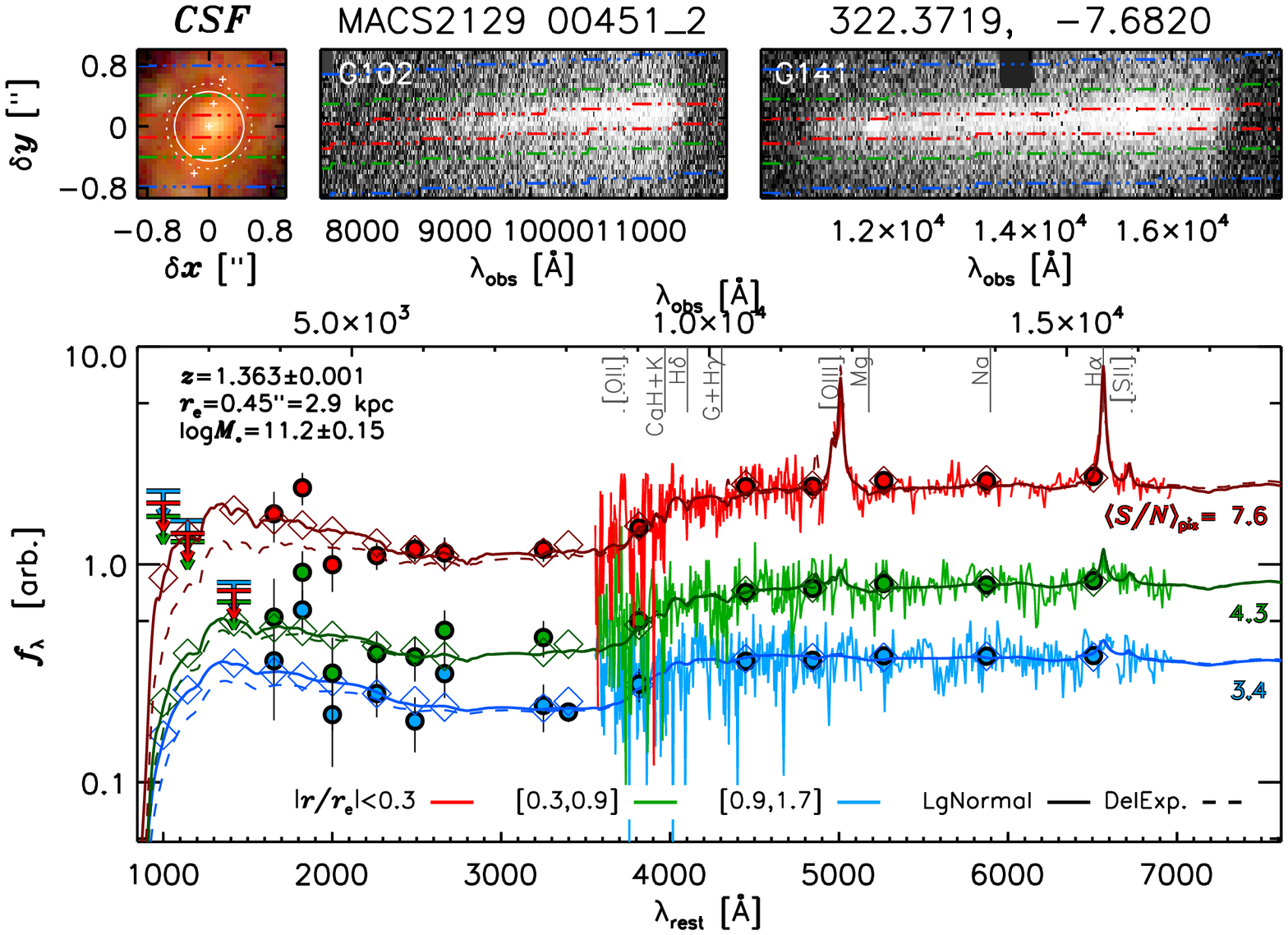}
\caption{Images, 2D spectra, and radially resolved 1D extractions for \ssf\ and \csf\ (others shown in the continuation of this plot). {\it Top-left} images are insets of those in Figure \ref{fig:mosaic} showing {\bfb 1D $r_{e}$ (solid) and 2D half-light radii (dashed circles),} the extraction regions, and the light-weighted $(x,y)$ centroids for each. Rightward of these are the G102/141 2D GLASS spectra, with extractions zones overplotted in red (\inn), green (\mid), and blue (\out). These are stepped as the spectra are not $(x,y)$ rectified. Black areas on the 2D spectra show masking due to contamination or missing data. The resulting 1D spectra are plotted in the main panel, color-coded by region, with CLASH+HFF photometry shown as in Figure \ref{fig:mosaic}. The best-fit {\tt PYSPECFIT} lognormal (delayed exponential) SED fit is plotted as a solid (dashed) line, with open diamonds showing the predicted locations of the photometry. In all but \psb, the lognormal provides as good a fit to the data as the delayed exponential SFH. The radial extent of each extraction zone is listed at bottom.}
\label{fig:spectra}
\end{figure*}

\renewcommand{\thefigure}{\arabic{figure} (cont'd)}
\addtocounter{figure}{-1}

\begin{figure*}[t!]
\centering
\includegraphics[width = 0.77\linewidth, trim = 0cm 0cm 0cm 0cm]{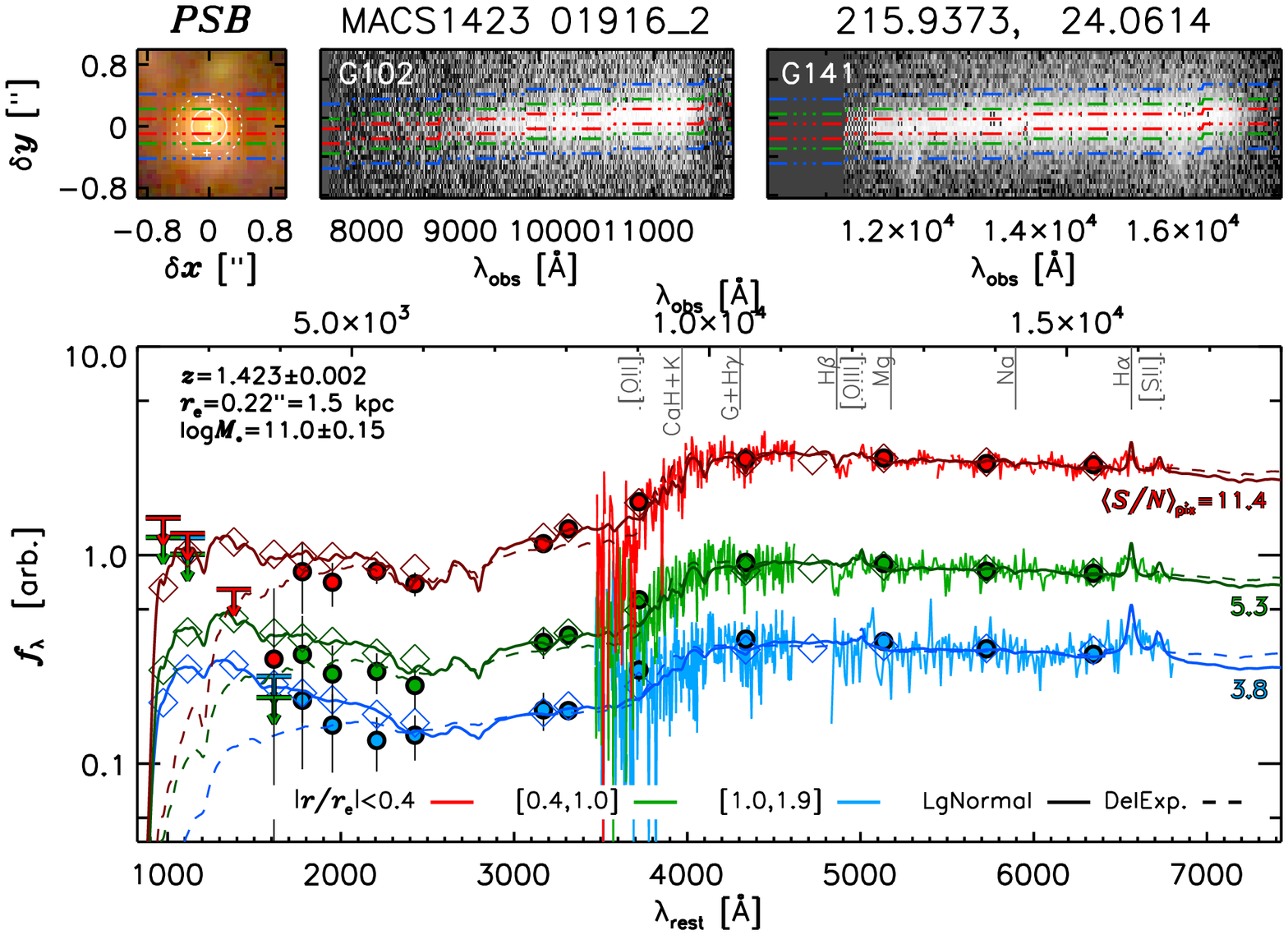}
\includegraphics[width = 0.77\linewidth, trim = 0cm 0cm 0cm 0cm]{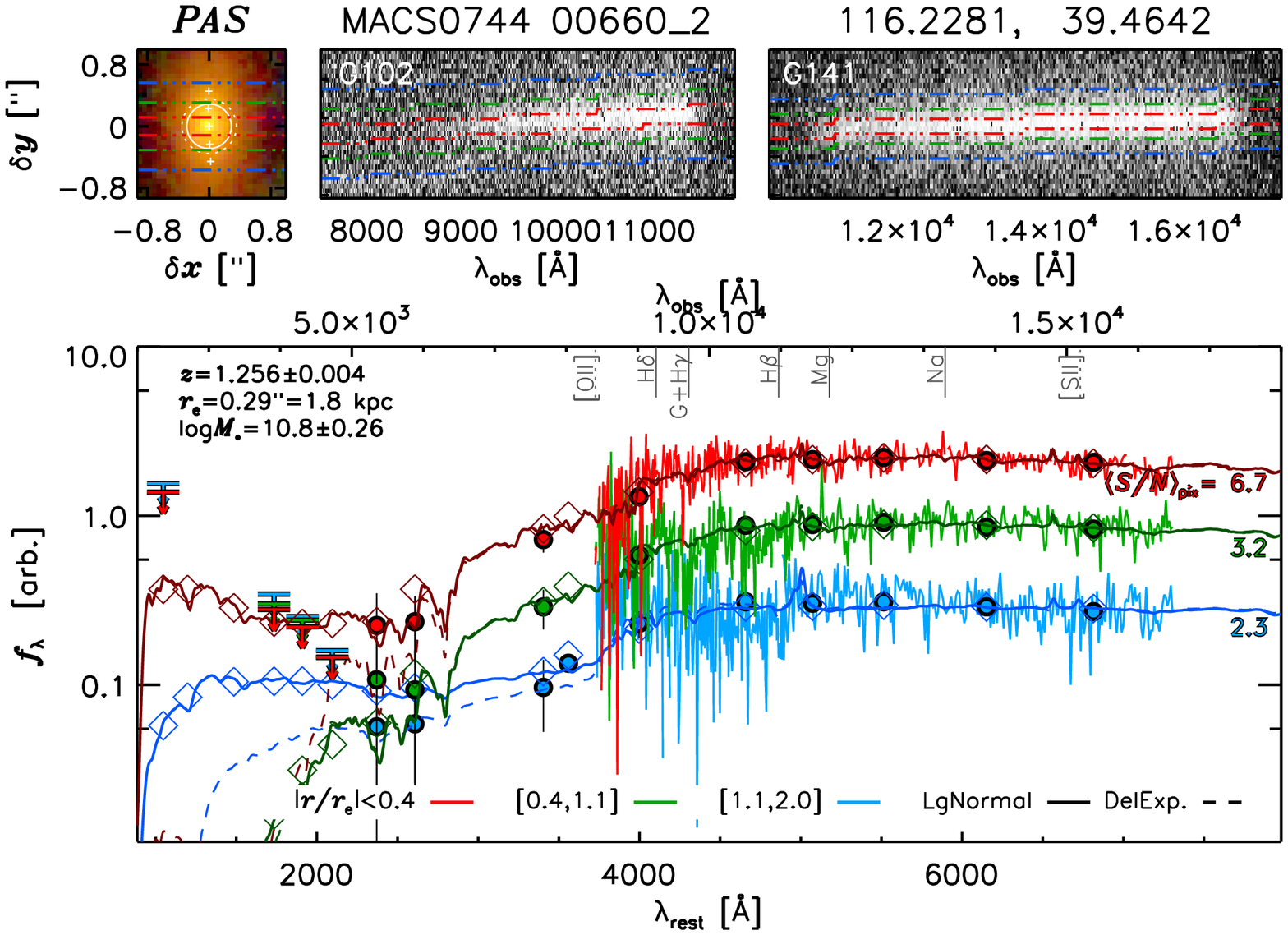}
\caption{Images, 2D spectra, and radial 1D spectra for \psb\ and \pas.}
\end{figure*}

\renewcommand{\thefigure}{\arabic{figure}}

\subsection{1D Spectra}
\label{sec:extraction}

We extract an optimal $S/N$ integrated spectrum and {\bfb three, folded,} radial {\bfb spectra} from each source. {\bfb To mimic typical} analyses, the former is used to derive {\bfb total} galaxy properties, the latter resolved ones. Summing the spatial extractions produces self-consistent results (Appendix \ref{sec:resWhole}). Obtaining and analyzing these requires estimating 1D half-light radii, $r_{e}$, optimal extraction spatial weights, and line spread functions (LSFs) using the GLASS postage stamps ({\tt DIRECT\_IMAGE}; stacks of all WFC3IR CLASH/HFF pre-imaging).

To avoid including near neighbors whose {\it spectra} are removed by the pipeline, this {\bfb requires some modeling. We collapse} the central 20$\times$20\,pix ($\sim$10$\times$10\,kpc) of each image along the {\bfb spectral axis (to obtain the target's spatial profile), or spatial axis (to obtain the LSF from its morphology). We then fit Moffat functions to these profiles---which reach beyond bulges, but not to outer disks (Figure \ref{fig:mosaic})---and extrapolate to extend them to the full 60\,pix stamp size.} The LSF is used in SED fitting (Section \ref{sec:models}); the spatial profile defines the optimal extraction weights \citep{Horne86}, and $r_{e}$, on which the radial extractions are based. 

Figure \ref{fig:spectra} shows each source's CLASH/HFF stamps, 1D $r_{e}$ and 2D half-light radii, 2D spectra, radial extraction zones, and corresponding 1D spectra. The radial bins are:
\bitem
	\item {\bf INNER:} $|r/r_{e}|\leq0.4$;
	\item {\bf MIDDLE:} $0.4<|r/r_{e}|\leq1.0$; 
	\item {\bf OUTER:} $1.0<|r/r_{e}|\leq2.0$;
\eitem
where the \mid\ and \out\ bins are reflected over the GLASS pipeline-output spectral centerline ({\tt YTRACE}; galactocentric radius $r\equiv y-{\tt YTRACE}=0$) {\bfb and averaged before fitting}. Bin widths are thus 0.8, 2$\times$0.6, and 2$\times$1.0\,$r_{e}$, \resp. While somewhat arbitrary and subject to pixelation, these were chosen to counter decreasing $S/N(r)$ while preserving spectral line shapes. {\bfb Each zone's} 1D spectrum is simply the {\bfb mean} of all unmasked pixels {\bfb (${\tt SCI}>0$, ${\tt CONTAM}/{\tt SCI}<0.01$; Appendix \ref{sec:AA})} in the {\tt CONTAM}-subtracted 2D {\tt SCI} spectrum at {\bfb each} $\lambda$. {\bfb Uncertainties are} the identical (quadrature) sums {\bfb from} the {\tt WHT} RMS-map. GLASS 2D spectra are cut from interlaced---not drizzled---mosaics, so {\bfb all} pixels are independent. We find no significant backgrounds.

Note: as defined, ``$r_{e}$'' is not the half-light radius returned by, e.g., SExtractor or GALFIT \citep{BertinSEX,PengGALFIT}. Rather, it is the $y$ extent containing half the light collapsed across $x$. This is appropriate for defining {\bfb spectral} bins by spatial $S/N$, but underestimates 2D half-light radii. This effect is small except in ID 01916\_2 (``\psb''; Figure \ref{fig:spectra}). 

Finally, we estimate LSFs independently in each zone. This accounts for meaningful galaxy chord profile variations and provides light-weighted bin centroids to assess position angle effects{\bfb : At G102/141 spectral resolution, PAs {\bfb near $45^{\degree}$} impose a {\bfb detectable} geometric $r$--$\lambda$ covariance. {\bfb Unlike} typical (physical) $v$--$\lambda$ covariance at higher resolution, this is due to the slitless grisms reproducing an image of the entire source at every wavelength, not the doppler shift. Such offsets can reach $\sim$3\,pix (30\% LSF FWHM; $\sim$60\,\AA; $\Delta z\sim0.005$; Figures \ref{fig:spectra}, \ref{fig:lsfs}) and so should be accounted for in line analyses.}

\subsection{Photometry}
\label{sec:photometry}

Each source is covered by 17-band CLASH imaging, with one having deeper, supplementary data in the 7 HFF bands (Table \ref{tbl:sample}). These reach into the rest-UV ($\sim$1500\,\AA), providing critical leverage on dust and recent star formation.

We {\bfb incorporate these data by} cutting stamps/RMS maps for each source from its full, PSF-matched CLASH/HFF mosaic \citep{Morishita17}, resampling and rotating these to the GLASS plate scale/orientation (Appendix \ref{sec:AA}). Resolved photometry is then the sum in each of radial zone from the cutouts. To match the spectra, \inn\ data is left as-is, while upper and lower \mid\ and \out\ counts are averaged before measuring. For the optimal extraction, the zones are summed after weighting by the spatial profile.

Since all images are convolved {\it a priori} to F160W resolution, the resultant spectrophotometry is {\bfb effectively} aperture-matched. Comparing synthetic F140W magnitudes from the GLASS spectra to broadband fluxes suggests an absolute accuracy of $\pm$$\sim$30\% (1.5\,$\sigma$). Yet, relative spectrophotometric agreement is more important in SED fitting. As we correct the photometry for galactic extinction \citep{Schlafly11},\footnote{\url{https://ned.ipac.caltech.edu/forms/calculator.html}} we ultimately {\bfb tie} the spectra to those data by applying a single multiplicative constant. Figures \ref{fig:mosaic} and \ref{fig:spectra} demonstrate that the subsequent agreement is excellent.

This process yields four, radially resolved, 17-band + G102/141 spectrophotometric datasets per galaxy ready for SPS modeling. {\bfb Extant spatially integrated data not used in the modeling serve as cross-checks (Section \ref{sec:resolved}).}

\subsection{Lensing Magnification}
\label{sec:mag}

{\bfb Lensing boosts apparent fluxes and sizes. Hence, estimating} absolute quantities {\bfb using those data} ($\Mstel$, $\sfr$, intrinsic size) requires knowledge of a source's magnification, $\mu$. This is {\bfb provided} by models {\bfb built} for the GLASS cluster sightlines by a number of teams \citep{Zitrin09,Zitrin13,Jauzac14,Johnson14,Richard14,Grillo15,Ishigaki15,Wang15,Hoag16,Caminha17}. Since $\mu$ is applied to derived quantities, this {\bfb step} is independent of the preceding.

{\bfb ID 00900\_1 (``\ssf'') lies in the HFF footprint, so} we download its default set of values from the public HFF lens models,\footnote{\url{https://archive.stsci.edu/prepds/frontier/lensmodels/\#magcalc}} discard the extremes, and adopt the mean (1.67) and scatter (0.23; 14\%) of the other inferences as $\mu$ and its error. 

{\bfb All other sources lack HFF lens models. For these,} we average the ``LTM-Gauss\_v2'' and ``NFW\_v2'' \citet{Zitrin09,Zitrin13} models for $\mu$. For uncertainties, we add their 3\,$\sigma$ formal errors ($\lesssim$0.05) in quadrature with the (much larger) HFF model dispersion for \ssf. Identical models are not available, but the ``LTM-Gauss\_v1'' best estimate is consistent with \ssf's HFF-derived $\mu$ at 2.4\,$\sigma$. In all cases, $\mu$ is a small correction ($\lesssim$0.3 dex; Table \ref{tbl:sample}), and ratios such as $\ssfr\equiv\sfr/\Mstel$ and $\Mstel$ or $\sfr$ surface densities are $\mu$-independent.


\begin{deluxetable*}{cccccccc}
	\tabletypesize{\footnotesize}
	\tablewidth{0pt}
	\tablecolumns{8}
	\tablecaption{Derived Integrated Properties from Optimal Extractions}
	\tablehead{
	\colhead{Tag} & 
	\colhead{$\log \Mstel/\Msun$\tablenotemark{a,b}} & 
	\colhead{$\log \sfr/\Msun\,{\rm yr^{-1}}$\tablenotemark{a,b}} & 	
	\colhead{$r_{e}$ [kpc]\tablenotemark{b}} & 
	\colhead{$r_{e,\,\rm 2D}$ [kpc]\tablenotemark{b,c}} & 
	\colhead{$A_{V}$ [mag]\tablenotemark{a}} & 
	\colhead{$(T_{0},\tau)$ [$\ln{\rm Gyr}$]\tablenotemark{a}} & 
	\colhead{$({\rm age}, \tau_{\rm exp})$ [Gyr]}
}
\startdata
\pas & 10.83$\pm$0.06 (0.17) & $<$-1.18\tablenotemark{d} & 1.76$\pm$0.11 & 2.15$\pm$0.13 & 0.09$\pm$0.09 & $(1.03\pm0.08,\,0.13\pm0.03)$ & $(1.48\pm0.25,\,0.04\pm0.05)$ \\
\ssf & 10.57$\pm$0.06 (0.06) &  1.40$\pm$0.06 (0.12) & 2.20$\pm$0.14 & 2.86$\pm$0.19 & 1.06$\pm$0.02 & $(1.64\pm0.00,\,0.10\pm0.00)$ & $(0.70\pm0.09,\,0.24\pm0.04)$ \\
\psb & 10.95$\pm$0.07 (0.12) &  0.54$\pm$0.10 (0.93) & 1.54$\pm$0.11 & 2.88$\pm$0.21 & 0.41$\pm$0.05 & $(1.21\pm0.01,\,0.10\pm0.00)$ & $(0.42\pm0.09,\,0.06\pm0.02)$ \\
\csf & 11.15$\pm$0.07 (0.10) &  1.59$\pm$0.11 (0.39) & 2.94$\pm$0.20 & 3.61$\pm$0.24 & 1.19$\pm$0.12 & $(1.32\pm0.35,\,0.68\pm0.31)$ & $(3.51\pm0.66,\,1.42\pm0.60)$ 
\enddata
\tablecomments{$^{\rm a}$\,Lognormal SFHs; parentheses show summed \inn\ + 2\,\mid\ + 2\,\out\ $\Mstel$ or $\sfr$ errors; {\bfb delayed exponential $\Mstel$ but not necessarily $\sfr$ are consistent (Figure \ref{fig:sfms}, Table \ref{tbl:derivedExp})}. $^{\rm b}$\,Magnification-corrected (Table \ref{tbl:sample}); uncertainties incorporate $\mu$ errors. $^{\rm c}$\,SExtractor half-light radius. $^{\rm d}$\,2\,$\sigma$ limit.}
\label{tbl:derived}
\end{deluxetable*}

\section{Spectral Synthesis Modeling}
\label{sec:models}

{\bfb We fit the full spectrophotometric SED (broadband fluxes + spectra) to infer the sample galaxies' spatially integrated and resolved properties---$\Mstel$, $\sfr$, $A_{V}$ for each optimal, \inn, \mid,\ and \out\ dataset, marginalized over all other quantities---and SFHs based on delayed exponential and lognormal parameterizations. We adopt the lognormals for our main analysis since they are smooth over all $t$, but using both allows us to assess SFH systematics (Section \ref{sec:resolved}) and check certain results (Appendices \ref{sec:AC}, \ref{sec:AD}). 

We use {\tt PYSPECFIT} for all fitting} \citep{Newman14}. This is a flexible, Bayesian python package wrapped around the {\tt multinest} Monte Carlo Markov Chain engine \citep{Feroz08}. We assume \citet[][]{BC03} templates, a \citet{Salpeter55} initial mass function, solar metallicity \citep[][]{Gallazzi05,Gallazzi14}, a \citet{Calzetti00} dust law, {\bfb and a $\chi^{2}$ likelihood function (see \citealt{Newman14}; Eq. B1)}. Appendix \ref{sec:AB} lists other fit parameters, {\bfb which include H$\alpha$, H$\beta$, [\ion{O}{3}], and [\ion{S}{2}] emission line equivalent widths}.

{\bfb The two SFH models are used in} two fitting passes. First is the delayed exponential \citep[e.g.,][]{Lee10}:
\beq
	\sfr(t) \propto \frac{t-t_{\rm start}}{\tau_{\rm exp}^{2}}\,\exp\left[-\frac{t-t_{\rm start}}{\tau_{\rm exp}}\right],
\label{eq:delayedExponential}
\eeq
where $t$ is cosmic time, $\tau_{\rm exp}$ is an $e$-folding timescale, and $t_{\rm start}$ is the beginning of the SFH. This last parameter is independent of $t$ (except that $t_{\rm start}\leq \tobs$), so the SFH can be normalized without knowing its redshift. This is not true of the lognormal SFH (below). Hence, while Appendix \ref{sec:AC} discusses $\tau_{\rm exp}$ and ``age'' ($=\tobs-t_{\rm start}$) trends, the core purpose of this fit is fine redshift estimation, an input to step two.

{\bfb Next, we fit a lognormal SFH} \citep{Gladders13b}:
\beq
	\sfr(t) \propto \frac{1}{t\,\sqrt{2\pi\tau^{2}}}\exp\left[{-\frac{(\ln t - T_{0})^{2}}{2\tau^{2}}}\right],
\label{eq:logNormal}
\eeq
where $T_{0}$ is the half-mass time (in $\ln$\,Gyr) and $\tau$ controls its width. This form has significant advantages over the delayed exponential, including a smooth derivative over all $t$---avoiding unrealistic $d\Mstel/dt$ at late epochs (Figure \ref{fig:expSfhs})---and better agreement with SFHs from simulations \citep[][]{Diemer17,Ciesla17}. Further, they reproduce many ensemble observations at $z\lesssim8$ \citep{Gladders13b, Abramson15, Abramson16, Dressler16}. Hence, unless stated otherwise, this SFH is the default source for derived quantities. That said, the lognormal's lack of discontinuities and the fact that its decline is, by definition, slower than its rise make it a poor description in some cases (\citealt{Diemer17,Iyer17}; Section \ref{sec:results} below).

Since the lognormal has $\sfr>0$ for all $t$, normalizing it to a total $\Mstel$ requires knowing a source's redshift (the upper bound to $\int\sfr\,dt$). Hence, all lognormal fitting is fixed to the redshift measured using the delayed exponential SFH.

SED fitting thus produces a redshift, and 4$\times$2 sets of $\Mstel$+$\sfr$+SFH parameters for each galaxy describing the optimal, \inn, \mid, and \out\ extractions given delayed exponential or lognormal SFHs. {\bfb Mass estimates are consistent, though SFRs may not be (Figure \ref{fig:sfms}; Section \ref{sec:resolved}).} Broad priors are placed on the SFH parameters and other outputs (e.g., emission line equivalent widths and $A_{V}$). Best-fit values are the medians of the resultant {\tt multinest} posterior distributions, with the 16th--84th percentiles as formal uncertainties. Full parameter covariances are also output and used when necessary---e.g., when reconstructing the SFHs themselves.

\section{Results}
\label{sec:results}

{\bfb In this section, after characterizing the sample's spatially integrated (Section \ref{sec:global}) and resolved (\ref{sec:resolved}) diversity at $\tobs$, we use the lognormal SFHs to chart each galaxies' development over multiple epochs (\ref{sec:longitudinal}). Projecting these systems back in time reveals many signs of inside-out (bulge-first) growth, hints of a link between high bulge mass fractions and non-constant SFHs, and a difference between longitudinal and cross-sectional size reconstructions (\ref{sec:spatResSFH}), encouraging further study of their relationship. 

All quantities are magnification-corrected and incorporate that uncertainty.}

\begin{figure}[t!]
\centering
\includegraphics[width = \linewidth, trim = 0cm 0.5cm 0cm 0cm]{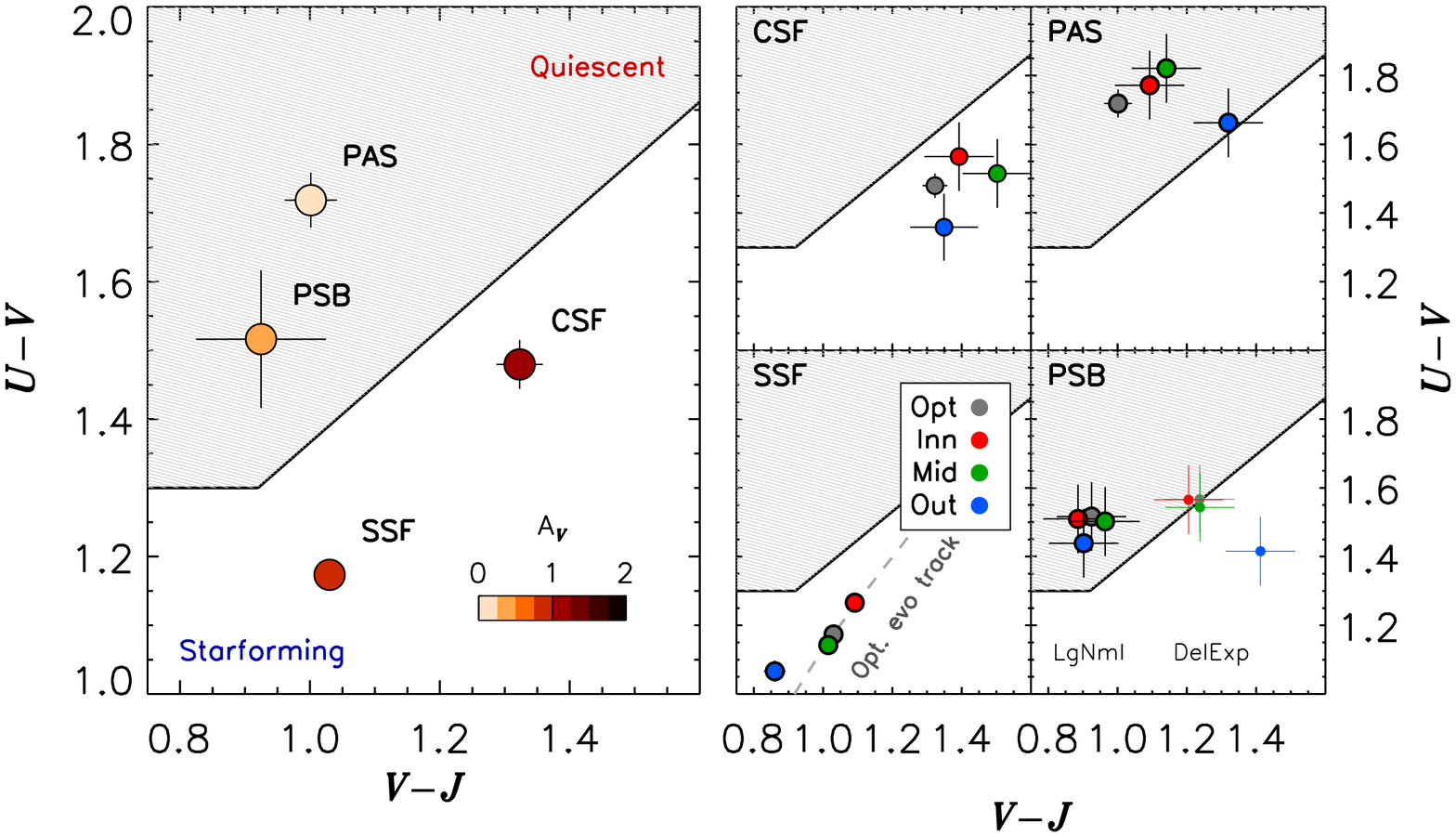}
\caption{{\it Left}: Integrated synthetic rest-{\it UVJ} colors from the best-fit SFHs coded by $A_{V}$. Consistent with their appearances (Figure \ref{fig:mosaic}), \csf\ and \ssf\ lie in the diagram's starforming region and are globally more extinguished than their quiescent counterparts (Figure \ref{fig:massProf}). {\it Right}: resolved colors from the radial spectral extractions. \csf\ and \ssf\ have $x$ or $y$ gradients such that \inn\ regions are redder than \out\ ones. The quiescent \pas\ and \psb\ show opposite/no gradients. Notably, \ssf's pieces lie roughly along the trajectory of its integrated SED fit, suggesting larger-$r$ regions developmentally lag \inn\ ones (``inside-out growth''; Section \ref{sec:spatResSFH}). {\bfb As we lack resolved rest-$J$ coverage, \psb's truncated history may drive SFH systematics to which the inferred colors are sensitive (Section \ref{sec:resolved}).}}
\label{fig:uvj}
\end{figure}

\begin{figure}[t!]
\centering
\includegraphics[width = 0.9\linewidth, trim = 0.5cm 0.25cm 0cm 0cm]{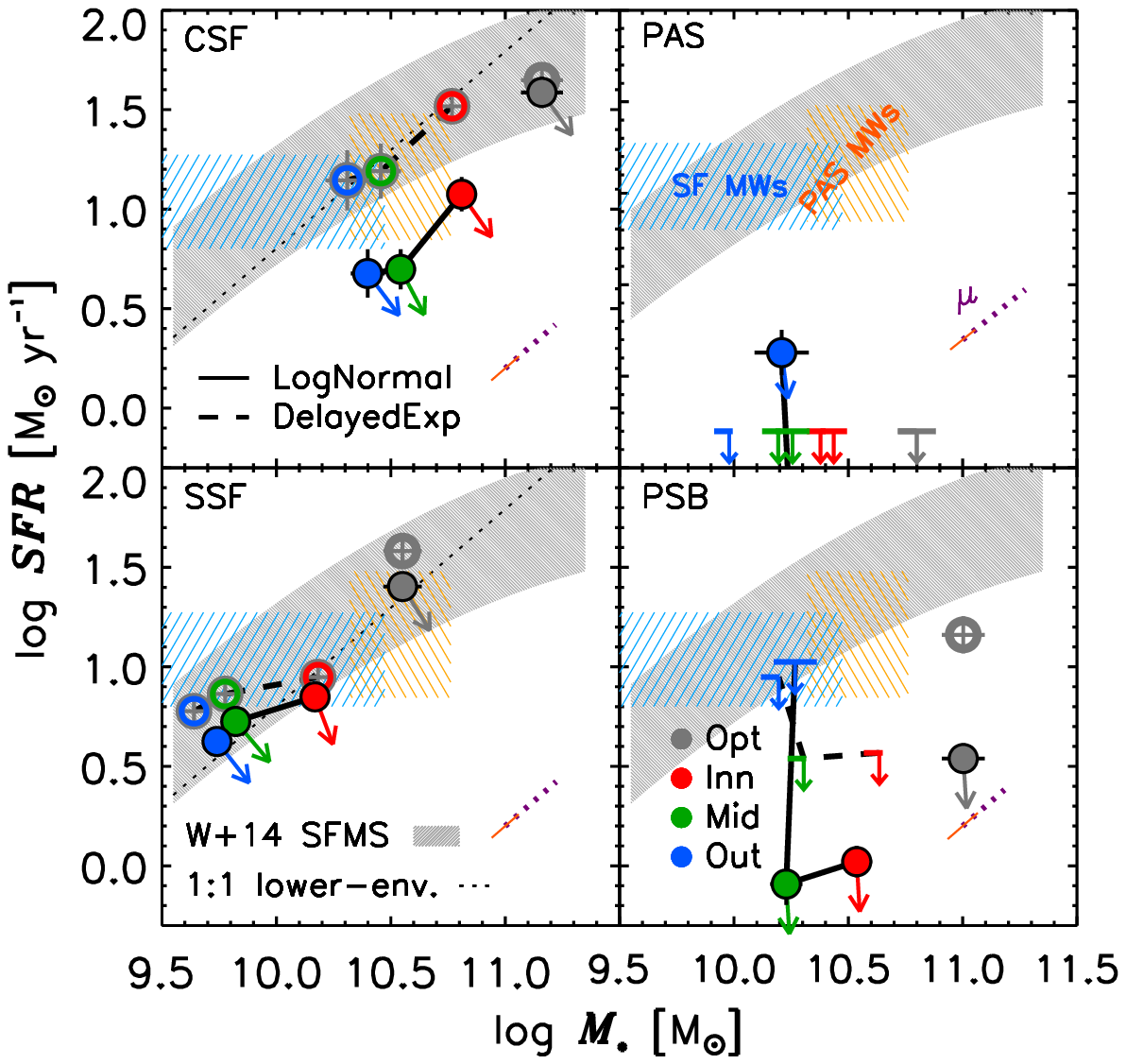}
\caption{The spatially resolved $\sfr$--$\Mstel$ plane. Starforming (passive) systems are plotted at {\it left} ({\it right}). Solid symbols/filled lines show lognormal results; open symbols/dashed lines delayed exponentials. Bars denote 2\,$\sigma$ upper limits; arrows $(\Mstel,\sfr)(t)$ unit vectors averaged over $\tobs\pm1$\,Gyr. Grey shading shows the $z\sim1.25$ SFMS \citep{Whitaker14}; blue/orange blocks locate progenitors of $z=0$ MW-mass starforming and passive galaxies \citep{Gladders13b}. Resolved and integrated metrics (colored/grey circles, \resp) support similar inferences with telling exceptions: The SFMS' bend allows galaxies to be on the locus while their components are not (\csf), suggesting SFMS-based inferences can yield biased insights into SF physics. {\bfb Dotted purple/thin orange lines show magnification effects/uncertainties.}}
\label{fig:sfms}
\end{figure}

\subsection{Integrated Properties at $\tobs$}
\label{sec:global}

\subsubsection{Masses and Star Formation Classes}
\label{sec:massClass}
{\bfb
In terms of monolithic properties (Table \ref{tbl:derived}), the sample spans $10.5\lesssim\log\Mstel\lesssim11.2$ and a range of SF states---$0.1\lesssim\sfr/\Msun\,{\rm yr^{-1}}\lesssim40$; i.e., likely progenitors of $z=0$ Milky Way (MW) to $\geq$M31-mass galaxies (see below).

Figure \ref{fig:mosaic} shows the sample's integrated spectrophotometry. Notable H$\alpha$ and UV flux combined with prominent disks/blue spiral arms reveal IDs 00451\_2  and 00900\_1 to be starforming. The other two sources---IDs 00660\_2 and 01916\_2---show either no or much less H$\alpha$/UV flux, which---combined with their spheroidal, redder appearance---reveal them to be passive. The sources' positions in the {\it UVJ} (Figure \ref{fig:uvj}, {\it left}; \citealt{Williams09}) and $\sfr$--$\Mstel$ planes (Figure \ref{fig:sfms}) reflect these statements: 00451\_2  and 00900\_1 lie outside the quiescent box and on the $z\sim1.25$ SF ``Main Sequence'' \citep[SFMS;][]{Noeske07,Whitaker14}; 00660\_2 and 01916\_2 lie in the box and off the SFMS.

While evenly split between starforming and passive galaxies, Figure \ref{fig:mosaic}'s spectra expose more sample diversity. First, despite being ``normal'' SFMS galaxies, 00451\_2  and 00900\_1 are quite distinct: the former is redder and has a shallower UV slope than the latter, suggesting it is more mature/farther from its peak SFR (Section \ref{sec:longitudinal}). This is consistent with 00451\_2's $\sim$4$\times$ greater mass, reduced EW(H$\alpha$+[\ion{N}{2}]) ($\sim$$\ssfr$), and the fact that its resolved components actually lie {\it off} the SFMS (Section \ref{sec:resolved}; the locus' bend keeps it defined as ``starforming''). Thus, we refer to these starforming systems as:

\bitem
	\item 00451\_2 $\equiv$ ``CSF'' --- {\it Continuously Starforming};
	\item 00900\_1 $\equiv$ ``SSF'' --- {\it Strongly Starforming}.
\eitem	
Notable [\ion{O}{3}] emission suggests \csf\ hosts an AGN, confirmed by its resolved spectra and mass--excitation diagram (\citealt{Juneau14}; Figures \ref{fig:spectra}, \ref{fig:AGN}).

The two passive galaxies are also distinct: 00660\_2 is unquestionably old, with a red continuum that rises from the essentially absent UV to [\ion{O}{3}], and strong G-band absorption. Meanwhile, 01916\_2 has a bluer, A star-like spectrum (similar continuum near [\ion{O}{3}] and H$\gamma$; weak G-Band; Figures \ref{fig:uvj}, \ref{fig:highResCheck}), and higher UV flux. Combined with residual H$\alpha$+[\ion{N}{2}] emission and the photometry's preference for the more rapidly decaying delayed exponential SFH (Section \ref{sec:resolved}), this suggests 01916\_2 was rapidly quenched within $\sim$\,1 Gyr of $\tobs$ (Table \ref{tbl:derived}; \citealt{DresslerGunn83, Poggianti99}). We therefore refer to these quiescent systems as:
\bitem
	\item 00660\_2 $\equiv$ ``PAS'' --- {\it Passive}; 
	\item 01916\_2 $\equiv$ ``PSB'' --- {\it Post Starburst}. 
\eitem	

From Figure \ref{fig:sfms}, \ssf\ and \pas\ appear set to become $\sim$MW-mass $z=0$ E/S0s, while \csf\ and \psb\ are modern $\geq$M31-mass progenitors (see also Section \ref{sec:G13tests}). Given their aforementioned spectral diversity, dissecting these galaxies in space and time  thus provides insights relevant to---if not representative of---important parts of parameter space.

Note: Given their consistency, we occasionally plot $\Mstel$ averaged over SFH choice (Figure \ref{fig:sizeMass}). However, by default, SFRs reflect lognormal inferences. This (1) ensures self-consistency with evolutionary projections (Section \ref{sec:longitudinal}); (2) captures representative mass growth rates (which presumably change on timescales longer than the $\sim$10$^{7}$\,yr on which \ion{H}{2} regions evolve); (3) avoids line flux and dust correction biases due to H$\alpha$+[\ion{N}{2}] and H$\beta$+[\ion{O}{3}] blending (but see \citealt{Wang17}); and (4) reduces ambiguities associated with active galactic nuclei (AGN), present in \csf. Nevertheless, using dust-uncorrected EW(H$\alpha$+[\ion{N}{2}]) as an $\ssfr$ proxy yields similar trends to those inferred from full SED fitting (Appendix \ref{sec:AD}, Figure \ref{fig:ewHa}).

}
\subsubsection{Size and Structure}
\label{sec:grossProperties}

\begin{figure}[t!]
\centering
\includegraphics[width = \linewidth, trim = 0.5cm 1.25cm 0cm 0.5cm]{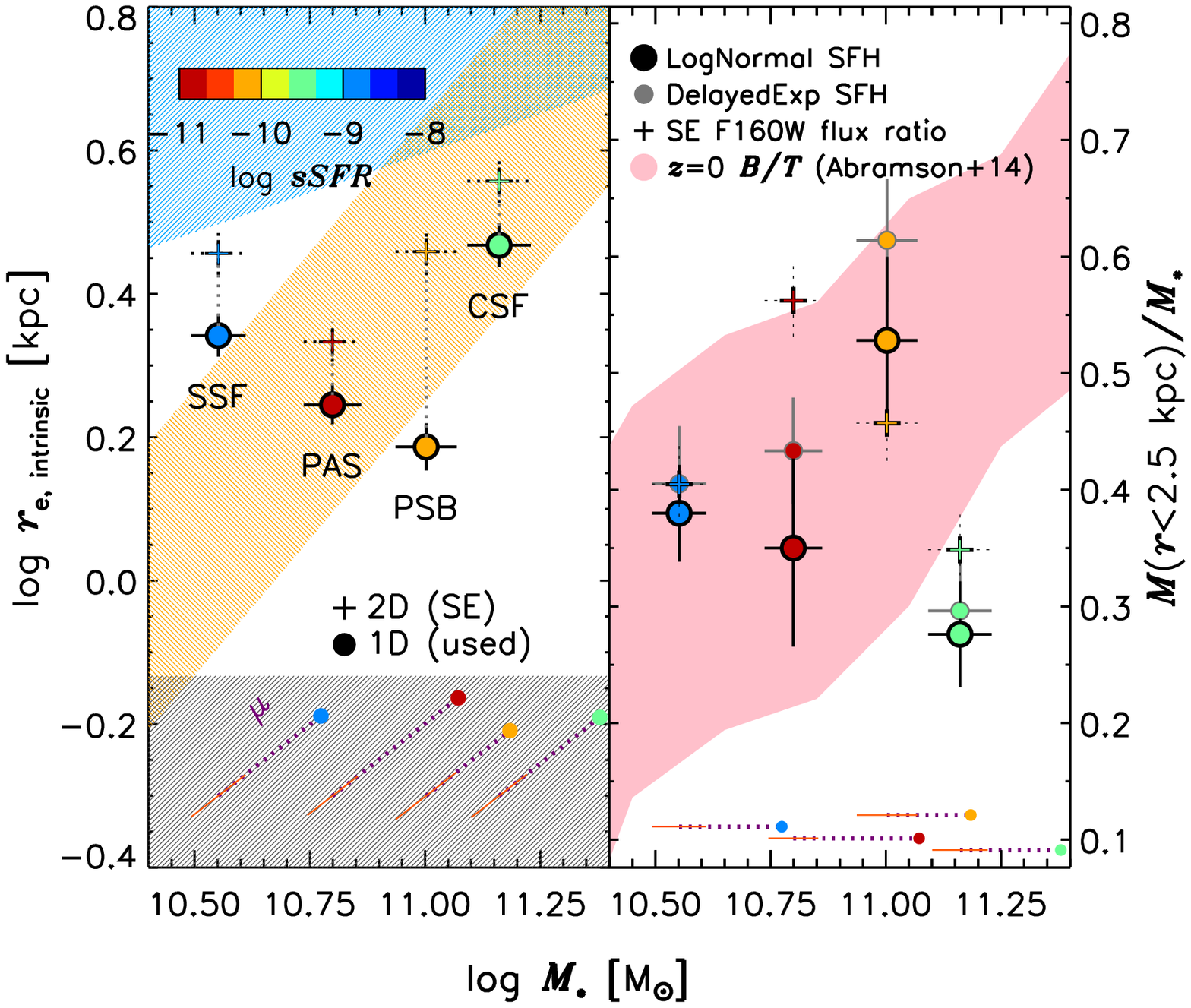}
\caption{{\it Left}: the size--mass plane. Points are {\bfb color-}coded by $\ssfr$ with circles and crosses showing 1D $r_{e}$ used in this analysis and 2D SExtractor half-light radii, \resp. Blue and orange shading shows \citet{vanderWel14}'s starforming and passive $z=1.25$ relations, with the HST resolution limit in black. {\it Right}: the $\Mstel(r<2.5\,\kpc)$ fraction ($\sim$$B/T$)--mass plane. Black/grey circles show lognormal/delayed exponential results, with \citet{Abramson14a}'s $z=0$, 25\%--75\% range shaded pink ({\bfb a good description at all} $z\lesssim2$; \citealt{Lang14}). {\bfb Each galaxy is plotted at the mean of its two (consistent) total $\Mstel$ estimates (Figure \ref{fig:sfms}, Table \ref{tbl:derivedExp}).} All galaxies have normal $B/T$ for their mass except \csf, which is unexpectedly disk-dominated.}
\label{fig:sizeMass}
\end{figure}

Figure \ref{fig:sizeMass}, {\it left}, shows the sample on the size--mass plane. Considering their 2D half-light radii ($\sim$2.1--3.6\,kpc), the passive \pas\ and \psb\ lie where they are expected to, while the starforming galaxies are small for their mass \citep[][though note we plot SExtractor and not GALFIT estimates]{vanderWel14}. {\bfb This could be due to chance or our selection criteria, as small galaxies' narrow spectra are less likely to be contaminated (Section \ref{sec:selection}).} Trends hold in 1D ($r_{e}\sim$1.5--3.0\,kpc) except \psb\ shrinks notably (see also Figure \ref{fig:spectra}).

Structure should also encode important evolutionary physics. Figure \ref{fig:sizeMass}, {\it right}, shows the sample's mass--mass-concentration---$\Mstel(r\leq 2.5\,\kpc)/\Mstel$---relation. The systems span $\sim$30--60\% in this quantity, which---to the extent that it reflects bulge-to-total ($B/T$) ratios---places most on the $z=0$--2 expectation \citep[][latter plotted]{Lang14, Abramson14a}. \csf\ is, however, in the lower quartile for its mass, which may be revealing (Sections \ref{sec:spatResSFH}, \ref{sec:discussionPhysics}).

We calculate mass concentrations by interpolating the stellar mass surface density in each radial zone ($\Sigma_{\Mstel}(r)$) onto a finer grid, {\bfb assuming circular} symmetry, and integrating to $2\,r_{e}$, e.g., $M_{*,\rm int}(<r)=2\pi\int_{0}^{r} r'\,\Sigma_{\Mstel}(r')\,dr'$. We force the outermost $r=2\,r_{e}$ bin to match the optimal extraction $\Mstel$ estimates (Table \ref{tbl:derived}). While not formally self-consistent, this can be repeated as we evolve the galaxies to different epochs to estimate half-mass radii and ``$B/T$'' {\bfb trajectories} in a way comparable to cross-sectional studies \citep{Morishita15}. Before re-normalizing, we obtain consistent total masses for two systems (\csf, \pas) with the others $\sim$0.3--0.4\,dex low, {\bfb perhaps due to their clumpier structure (Figure \ref{fig:mosaic})}. SFH systematics are $\Delta B/T\sim$0.1, comparable to offsets assuming $B/T$ is the ratio of the F160W flux at $r<2.5$\,kpc to SExtractor's {\tt FLUX\_AUTO}. As such, Figure \ref{fig:sizeMass}, {\it right}, is most robustly interpreted to mean that these galaxies have normal bulge fractions for their mass {\it except} \csf, which is unexpectedly disky.

\subsection{Spatially Resolved Properties at $\tobs$: \\Colors, Mass and SFR Densities, and Dust}
\label{sec:resolved}

Figure \ref{fig:uvj}, {\it right}, shows color gradients in $U-V$, $V-J$, or both in the starforming systems \csf\ and \ssf, such that \inn\ regions are at least redder than \out\ ones. These trends are due to each zone's intrinsic stellar populations, not dust (Figure \ref{fig:massProf}, {\it bottom}). This is expected (bulges are old) but it is interesting that \ssf's zones fall roughly along the time trajectory of a single lognormal SFH (dashed line); i.e., the \mid\ and \out\ regions simply lag \inn\ by some delay. We revisit this below and in Section \ref{sec:resSFH}.

The passive systems---\pas, \psb---are either uniformly red, or have gradients in the opposite sense, reddening with increasing $r$ at least in $V-J$. This may be due to dust (Figure \ref{fig:massProf}, {\it bottom}), or may reflect SFH systematics. These are significant in \psb, where the lognormal and delayed exponential models disagree at rest-$J$. Because delayed exponentials can fall faster than the lognormal, better mimicking what in reality may be a discontinuity, these fit \psb's UV photometry better at all $r$ (Figure \ref{fig:spectra}). Regardless, the exponential UV fit quality and the lognormal color inferences agree in implying that whatever quenched \psb\ acted quasi-globally, such that much of the galaxy shut down at about the same time.

\begin{figure}[t!]
\centering
\includegraphics[width = 0.825\linewidth, trim = 1.5cm 0cm 0cm 0cm]{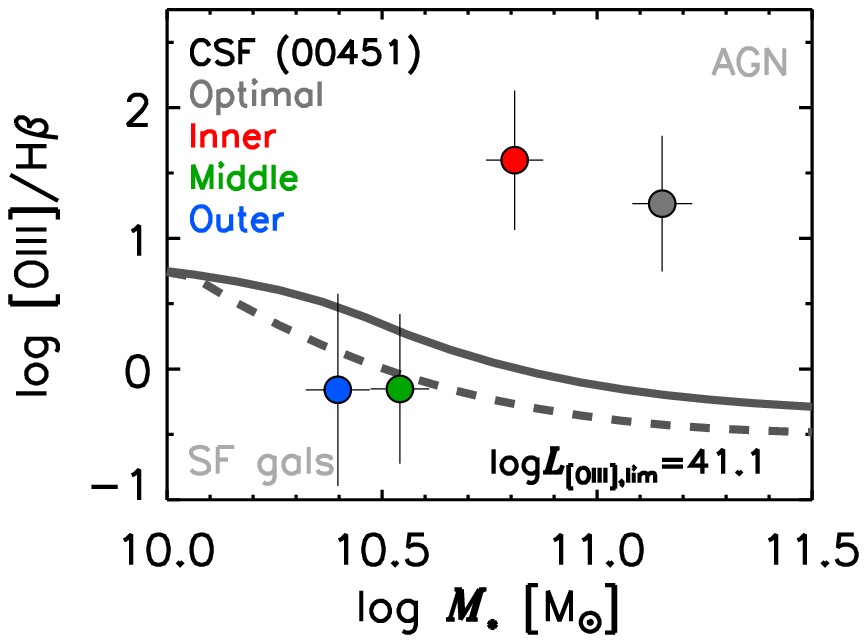}
\caption{\csf's resolved mass--excitation diagram \citep{Juneau14} at GLASS' $2\,\sigma$ $F_{\rm lim}\simeq2\times10^{-17}\,{\rm erg\,s^{-1}\,cm^{-2}}$ \citep[][i.e., $\log(L_{\rm[OIII]}/{\rm erg\,s^{-1}})_{\rm lim}=41.1$]{Schmidt14}. \inn/optimal estimates {\bfb imply AGN activity (above the solid line). Meanwhile, \mid\ and \out\ regions lie with normal starforming galaxies (below the dashed line). These maintain a combined $\sfr\sim$10--25\,$\Msun\,\yr^{-1}$, apparently unquenched by the AGN (Figure \ref{fig:sfms}).}}
\label{fig:AGN}
\end{figure}

{\bfb
While it confirms the global nature of \psb's SFH truncation, Figure \ref{fig:spectra}'s radially resolved spectra conversely reveal the fact that the \mid\ and \out\ regions of \csf\ show no signs of the AGN activity inferred from its integrated data (Figure \ref{fig:mosaic}; see also Figure \ref{fig:AGN}). Further, their $\sfr$s remain $\sim$3--6\,$\Msun\,\yr^{-1}$, \resp.}

That said, Figure \ref{fig:sfms} does suggest that these $\sfr$s are below {\bfb average} for objects with \csf's \inn, \mid, or \out\ regions' masses assuming lognormal SFHs (if not delayed exponentials; see below). Indeed, these components lie 1.5--2\,$\sigma$ below the SFMS even though \csf\ {\it as a whole} appears normal. This is due to the bend in the SFMS at $\log\Mstel\gtrsim10$ \citep[e.g.,][]{Salim07,Whitaker12,Whitaker14,Abramson14a,Schreiber16} and affects the bluer \ssf\ to a lesser extent. Regardless, this---and similar but opposite phenomena in $z\approx0$ ``Green Valley'' systems \citep[][]{Dressler15,Vulcani15a}---{\bfb highlights key complications in inferring physics from the} SFMS{\bfb: It is hard to see how \csf\ could be normally starforming while its pieces are quenched. As now seen with emission lines \citep{Sanders17}, integrated inferences are not always the sum of their parts, a dichotomy requiring spatially resolved data to understand.} 

\begin{figure}[t!]
\centering
\includegraphics[width = 1\linewidth, trim = 0cm 0cm 0cm 0cm]{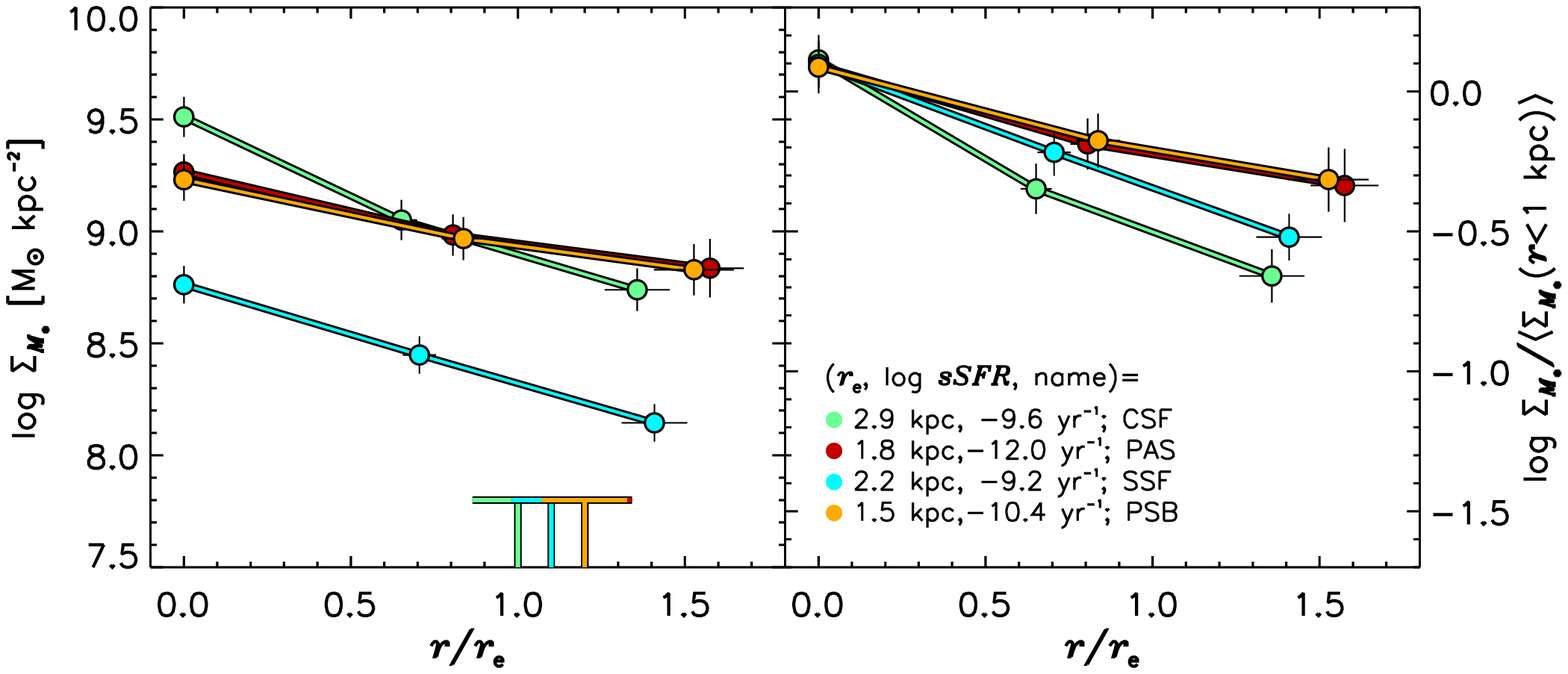}
\includegraphics[width = 1\linewidth, trim = 0cm 0cm 0cm 0cm]{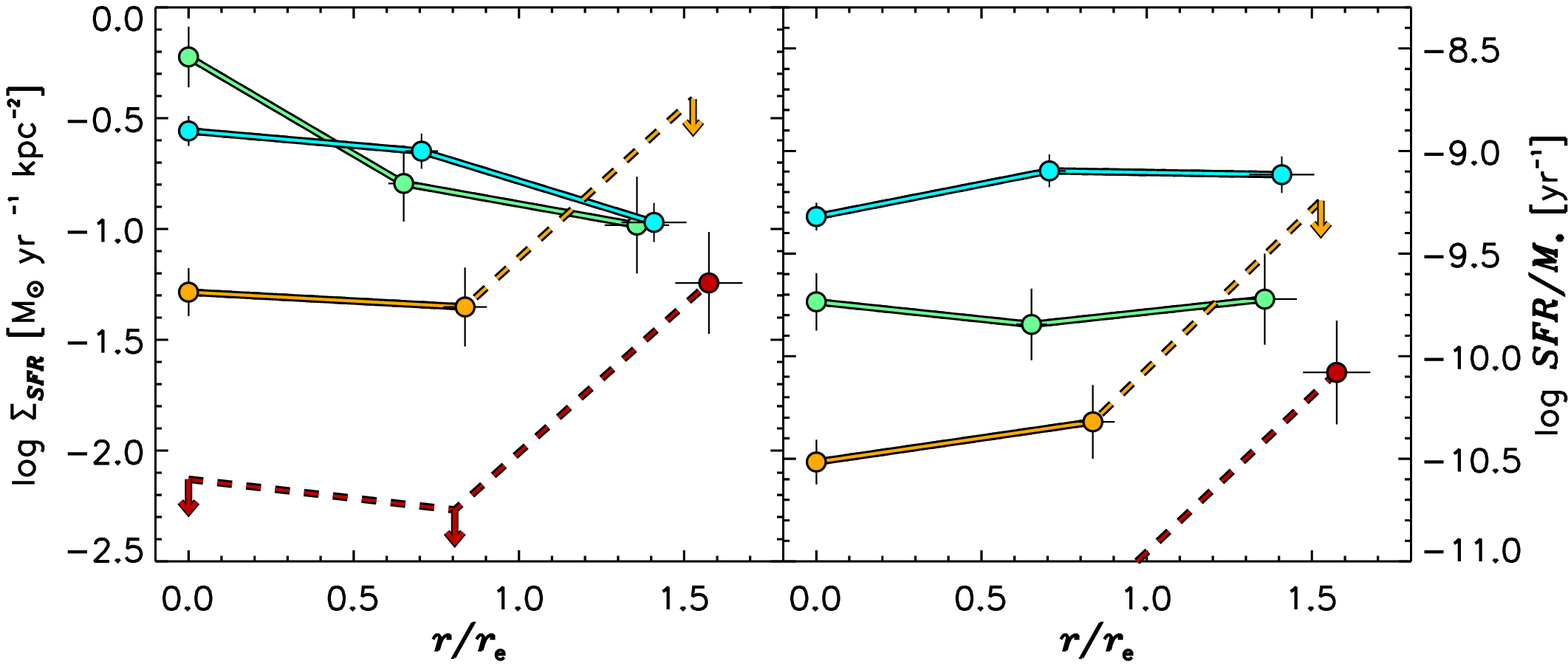}
\includegraphics[width = 1\linewidth, trim = 0cm 0cm 0cm 0cm]{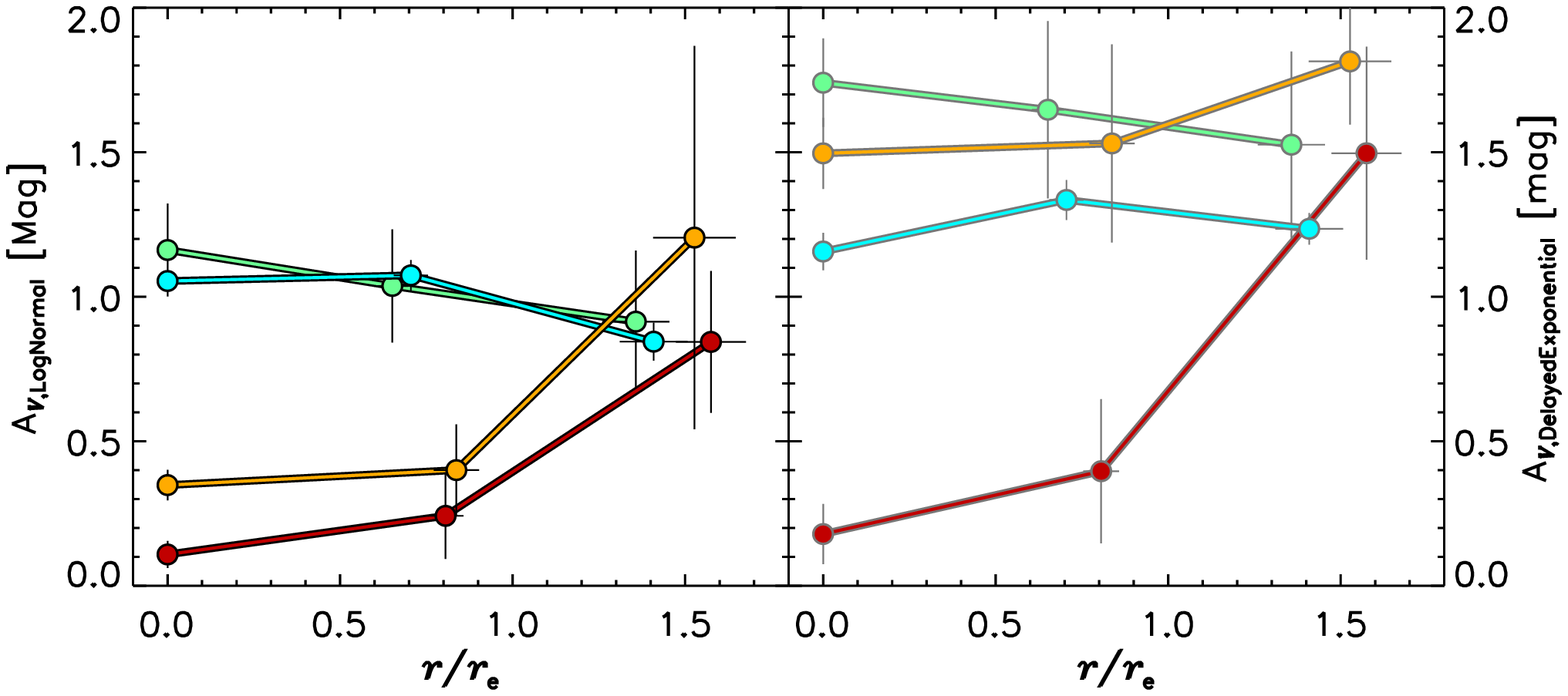}
\caption{{\it Top}: Lognormal absolute ({\it left}) and relative ({\it right}) mass surface density ($\Sigma_{\Mstel}$) profiles. The latter is normalized to the mean in each source's central kpc. {\it Middle}: Lognormal $\sfr$ surface density ($\Sigma_{\sfr}$) and $\ssfr$ profiles (areas cancel). {\it Bottom}: $A_{V}$ (dust) profiles from lognormal ({\it left}) and delayed exponential SFHs ({\it right}). While the outermost $\ssfr$s of the sample differ by $>$10$\times$, they have consistent $\Sigma_{\sfr}$ at those radii (and all $r$ for the starforming systems), favoring, e.g., gas surface density over mass as an $\sfr$ governor. Note also the significant systematic offsets in the $A_{V}$ inferred using the two SFHs. This may point to physically meaningful, testable differences between the two SFHs (Section \ref{sec:resolved}). {\bfb Bin $x$-offsets reflect pixelation effects.}}
\label{fig:massProf}
\end{figure}

Figure \ref{fig:massProf} quantifies further radially resolved sample trends. The top row shows the galaxies' surface mass density profiles: $\Sigma_{\Mstel}(r)\equiv\Mstel(r)/{\rm Area}(r)$, where Area$(r)$ is based on the number of $0\farcs065\times0\farcs065$ pixels in the SExtractor segmentation map in each radial zone. The {\it left} panel shows $\Sigma_{\Mstel}$ in absolute units, the {\it right} normalized to inner 1\,kpc means. Interestingly, the starforming \csf\ and \ssf\ have {\it steeper} $\Sigma_{\Mstel}$ profiles compared to their passive counterparts (though \ssf\ is the least dense system). This remains true, if less significant, when viewed in terms of physical radii instead of $r/r_{e}$ (Figure \ref{fig:ewHa}) and suggests either that (1) the passive systems have already undergone some amount of, e.g., minor-merger-driven envelope building \citep[e.g.,][]{Newman12a, Nipoti12, Morishita16}, or (2) the starforming systems are growing ``inside-out,'' with their disks following their bulges \citep[e.g.,][]{Eggen62, Silk81, Kepner99, Tacchella15}.

{\bfb Figure \ref{fig:massProf}, {\it middle}, provides some leverage on the latter issue.} Instead of stellar mass, here we plot $\sfr$ surface density---$\Sigma_{\sfr}(r)\equiv\sfr(r)/{\rm Area}(r)$ ({\it left})---and $\ssfr$ gradients ({\it right}). \csf\ exhibits a fairly constant $\ssfr(r)$, while \ssf's rises by $\sim$40\%--80\% with increasing $r$; i.e., its center is perhaps half as active as its disk \citep[see also][]{Nelson16,Abdurrouf17}. This is consistent with \ssf's {\it UVJ} color gradient suggesting its bulge is relatively developmentally advanced. However, this is not true of \csf, whose star formation appears more uniform{\bfb, suggesting its higher mass was built up more gradually at all $r$ (Section \ref{sec:spatResSFH})}. Hence, these starforming galaxies had quite different pasts. 

Nonetheless, \csf\ and \ssf\ do share an interesting trait: While their \out\ $\ssfr$s differ by $\gtrsim$3$\times$, they correspond to precisely the same SFR density---$\Sigma_{\sfr}/\Msun\,\kpc^{-2}\simeq0.1$---and indeed nearly identical SFRs (Figure \ref{fig:sfms}). 
Taken together, Figure \ref{fig:massProf}'s middle panels imply that quantities beyond stellar mass are important to characterizing or controlling star formation in starforming galaxies: \csf's relatively heavy disk supports the same {\it absolute} level of star formation as \ssf's lighter-yet-similarly sized disk. Without resolved data, this phenomenon would not have been detected.

This signal---where $\ssfr$ but not $\Sigma_{\sfr}$ declines at fixed-$r$ with increasing mass---is not inconsistent with \citet[][]{Nelson16}'s {\bfb stacked HST} H$\alpha$ map {\bfb results} (see their Figures~12, 13), though sample size prevents us from saying more.

Finally, Figure \ref{fig:massProf}, {\it bottom}, shows lognormal- and delayed-exponential-inferred $A_{V}$ trends ({\it left}/{\it right}, \resp). At {\it left}, the globally higher dust content inferred from \csf's and \ssf's integrated extractions hold at all $r$ probed, in qualitative agreement with gas-phase Balmer decriment estimates by \citealt{Nelson16b}. Yet, using either SFH, we infer that \pas, while clear of dust at $r\lesssim r_{e}$, has comparable $A_{V}\sim1$\,mag to the starforming galaxies at $\sim$1.5\,$r_{e}$. {\bfb This trend is opposite to that inferred for a more massive $z\sim2.15$ system by \citet{Toft17},} but could be a hint of ancient spiral arms and thus a sign of inside-out growth/quenching (as those authors' $\ssfr$ inference suggest), or conversely the contents of recently accreted satellites. Both scenarios are supported by the delayed exponential-derived age gradients, where \pas' \out\ region appears $\sim$1.5\,Gyr younger than its \inn\ one (Figure \ref{fig:expSfhs}, {\it top}; also similar to \citealt{Toft17}'s findings).

\begin{deluxetable}{cccc}
	\tabletypesize{\footnotesize}
	\tablewidth{0pt}
	\tablecolumns{4}
	\tablecaption{ID 01916\_2 GLASS/MOSFIRE Spectral Index Comparisons}
	\tablehead{
	\colhead{Feature\tablenotemark{a}} & 
	\colhead{EW$_{\rm GLASS\,inf}^{\rm Del.\ Exp.}$ [\AA] } & 
	\colhead{EW$_{\rm GLASS\,inf}^{\rm Lognormal}$ [\AA] } & 
	\colhead{EW$_{\rm MOSFIRE}$ [\AA]\tablenotemark{b}}
}
\startdata
	H$\delta_A$ 	& 9.3 & $7.4$ & $5.3\pm1.1$\\
	H$\delta_F$ 	& 6.4 & $5.1$ & $4.7\pm0.8$\\
	H$\gamma_A$ 	& 6.2 & $4.8$ & $1.2\pm1.0$\\
	H$\gamma_F$ 	& 4.8 & $4.5$ & $2.7\pm0.7$\\
	G4300 		& -0.5 & $1.2$ & $2.7\pm1.1$
\enddata
\tablecomments{$^{\rm a}$\,Lick IDS definition (\url{http://astro.wsu.edu/worthey/html/index.table.html}). $^{\rm b}$\,Bootstrapped uncertainties. Spectra in Figure \ref{fig:highResCheck}.}
\label{tbl:mosGlass}
\end{deluxetable}

The remaining system---\psb---{\bfb presents other challenges}. The lognormal fits imply a centrally clean $A_{V}$ trend similar to the older \pas. However, the delayed exponentials imply a flatter, $\sim$1\,mag more extinguished profile similar to the starforming systems'. Indeed, for all but \pas, delayed exponential $A_{V}$ are systematically higher than lognormal inferences. 

{\bfb This has to do with the geometry of the SFHs, specifically the exponential's delay: By definition, the model uses less than the maximal amount of time to produce the $\Mstel$ required to match a system's observed flux. As such, the delayed exponentials' {\it mean} $\sfr(t)$ are, by construction, higher than the lognormals', which start at $t=0$. This naturally leads to higher $\sfr(\tobs)$ (Figure \ref{fig:sfms}). Of course, despite the requisitely high SFRs, the delayed exponentials cannot simultaneously overshoot the UV photometry (Figure \ref{fig:spectra}). To ensure this, the fitter does what it can to diminish UV but not OIR flux: add dust, leading} to the offsets in Figure \ref{fig:massProf}, {\it bottom-right} compared to {\it bottom-left} for the systems with even trace star formation.

As to which model is {\bfb ``right'' for \psb, while the delayed exponential fits its rest-UV photometry better than the lognormal, the mathematics of the latter prevent it from decaying rapidly enough to mimic what in reality may be a truncation of a more leisurely falling SFH. Hence, UV fluxes may not provide the answer. {\bfb Also, while [\ion{O}{3}]/H$\beta$ ratios can be constrained using the lines' covariance, the H$\beta$ fluxes themselves are not robust enough to support dust-corrected nebular $\sfr$s as a cross-check (the above SED-derived $A_{V}$--$\sfr$ degeneracy precludes, e.g., \citealt{Wuyts13}'s continuum-based corrections). As the models disagree on rest-$J$ fluxes, incorporating {\it Spitzer} photometry may be discriminating at a spatially unresolved level, but we can turn to another sensitive test now.}}

\begin{figure}[t!]
\centering
\includegraphics[width = \linewidth, trim = 0cm 0cm 0cm 0cm]{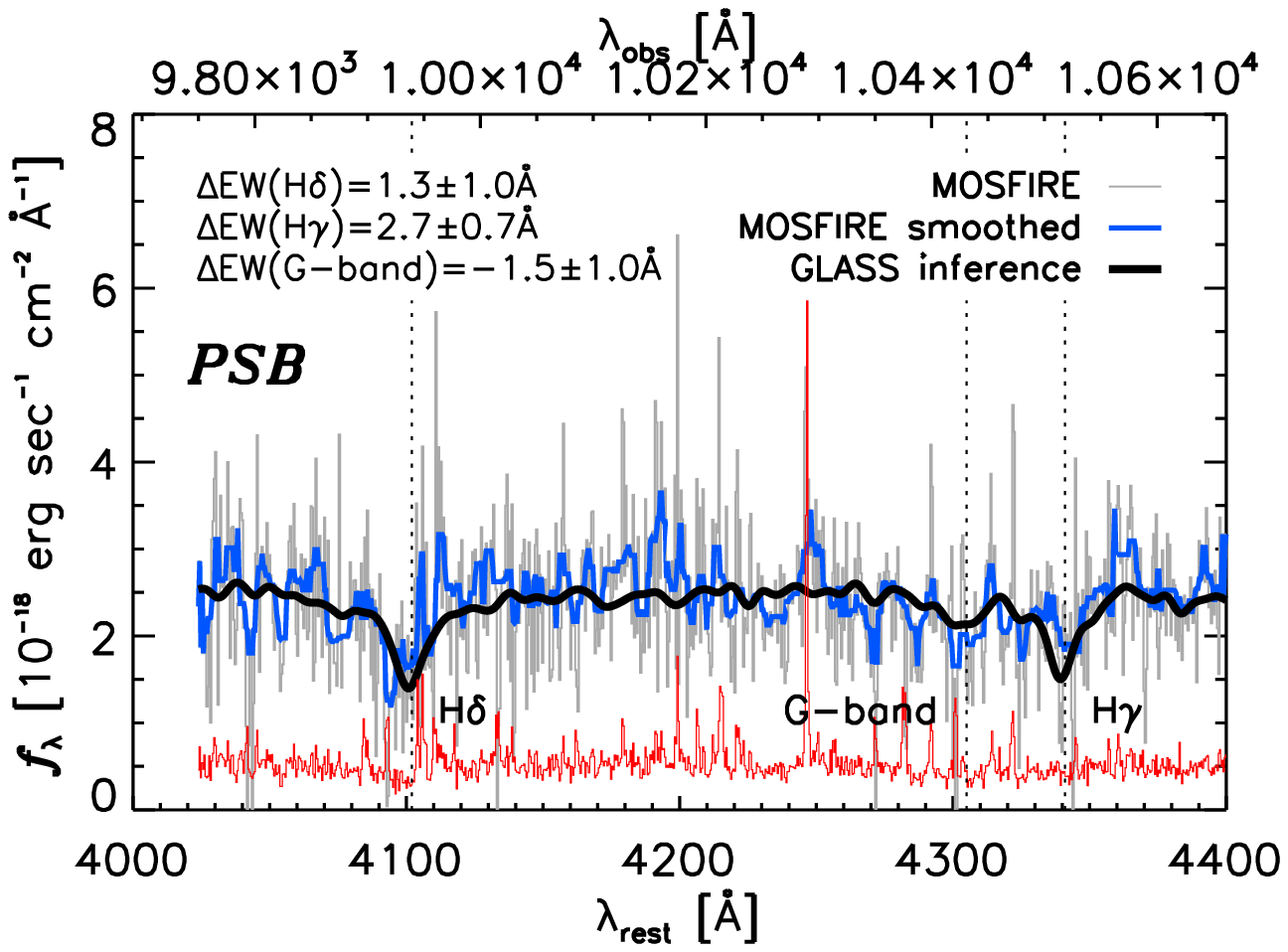}
\caption{\psb's optimal lognormal model spectrum (black; broadened to $\sigma_{v} = 190\pm40\,\kms$) compared to MOSFIRE data (grey). RMS errors are in red; blue shows the data median-smoothed over a 10\,\AA\ window. Given the implied line strengths (Table \ref{tbl:mosGlass}), the lognormal SFH reproduces these high-spectral-resolution data somewhat better than the delayed exponential, though the latter better match \psb's UV fluxes (Figure \ref{fig:spectra}).}
\label{fig:highResCheck}
\end{figure}

{\bfb Independently, a MOSFIRE $Y$-band spectrum (PI Brada\v{c}; 27 April 2017) was obtained for \psb\ as a slitmask filler} covering rest $\lambda=4000$--4400\,\AA, a region containing three key age-sensitive spectral features: H$\delta$, H$\gamma$, and the G-band. Figure \ref{fig:highResCheck} shows that this spectrum's much higher-resolution features are indeed well predicted by the fit to the low resolution spectrophotometry. Quantitatively, comparing these data to the delayed exponential and lognormal SFH-inferred templates yields somewhat better agreement in these lines' equivalent widths with the lognormal (Table \ref{tbl:mosGlass}). Indeed, the observed Balmer EWs are smaller---and G-band larger---than either model would predict, suggesting an older stellar population (though H$\gamma$ especially may suffer infilling from residual emission). {\bfb Note: since EWs are flux ratios, absent radical asymmetries, they are insensitive to slit losses.} 

{\bfb Notwithstanding the abrupt growth implied by the delayed exponentials (Figure \ref{fig:expSfhs}), the above is} spectroscopic evidence that {\bfb such a model may} not be a better description of the actual {\bfb history} of \psb\ than a lognormal, even if the latter cannot fall rapidly enough to match the system's observed rest-UV flux. For these reasons---and to further the goal of {\bfb illustrating} what analyses {\it could} be done with resolved SFHs covering $>$Gyr timescales---we limit all following discussion to the spatially resolved lognormals.

\begin{figure*}[t!]
\centering
\includegraphics[width = \linewidth, trim = 0.25cm 0.25cm -0.25cm 0cm]{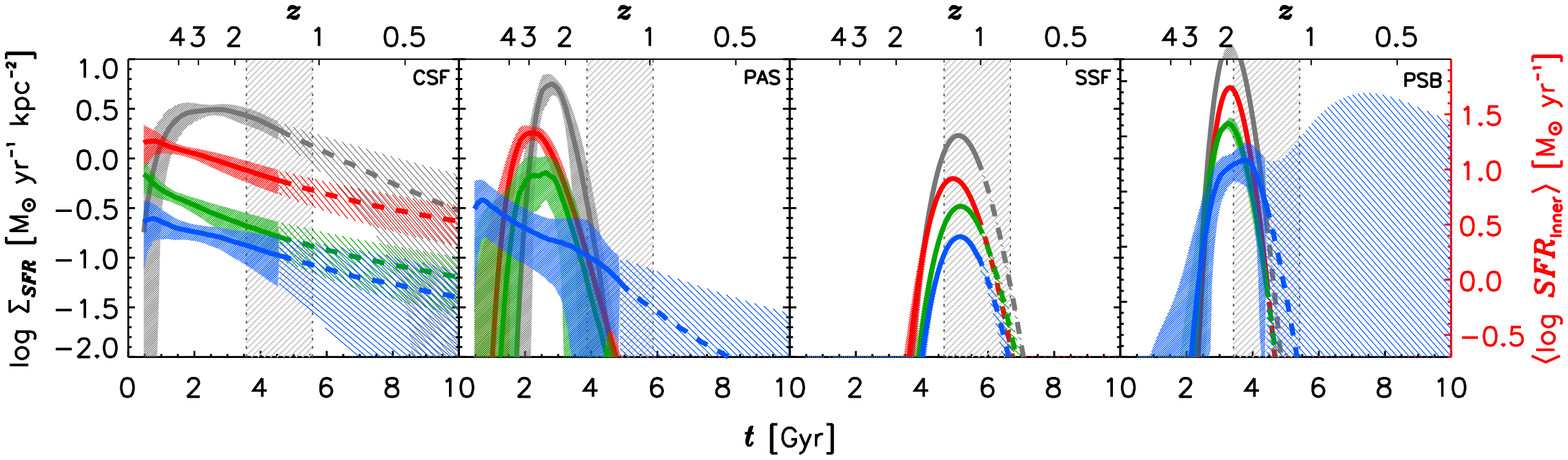}
\caption{The sample's integrated and resolved SFHs. Colors correspond to previous figures with optimal extraction results (scaled to \inn\ surface areas) in grey. Epochs prior to $\tobs$ are shown as solid lines/heavy shading; forecasts are dashes/light shading. {\bfb Vertical bands show the $\tobs\pm1$\,Gyr interval over which we draw longitudinal inferences (Section \ref{sec:spatResSFH}, Figure \ref{fig:evostuff}).} Lines are medians; envelopes show 16th--84th pctle.\ spreads in $\sfr(t)$ from the $(T_{0},\tau)$ {\tt PYSPECFIT} covariances. All galaxies are seen after their peak, and all but \csf\ show signs of inside-out/bulge-first growth: \inn\ SFRs approach/are exceeded by \mid\ and \out\ SFRs between $t_{\rm peak}$ and $\tobs$. Conversely, the relatively bulge-free \csf\ (Figure \ref{fig:sizeMass}) had a nearly constant SFH across its face. At least the inner $\sim r_{e}$ of \psb\ shut down rapidly and uniformly less than a Gyr before $\tobs$ (Table \ref{tbl:derived}). \ssf\ is growing strong---proportionately more so at larger-$r$ (Figures \ref{fig:uvj}, \ref{fig:sfms})---and \pas\ has been quenching for $\sim$twice as long as \psb. Right $y$-axis shows absolute $\sfr_{\rm INNER}$ at the sample's mean area and magnification for reference.}
\label{fig:sfhs}
\end{figure*}

\subsection{Temporally Resolved Sample Properties: \\Data ``Longitudinalization''} 
\label{sec:longitudinal}

{\bfb Section \ref{sec:resolved} resolved the galaxies in space, taking them from point- $(\Mstel,\sfr, A_{V})$ to distribution-like descriptions $(\Sigma_{\Mstel}, \Sigma_{\sfr}, A_{V})(r)$. This showed that systems with different $\ssfr(r)$ can have similar $\sfr$ surface densities, galaxies can lie on the SFMS while their resolved components do not, and significant $\sfr$s can persist at large-$r$ despite AGN activity.

These facts increase the burden on models aiming to describe these systems at $\tobs$. However, the problem of validating those models' {\it causal} statements remains open. Addressing this requires empirical inferences about how these specific systems---not the galaxy population---evolved. We present a suite of such inferences here, resolving the sample in time to constrain what it most likely {\it did} and {\it will} look like.}

\begin{figure*}[t!]
\centering
\includegraphics[width = 0.75\linewidth, trim = 0cm 0cm 0cm 0cm]{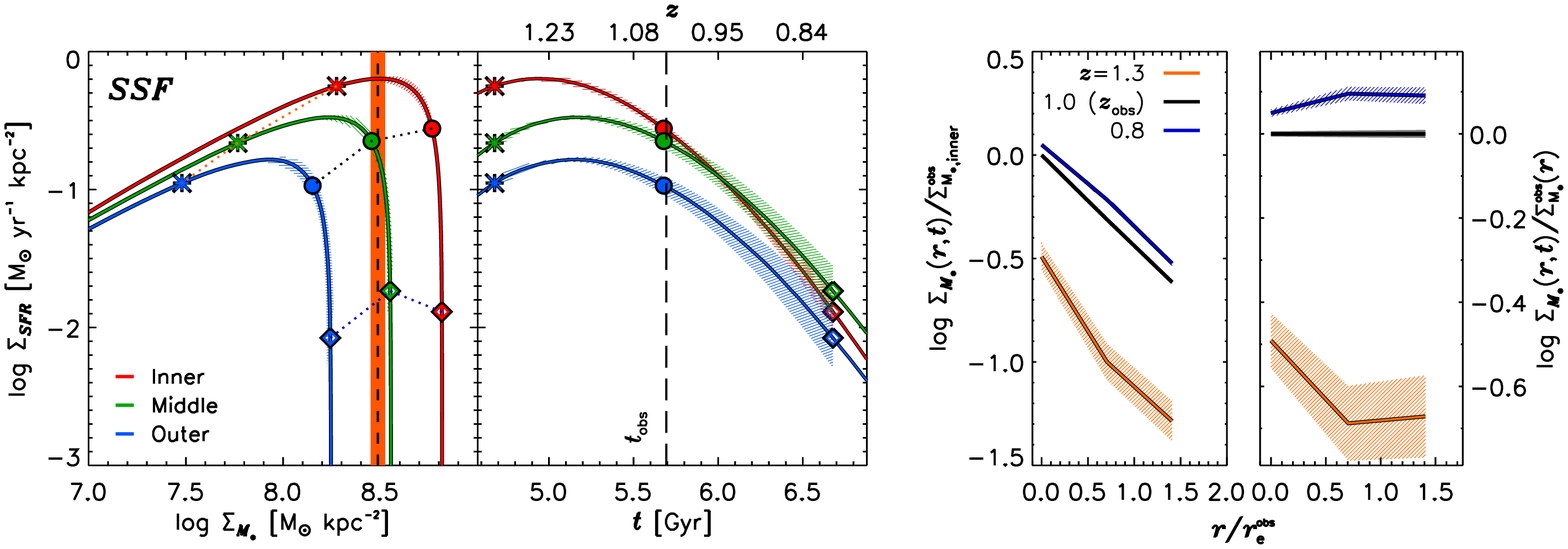}
\includegraphics[width = 0.75\linewidth, trim = 0cm 0cm 0cm 0cm]{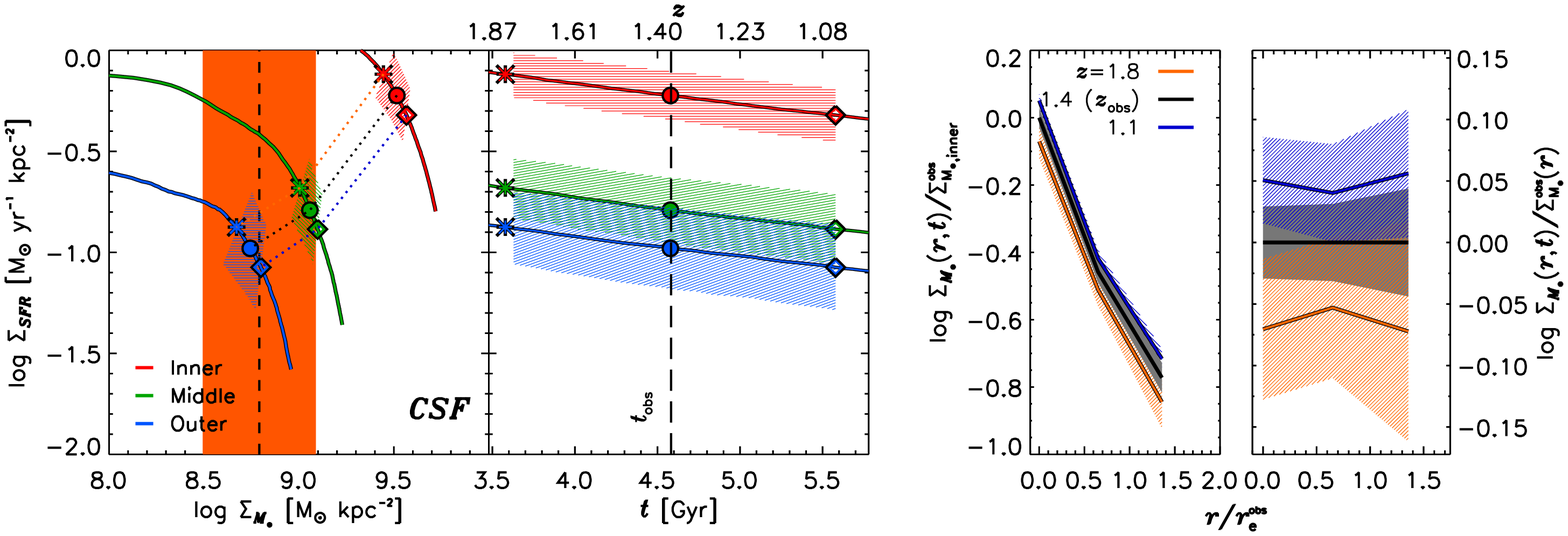}
\includegraphics[width = 0.75\linewidth, trim = 0cm 0cm 0cm 0cm]{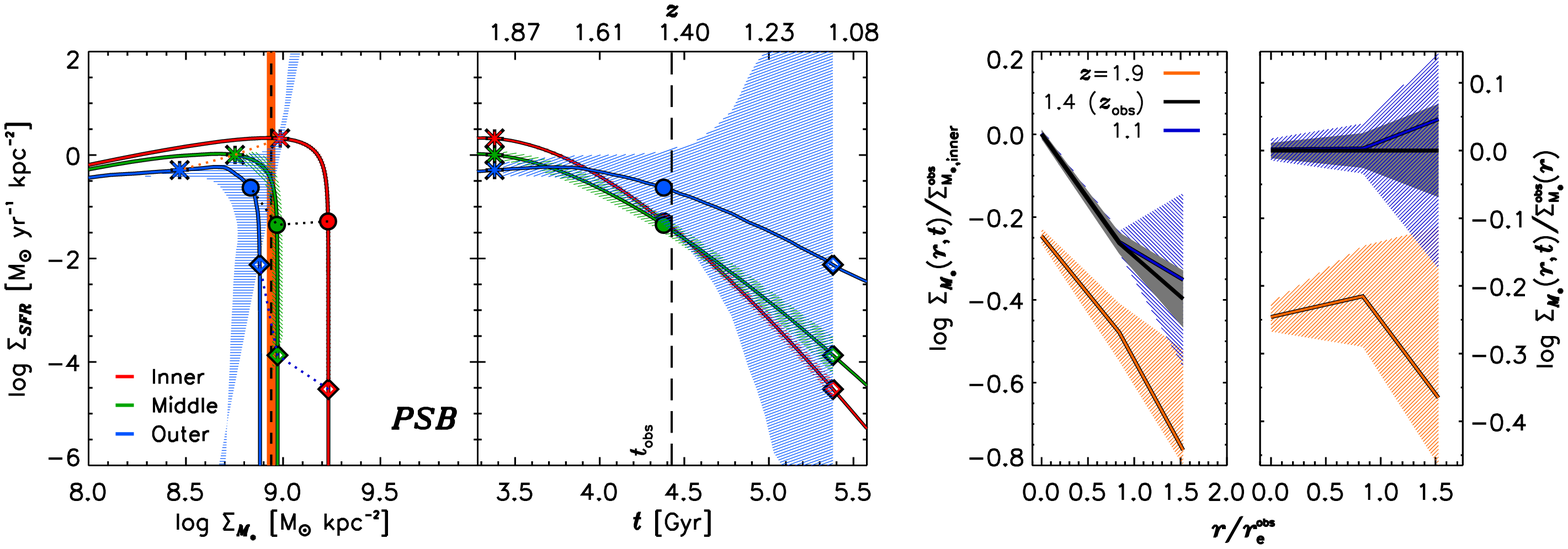}
\includegraphics[width = 0.75\linewidth, trim = 0cm 0cm 0cm 0cm]{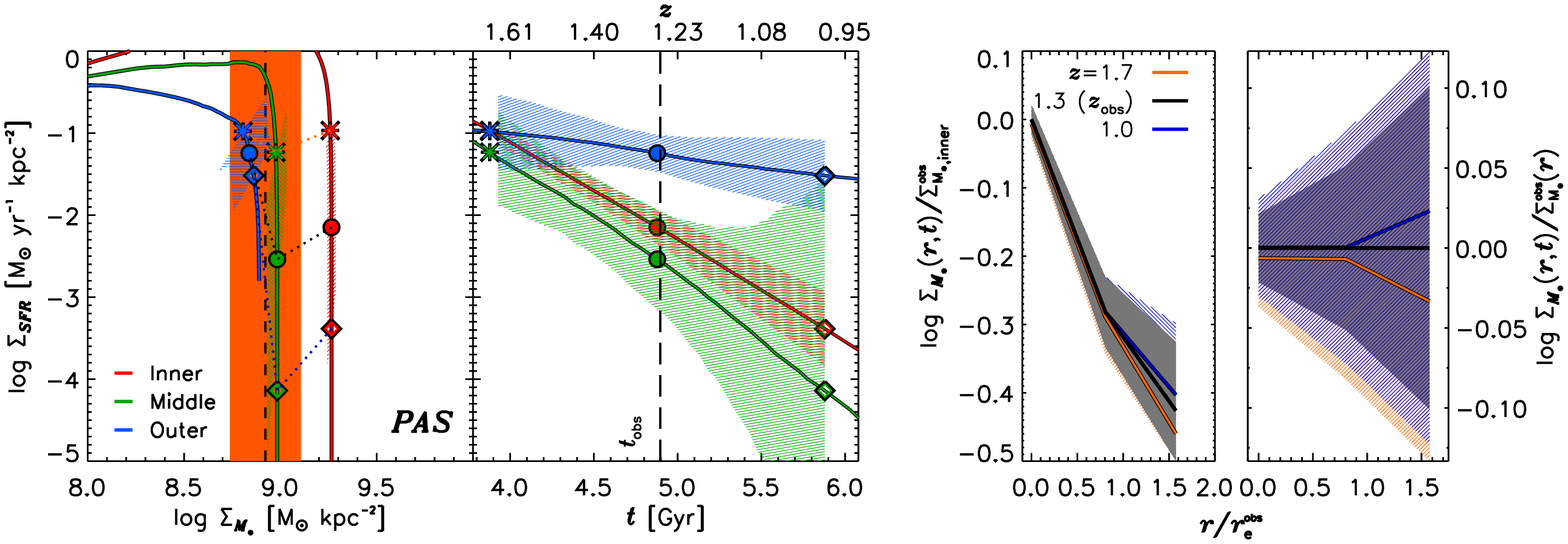}
\caption{The resolved mass growth history of each of the sample galaxies (rows). {\it From left}: (1) $\Sigma_{\sfr}(r)$ as a function of radial $\Mstel$ surface density; (2) $\Sigma_{\sfr}(r)$ as a function of time {\bfb since the Big Bang}; (3) the $\Sigma_{\Mstel}(r)$ at $\tobs\pm1$\,Gyr normalized to \inn\ values at $\tobs$ (showing shape evolution); (4) $\Sigma_{\Mstel}(r)$ at the same epochs normalized to the observed profile (showing total $\Mstel$ growth). Points in the left two panels denote epochs in the right two panels, and orange vertical bars show \inn\ $\Sigma_{\Mstel}$ at the time of peak \inn\ SFR. Shading shows 1\,$\sigma$ ranges in all cases. All galaxies but \csf\ show signs of bulge-first growth, visible in the resolved $\Sigma_{\sfr}(t)$, the change in rank-ordering of $\Sigma_{\sfr}(\Mstel; t)$, or the profile evolution itself. Also with the exception of \csf\ (Section \ref{sec:discussionPhysics}), the \inn\ $\Sigma_{\Mstel}$ at peak-SFR seems to be an asymptote for the \mid\ and \out\ regions, even though the \inn\ regions themselves may grow $\sim$2$\times$ since passing this mark. This may reflect an association between bulge-building and quenching \citep[][Figure \ref{fig:feedback}, below]{Zolotov15,Tacchella16} or perhaps a coincidence given the masses/mass profiles of these systems.}
\label{fig:evostuff}
\end{figure*}

\subsubsection{Spatially Resolved Star Formation Histories}
\label{sec:resSFH}
{\bfb
Figure \ref{fig:sfhs} shows the sample's spatially integrated and resolved lognormal SFHs. These confirm Section \ref{sec:global}'s inferences: \csf\ has continuously formed stars at a relatively constant rate across its face, declining by only $\sim$3$\times$ since its SFH peaked at $z>4$. It is thus much more mature than its starforming counterpart \ssf, which is within $\sim$1\,Gyr of its peak. A similar story holds for the passive systems: comparing peak times, \pas\ has been ``quenching'' for perhaps twice as long as \psb, though both are now farther from their peak/have SFRs that are falling faster than their starforming peers'.

As suggested by their masses, all four systems are in the declining phase of their SFHs. Yet, Figure \ref{fig:sfhs} suggests this would not be true of their $z\sim2$ progenitors: Except for \ssf---which grew substantially since then---these would be within a factor of $\sim$2 of their current mass but at {\it maximal} SFR (see Figure \ref{fig:crossCheck}). We return to this point in Section \ref{sec:spatResSFH}.

More resolved trends are shared by the three non-constant SFH systems: Compared to peak, {\it observed} SFRs in \ssf, \pas, and \psb's \inn\ regions are closer to or below their \mid\ and \out\ values. This is the formal signature of inside-out growth and is consistent with these systems' locations in Figure \ref{fig:sizeMass}, where they have significantly higher $B/T$ than the constantly starforming \csf. We return to this point in Section \ref{sec:discussion}, but for now note that this phenomenon has implications for the inferred progenitors of these galaxies in higher-$z$ cross-sectional datasets (Section \ref{sec:spatResSFH}).}

\subsubsection{Spatially Resolved Mass Profile Histories}
\label{sec:spatResSFH}

With spatially resolved SFHs, we can reconstruct the mass profile evolution of the sample galaxies.

Figure \ref{fig:evostuff} presents these inferences. Each row shows results for one galaxy in order of decreasing $\ssfr_{\rm obs}$ (\ssf\ to \pas). From {\it left}, the panels show: 1) $\Sigma_{\sfr}$ vs.\ $\Sigma_{\Mstel}$ in each radial zone; 2) $\Sigma_{\sfr}$ vs.\ {\it time} in each zone; 3) $\Sigma_{\Mstel}(r;\,t)$ profiles at $\tobs\pm1$\,Gyr relative to observed \inn\ values {\bfb (profile shape evolution)}; 4) $\Sigma_{\Mstel}(r;\,t)/\Sigma_{\Mstel}(r;\,\tobs)$ {\bfb (total mass growth)} at the same intervals (where we are willing to extrapolate the SFHs). Points {\bfb in the left panels mark the epochs in the right panels}.

{\bfb Starting at {\it right},} all galaxies besides \pas\ are inferred to have gained significant $\Mstel$ at all $r$ in the Gyr preceding $\tobs$. Further---clearest in \ssf---the two starforming systems should also grow marginally in the {\it subsequent} Gyr ($\lesssim$25\%). 

\ssf\ also shows explicit signs of {\bfb inside-out/}bulge-first growth: proportionately, \out\ regions gain more mass in this 2\,Gyr period than \inn\ ones ($\sim$0.8 vs.\ $\sim$0.6 dex). Hints of this trend appear in \psb\ at reduced significance.

These statements are obviously consequences of trends in the second-left panel, showing Figure \ref{fig:sfhs}'s SFHs zoomed to the $\tobs\pm1$\,Gyr window. In all but \csf, \inn\ $\Sigma_{\sfr}$s fell faster than those at larger-$r$, with \inn/\mid\ crossings close to $\tobs$ in \ssf\ and \psb, and over a Gyr ago in \pas.

\csf\ is special in this regard, however: its {\bfb resolved} SFHs have maintained rank-ordering over all time. Section \ref{sec:discussionPhysics} speculates on the potentially {\bfb illuminating} point of how this system escaped inside-out growth, {\bfb but the left-most panel of Figure \ref{fig:evostuff} bears on this question.

Replacing time with $\Sigma_{\Mstel}$ on the abscissa greatly accentuates the inside-out signature discussed above: In all but \csf---but especially \psb\ and \ssf---\inn\ regions fall much faster than at least \out\ ones, moving from a high to low relative position on the $y$-axis.

Indeed, {\it from the $\Mstel$ perspective}, what is a {\bfb gradual} decline in SFRs with {\it time} transforms into an abrupt quenching event as galaxies appear to hit a $\Sigma_{\Mstel}$ wall at all $r$. At root, this phenomenon---the ``L-shaped track'' of \citet{Barro16}---is a reflection of monotonically declining $\ssfr$s in each zone leading to asymptotic final masses. This is exacerbated by using typically logarithmic $x$-axes (as we do here), de-emphasizing meaningful, if undramatic, linear increases. Indeed, the aforementioned ``modest'' 25\% future mass growth in the starforming systems---current $\sfr\geq$25\,$\Msun\,\yr^{-1}$---can correspond to $\sim$$2\times10^{10}\,\Msun$ in \csf's case. 

This fact highlights a distinction between ``activity'' as defined by future {\it fractional} mass growth (systems that can $>$double $\Mstel$ in a Gyr) vs.\ current star formation: if the former is used, then even galaxies making tens of $\Msun$ of stars per year are arguably quenched, with ``active'' galaxies only those with sustained high $\ssfr$s; i.e., {\it rising SFHs}.

The point spacings on the curves in Figure \ref{fig:evostuff}, {\it left}, clearly illustrate this: \ssf's \out\ region grew by nearly 10$\times$ in $\Mstel$ over the 2\,Gyr ($\sim$0.4$\,t_{\rm Hubble}$) probed. Contrast this with \csf's \out\ region, which grew by at most a factor of 2 despite having a higher mean $\sfr$ over this interval (Figure \ref{fig:crossCheck}).

This leads to the question of why SFHs {\it stop rising}.} Given the vast literature on this topic---``quenching''---we need not dwell on it here {\bfb (\citealt{Kelson16} give an explicit mathematical framing in the above terms)}. However, features in these data may prove useful to that discussion (Section \ref{sec:discussionPhysics}). For now, we note simply that \csf\ may provide a meaningful counterexample to the idea that $\Sigma_{\Mstel}$ is the more causally informative abscissa: the sample's most massive system, it also has the highest $\sfr$ (Table \ref{tbl:derived}), the same $\Sigma_{\sfr}$ as \ssf---a system with $\sim$3$\times$ higher $\ssfr(\tobs)$ (Figure \ref{fig:massProf})---and has likely grown more since its SFH peaked (orange bands). As such, {\bfb while it graphically emphasizes their effects, quantities besides $\Sigma_{\Mstel}$ seem key to shaping SFHs for at least \csf-like (i.e., {\it disky}) galaxies \citep[see also][]{Lilly16}.}

\begin{figure*}[t!]
\centering
\includegraphics[width = 0.8\linewidth, trim = 0cm 0cm 0cm 0cm]{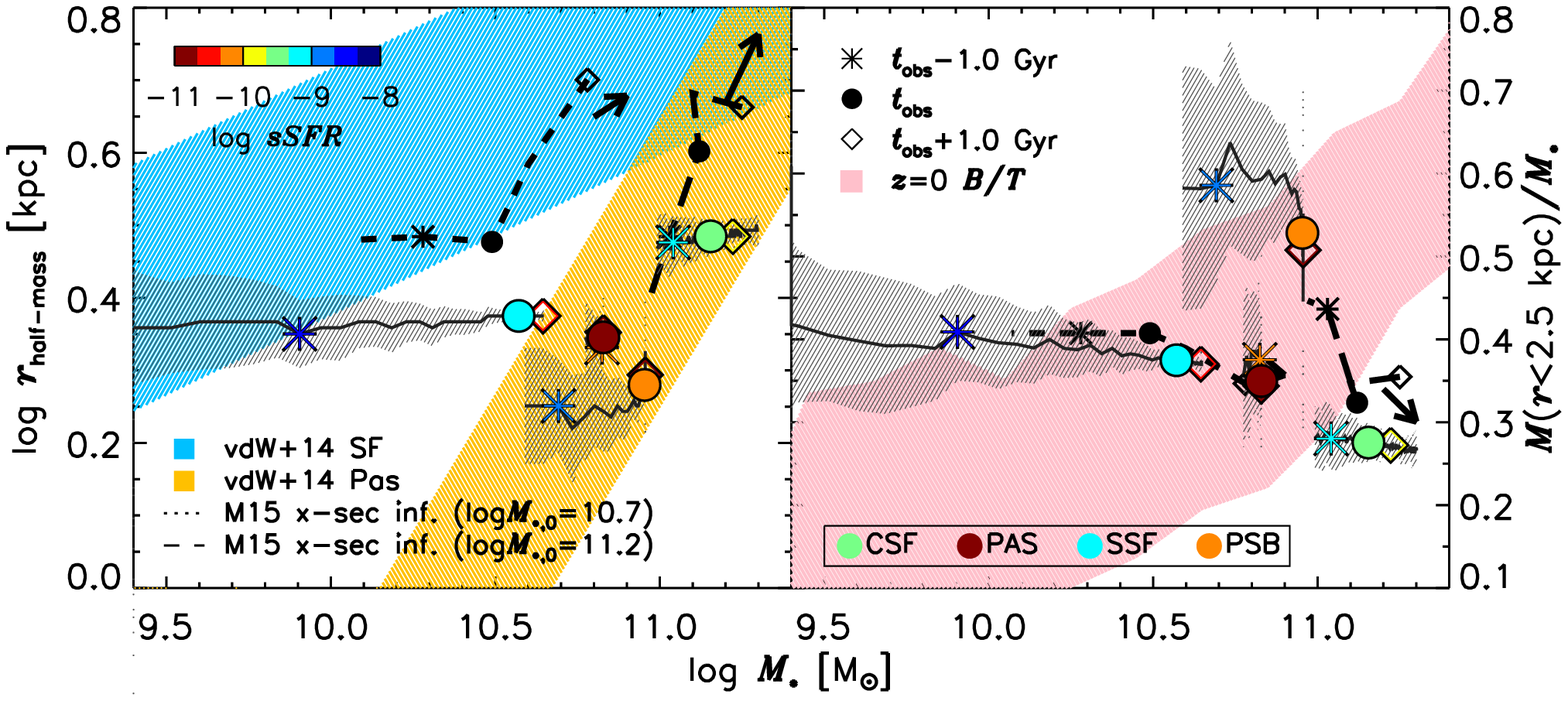}
\caption{As Figure \ref{fig:sizeMass} but showing the longitudinally reconstructed half-mass-radius-- ({\it left}) and mass-concentration/$B/T$--mass evolutionary trajectories of the sample galaxies. Black shading shows 1\,$\sigma$ envelopes across $2\geq z\geq0.5$, with points denoting 1\,Gyr before (stars) and after (diamonds) $\tobs$ (filled circles). Point colors show $\ssfr(t)$ at the corresponding epochs. Short and long black dashes, \resp, show \citet{Morishita15}'s cross-sectionally inferred trajectories for galaxies with the space densities of modern MW- and $\gtrsim$M31-mass systems over the same $z$-range, highlighting the same intervals (referenced to $\langle z\rangle=1.27$); arrows denote their time-averaged vectors. Population loci are shaded as in Figure \ref{fig:sizeMass}. \ssf, \csf, and \psb\ have all grown significantly in $\Mstel$ since $z=2$, but not necessarily in size \citep[cf.][]{Genel17}. This behavior differs from cross-sectional inferences---even for \csf, whose mass evolution is well-captured---but $B/T$ inferences are more consistent, with galaxies tending to descend onto the fixed population locus with time from earlier states of relatively high $B/T$ (inside-out growth). At {\it left}, note that three systems will end up as small red galaxies at $z\sim0.5$, though these will have gotten there in very different ways (Section \ref{sec:spatResSFH}), highlighting the importance of path-dependence/{\it temporal diversity} underlying single-epoch parameter projections in galaxy evolution.}
\label{fig:sizeMassEvo}
\end{figure*}

The final inference we draw from the resolved SFHs is how these galaxies moved in the size-- and $B/T$--mass planes introduced in Figure \ref{fig:sizeMass}. We plot these projections in Figure \ref{fig:sizeMassEvo}.

In both panels, we show the 1\,$\sigma$ trajectory envelopes for each system at $z=0.5$--2. These are reconstructed by integrating the azimuthally-converted $\Sigma_{\Mstel}(t)$ at each timestep as inferred from the radially resolved SFHs (Section \ref{sec:grossProperties}). ``$B/T$'' is again taken as the mass within $r=2.5\,\kpc$ over the total system mass at any epoch, but sizes here are half-{\it mass} radii as opposed to $r_{e}$ (though they differ little at $\tobs$; Figure \ref{fig:massProf}). Point styles show $\tobs\pm1$\,Gyr as in Figure \ref{fig:evostuff}.

Three points emerge: First, despite growing by $\sim$2.5--10$\times$ in $\Mstel$, no galaxy has evolved significantly in size or $B/T$ since at least $z=2$, {\bfb assuming all size evolution is due to {\it in situ} SF}. Second, these (flat-ish) trajectories {\it do not} follow the galaxy {\it population} loci in any of those epochs, or indeed inferences based on fixed 2- or 3-D density ($r\propto\sqrt{\Mstel}$ or $\sqrt[3]{\Mstel}$), or velocity dispersion ($r\propto\Mstel$). Third, {\bfb at least in this sample, and assuming {\it in situ} growth,} longitudinal {\it size} trajectories diverge markedly from abundance-matched cross-sectional inferences \citep[][]{Morishita15}. Intriguingly though, $B/T$ tracks are more similar (at least for \ssf). \csf's mass growth ($dx/dt$) is also remarkably well predicted. Section \ref{sec:G13tests} elaborates.

Our radial resolution and the fact that we must convert chordal {\bfb (section-based)} into radial {\bfb (annulus-based)} densities are important caveats in this context (Section \ref{sec:discussionTheory}). Yet, assuming we are not grossly in error---{\bfb supported at least in that we predict these galaxies to have} reasonable $(r,\Mstel,\ssfr)$ at $z\sim0.5$---these inferences support a few robust statements.

Regarding motion with respect to loci, echoing results in $\sfr$--$\Mstel$ (cf.\ \citealt{PengLilly10, Speagle14, Abramson16}), {\bfb even small regions in size--mass} space clearly contain galaxies that got there in very different ways. For example, at $z\sim0.7$, three systems will appear as relatively small red galaxies (absent merging). Yet, \pas\ will have been there since $z\sim2$, when \psb\ was a similarly small but high-$\ssfr$ ``blue nugget'' and \ssf\ was an average, much lower mass starforming galaxy. Also, if anything, \csf\ will have traversed {\it across} the red (or parallel to but $\sim$2\,$\sigma$ below the blue) galaxy locus since $z\sim2$, not along it, as {\bfb suggested by} cross-sectional inferences. This point emphasizes {\it temporal diversity} underlying single-epoch parameter projections in galaxy evolution.

That said, knowledge of $B/T$ {\bfb and $\ssfr$} seem capable of breaking some of the above degeneracies. Consistent with the blue-nugget interpretation \citep{Barro13,Barro16}, \psb\ should have come from a substantially (if not abnormally) bulge-dominated $z\sim2$ progenitor, whereas \ssf's progenitor was much diskier (if still more bulge-dominated than most similar-mass galaxies). \pas's progenitor would have had a similar $B/T$ to \ssf's, but significantly lower $\ssfr$. 

In all cases, if anything, $B/T$ {\it decreases} with time relative to the (quasi-static; \citealt{Lang14}) population locus, if not absolutely. {\bfb While another expression of inside-out growth, this} signals that progenitors of modern galaxies with substantial disks ($B/T\sim0.4$) may descend from lower-mass galaxies with relatively {\it high} $B/T$. Interestingly---fortunately---this signal is also reflected by at least \citet{Morishita15}'s {\bfb cross-sectional, abundance-matched tracks}. As discussed regarding \csf, it further suggests atypically disk-dominated galaxies as a special population. We address this point next.


\section{Discussion}
\label{sec:discussion}

{\bfb To close, we interpret our results in terms of their (1) astrophysical and (2) methodological/theoretical implications. Astrophysically, we focus on the relationship between $B/T$ and SFH shape, specifically \inn\ $\ssfr$s rising above global averages at the time of peak $\sfr$, suggesting feedback from bulge building may play a role in quenching as numerical models posit. We also compare our results to predictions from previous semi-empirical models, which may similarly suggest that particularly disky galaxies---at least as defined relative to equal-mass systems---are a special class. Methodologically, we discuss this analysis' limitations, and what future studies can (and cannot) do to overcome them.}

\subsection{Astrophysical Implications}
\label{sec:discussionPhysics}

\subsubsection{Effects at $\tobs$}

{\bfb Figures \ref{fig:sfms}, \ref{fig:AGN}, and \ref{fig:massProf} touch on basic astrophysical issues. These show that spatially resolved properties at $\tobs$ can lead to different physical inferences than integrated measurements. Specifically, (1) galaxies and their components can be classified differently with respect to population loci (e.g., the SFMS); (2) systems with distinct integrated or resolved $\ssfr$ can have similar $\Sigma_{\sfr}$; and (3) even at $\log\Mstel\sim11.2$, AGN do not rapidly quench SF at large-$r$ \citep[see also][]{Aird12, Harrison14, Poggianti17}. Regarding (1), if SFHs are lognormal, our two starforming systems go from (somewhat) underperforming pieces to normal wholes at their respective $\Mstel$. As all of these entities are evolving through---not along---the SFMS at $\tobs$, this fact furthers doubts about the profoundness of this locus.}

\subsubsection{The Epoch of Peak \inn\ SFR}
\label{sec:peakSFR}

Higher-level issues emerge from interrogating Figure \ref{fig:evostuff}. In the left-most panels, $\Sigma_{\Mstel}$ at the time of peak \inn\ SFR (i.e., SFH turnover) is overplotted as vertical orange bands. These span about a factor of 3 across the four systems---$\log\Sigma_{\Mstel}\sim8.5$--9. In all but \csf\ (see below), while the \inn\ region itself goes on to roughly double its mass, this appears to serve as a sort-of {\bfb boundary} beyond which the rest of the galaxy will not pass. This may be a coincidence {\bfb arising from the small sample size or these specific} mass profiles (third panel), but it {\bfb may also support} recently popular density-related quenching mechanisms, where strong outflows at peak \inn\ SFRs signal the onset of galaxy-wide quenching. We find evidence for this scenario{\bfb---which we interpret as a link between high-$B/T$ and falling SFHs---but also additional quenching processes.}

\begin{figure}[t!]
\centering
\includegraphics[width = \linewidth, trim = 0.5cm 0cm 0cm 0cm]{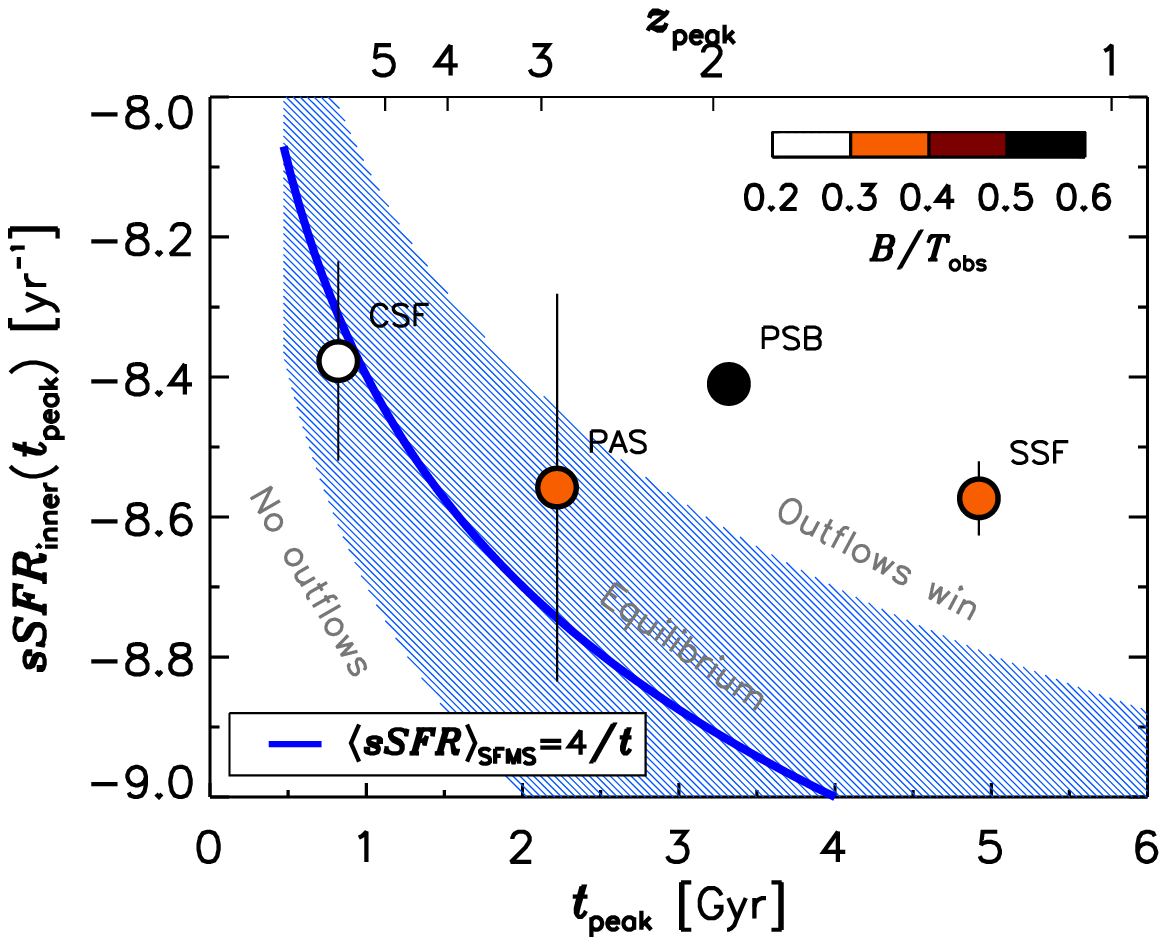}
\caption{Higher $B/T$ sample galaxies (point colors) tend to have \inn\ SFHs that peaked above the SFMS. Meanwhile, \csf\ has the lowest $B/T$ and never rose above it, given the locus' elevated position at that earlier time. If the SFMS reflects an inflow/outflow equilibrium \citep[e.g.,][]{Dave11}, this suggests a link between bulge building/strong central SF and a feedback-related mechanism that turns SFHs over \citep[outflows beat inflows; e.g.,][]{Zolotov15,Tacchella16}. If so, disk galaxies may be the objects with the flattest/most constant SFHs (Figures \ref{fig:sfhs}, \ref{fig:btPredict}; Sections \ref{sec:spatResSFH}, \ref{sec:peakSFR}).}
\label{fig:feedback}
\end{figure}

Bearing the caveats that come with this sample size, Figure \ref{fig:feedback} presents our test. It shows each galaxy's \inn\ $\ssfr$---i.e., SFR per unit of mass/outflow restoring force---at the time of peak \inn\ SFR plotted against the contemporaneous $\langle\ssfr\rangle$ of starforming galaxies \citep{MadauDickinson14,Kelson14} with points coded by $B/T_{\rm obs}$. 

{\bfb Hints indeed emerge that systems with high $B/T$---especially \psb\ and \ssf---had elevated $\ssfr$ compared to the mean starforming galaxy at the epoch just before their SFHs started falling (i.e., quenching).} Insofar as the SFMS reflects an equilibrium---where inflows balance outflows \citep{Dutton10, Dave11}---{\bfb this suggests} that \inn\ SF in these galaxies was strong enough to allow feedback via, e.g., supernovae winds to quench their \inn\ regions, with the rest of the galaxy to follow. This would support the scenarios presented in \citet{Zolotov15} and \citet{Tacchella16}, though here the enhanced $\ssfr$ represents a single, unique, {\bfb super-SFMS} peak, not the last in a series of pre-quenching surges. As mentioned (Section \ref{sec:spatResSFH}), this is also consistent with at least \psb's inferred motion in the $r_{e}$--$\Mstel$--$B/T$ plane, where it went from being a small, high-$\ssfr$, ``overly'' bulge-dominated blue nugget \citep{Barro13} to a galaxy with more normal $\ssfr$ for its $B/T$ and size.

{\bfb \csf\ also {\it partially} supports this scenario by being a good exception to it. The centrally densest galaxy in the sample (Figure \ref{fig:massProf}, {\it top-left}), it also has the highest $\sfr$, the lowest $B/T$ ($\sim0.3$), and the flattest/highest-$\tau$ SFH (Figure \ref{fig:sfhs}). While seemingly at odds with the above paradigm, these properties are in fact linked to it through {\it time}. Figure \ref{fig:feedback} shows that \csf\ peaked early enough} that, though it had a high absolute $\sfr$---essentially the same $\ssfr(t_{\rm peak})$ as the other systems---{\it given the prevailing inflow conditions at the time}, it need never have left equilibrium. Hence, \csf's inner region need never have blown itself out (unlike \psb\ and \ssf's), and it could proceed to grow calmly at all $r$ through to $\tobs$ some 4\,Gyr ($\sim$4\,$t_{\rm Hubble}$) later, if not today (Figure \ref{fig:glassVmike}). 

{\bfb This helps explain} why \csf's $\ssfr(r)$ can be $\sim$1/3 of \ssf's even though they have similar $\Sigma_{\sfr}(r)$ (Figure \ref{fig:massProf}): it can sustain the ``normal'' star formation implied by the latter because its higher mass was built slowly over Gyr from a gentle SFH, not rapidly via a violent one leading to equilibrium-breaking outflows. This scenario---that both the {\it time} and {\it timescale} of SF are critical---is precisely the mathematical premise of \citet[][hereafter G13]{Gladders13b}, and the above is qualitatively consistent with previous arguments linking bulge growth to its SFHs' shapes (see \citealt{Abramson16}, Section 3.3; Section \ref{sec:G13tests} below). If correct, they highlight the centrality of path-dependence in galaxy evolution, raising the need for future analyses such as this (Section \ref{sec:discussionTheory}).

{\bfb This said, if the above is true, some additional density-independent but time-dependent process will be necessary to shut \csf\ down (whenever that happens; see, e.g., \citealt[][]{Larson80,Croton06, Peng15, Voit15, Oemler16}).} 

How \pas\ fits into this story is unclear given uncertainties and the SFMS' width. While it seems more likely to have been in the ``overly productive'' regime of the later-peaking \ssf\ and \psb, a robust conclusion---as well as more generally representative empirical inferences---requires additional data.

The last of Figure \ref{fig:evostuff}'s points to emphasize is that, irrespective of their $\ssfr$ at any $t$, to the extent that lognormal SFHs reflect reality, the mode galaxy grows by $\sim$2.5$\times$ after peak $\sfr$. While it may take gigayears to attain that asymptote---over which it will display substantial linear SFRs---this corresponds to just {\it 0.4 dex} on a $\log\Mstel$ abscissa. As such, for any SFH with a declining tail, the quenching process will almost always {\it appear} rapid. Corollary to the SFMS' case, this phenomenon would question how informative the bimodal quenched/starforming dichotomy is: Even robustly starforming galaxies---those at peak SFR!---might reasonably be ``quenched'' from the perspective of future $\Mstel$ growth. While perhaps useful for modeling, e.g., the evolution of mass functions, it is difficult to see how such a division illuminates the processes controlling star formation. 

Regardless, in any framework, the above suggests that the best vehicles to learn about the latter are systems with rising SFHs, which are by definition low-mass. \ssf's progenitors---$\log(r_{e},\Mstel,\sfr)\sim(0.4, 9, 1)$ with high $B/T$---seem good $z\sim2$ follow-up candidates {\bfb by which to infer} the development of some of today's MW-mass galaxies (Figures \ref{fig:sfms}, \ref{fig:massProf}, \ref{fig:sizeMassEvo}, \ref{fig:glassVmike}).

\subsubsection{Relationship to Previous Model Predictions}
\label{sec:G13tests}

In the size--mass context, Figure \ref{fig:sizeMassEvo} {\bfb suggests the} obvious comparison between these galaxies' longitudinally inferred trajectories and, e.g., \citet[][hereafter M15]{Morishita15}'s abundance-matched cross-sectional inferences for the {\it mean} system of similar-mass. Clearly, {\bfb these methods yield different size projections}: \ssf\ and \csf\ have identical masses to M15's means at $\tobs$, but they are both observed to be and inferred to remain smaller than what that study would predict. This is despite the fact that M15 captures \csf's size at $\tobs-1$\,Gyr.

{\bfb Certainly, part of the discrepancy could simply be that our sample is not average. Further, M15's cross-sectional method captures {\it ex situ} growth from mergers---which we cannot---and our rather coarse radial resolution limits size-growth sensitivity. However, if substantiated by larger analyses, such tension would point to a meaningful divergence between the evolution of the mean size of galaxies and the size-evolution of the typical (i.e., {\it mode}) galaxy. This said, $B/T$ inferences seem more robust to the above issues,} agreeing better qualitatively---and quantitatively, at least at $\tobs$---than sizes. {\bfb We do not know why this is true, but it is intriguing.}

If {\bfb borne-out in the} future, Figure \ref{fig:sizeMassEvo}, {\it left}'s $r$--$\Mstel$ reconstructions would also disfavor a suite of analytic prescriptions \citep{Bezanson09, vanDokkum15, Abramson17, Lilly16} wherein galaxies grow at roughly fixed $\sigma_{v}$, $\rho_{\Mstel}$, $\Sigma_{\Mstel}(r<r_{e})$, or $\propto\Mstel^{1/3}\,(1+z)^{-1}$, corresponding \resp\ to tracks with slope 1, $1/3$, $1/2$, and $\sim0.7$ in that  plane. These are far steeper than our inferences. Conversely, our tracks are not {\it too} dissimilar from the {\it Illustris} simulation's flatter $r(\Mstel)$ trajectories \citep[][see their Figure 5]{Genel17}. In all cases, future longitudinal comparisons should have great potential here (Section \ref{sec:discussionTheory}).

Beyond such semi-empirical comparisons, {\bfb we can} explore how our inferences compare to the G13 global lognormal SFH model itself. This was the source of our SFH parameterization, and \citet{Abramson16} specify ways to falsify it using these kinds of data (see their Appendix A).

The first test states that G13 would be ruled out if:
\begin{quote}
	Large, abrupt discontinuities are required to explain a significant 
	fraction of SFHs for systems dominating the universe's mass and 
	star formation budget over a significant fraction of cosmic time 
	(smooth, continuous parameterizations cannot describe most SFHs).
\end{quote}
Now, of the four galaxies in this sample, one---the poststarburst \psb---has SEDs that significantly disfavor lognormal SFHs in the UV at all $r$ due to what was likely an abrupt SFH discontinuity $<$1\,Gyr prior to $\tobs$. {\bfb One} {\it could} thus posit that 25\% of galaxies---perhaps 50\% of passive ones---at $z\gtrsim1$ and $\log\Mstel>10.5$ may therefore indeed require such events. If more data supports this conclusion, it would seriously challenge G13's limited view of the role of SFH discontinuities.\footnote{Strong Balmer absorption is common in red galaxies at these and higher $z$ \citep[e.g.,][]{vanDeSande13,Whitaker13,Newman14,Belli17}, but it remains to be shown that these cannot be produced by smoothly rising and falling histories at these relatively early epochs.}

{\bfb On the other hand, \psb's MOSFIRE-derived Balmer lines are, if anything, weaker than the best-fit lognormal implies, suggesting} any discontinuity occurred farther from peak-SFR than we {\bfb conclude} from the HST data alone (Figure \ref{fig:highResCheck}; Table \ref{tbl:mosGlass}). This would make the natural lognormal fall-off responsible for a larger {\bfb portion} of \psb's decline. Also, {\bfb by assuming solar metallicity, we may have underestimated \psb's true metal content and the attendant} UV line blanketing \citep{Gallazzi14}. Third, we may simply have caught an unrepresentative system in our sample. Finally, {\bfb two additional tests} suggest that, if it is wrong about the relevance of discontinuities, G13 still captures key {\bfb aspects} of these data's story.

\begin{figure}[t!]
\centering
\includegraphics[width = \linewidth, trim = 0.5cm 0cm 0cm 0cm]{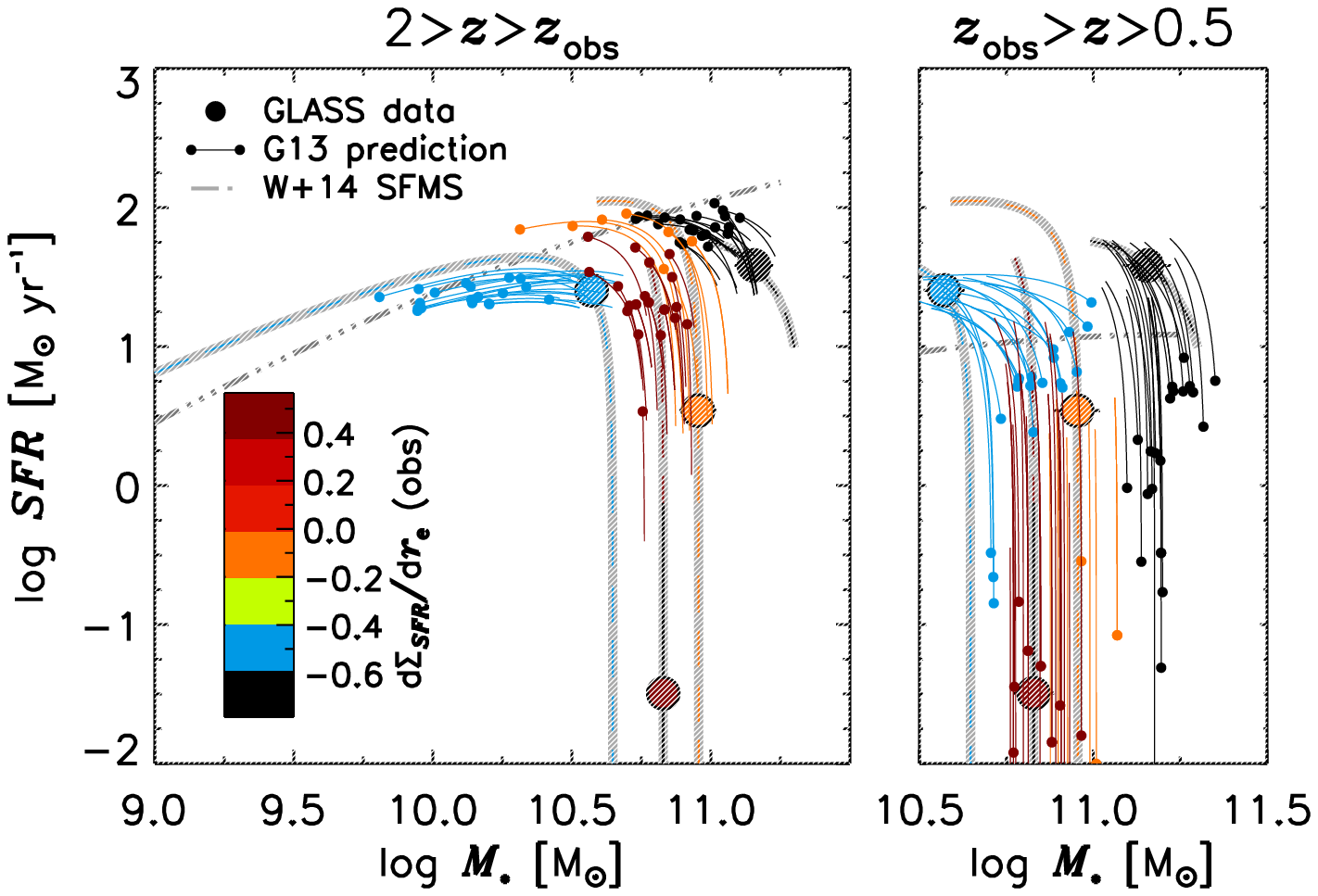}
\caption{G13 model SFHs (colored lines/small points) passing through the sample at $\tobs$ (large points) projected to show progenitor/descendant envelopes at $z=2$ and 0.5 ({\it left}/{\it right}, \resp). Colors show $\Sigma_{\sfr}(r;\,\tobs)$ data slopes (Figure \ref{fig:massProf}; bluer\,$\mapsto$\,more negative). G13 generally captures the data-inferred SFHs, though \ssf\ evolves faster than predicted. Both agree that all systems were SFMS galaxies at $z=2$ \citep[grey dot-dashes;][]{Whitaker14} except possibly \pas, which may already have shut down. At $z\sim0.5$, \csf\ has a $\sim$50\% chance of remaining a massive SF galaxy as data reconstructions suggest, while $\sim$20\% of \ssf's tracks suggest it will quench by then. More negative SFR gradients may imply slower evolution---another manifestation of bulge-first growth---but more data are needed to confirm.}
\label{fig:glassVmike}
\end{figure}

Figure \ref{fig:glassVmike} compares the sample SFHs to G13 model tracks passing within 2\,$\sigma$ of {\bfb the galaxies'} observed $(\Mstel,\sfr)(\tobs)$ coordinate/limit. The {\it left} panel projects the models back to $z=2$, the {\it right} forward to $z=0.5$. Color coding reflects the slope of the data's observed SFR surface density profiles.

{\bfb In this independent cross-check,} G13's high-$z$ predictions match the data inferences to within about a factor of 2 in both axes in all but \ssf, which has grown much faster than the model tracks. Given that G13 was based on no SED fits and contained no data-derived SFH information whatsoever, this level of agreement is encouraging.

Looking ahead to $z=0.5$, the model also does reasonably well, if perhaps in a more probabilistic sense. {\bfb Around} 20\% of model SFHs passing near \ssf\ at $\tobs$ suggest it will become a passive, $\log\Mstel\sim10.7$ low-$z$ galaxy; $\sim$1.3\,$\sigma$ agreement with its (median) data-inferred trajectory. Similarly, about half of \csf-matching trajectories suggest it will remain a massive, starforming system. Clearly, all inferences agree \psb\ and \pas\ will become $\sim$M31- and MW-mass passive systems. 

Note: these statements are converse to those in Section \ref{sec:grossProperties}: Figure \ref{fig:sfms} shows where $z=0$-selected MW-mass progenitors lie at mean sample redshift; Figure \ref{fig:glassVmike} shows where the descendants of $z>1$ selected galaxies lie closer to today.

In terms of informing future G13-like modeling, the spatially resolved data suggest a $d\ssfr/dt$--$d\Sigma_{\sfr}/dr(\tobs)$ anti-correlation, such that galaxies with the least-negative SFR gradients are evolving the fastest {\bfb (yet another restatement of inside-out growth)}. However, \ssf's data buck this trend, so robust statements await larger samples.

\begin{figure}[t!]
\centering
\includegraphics[width = 0.95\linewidth, trim = 0.25cm 0cm -0.25cm 0cm]{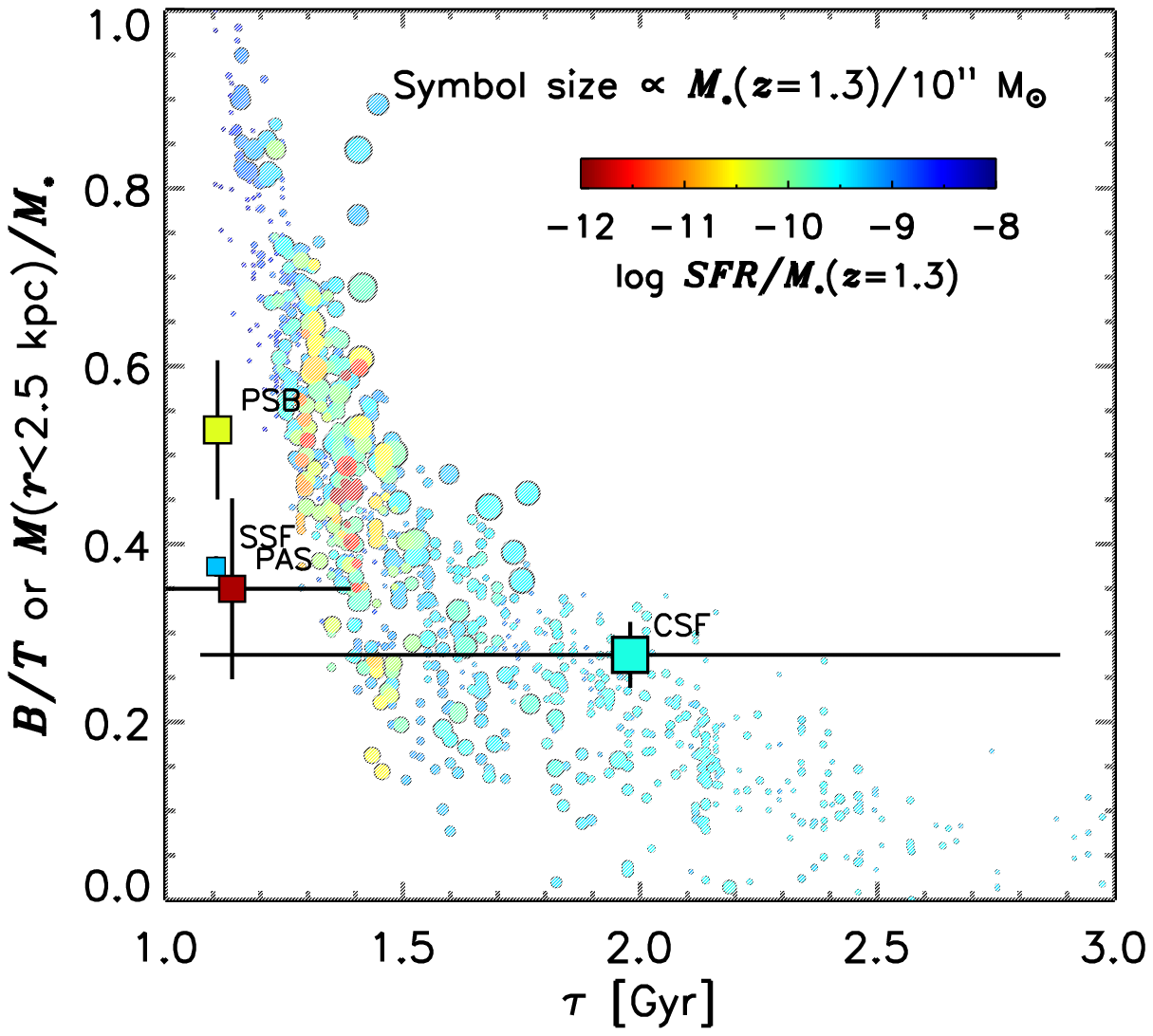}
\caption{GLASS sample data inferences (squares) vs.\ G13 $B/T$--$\tau$ predictions (circles; see main text/\citealt{Abramson16}, Figure 14 for details). Colors show model $z=1.27=\langle z_{\rm GLASS}\rangle$ and observed $\ssfr$s. The data's trend for higher {\it measured} $B/T$ to correspond to smaller {\it inferred} $\tau$ and the corresponding $\ssfr$s is well predicted, though and \ssf\ and \psb\ are quantitatively low. This amplifies Section \ref{sec:peakSFR}'s discussion of a bulge-growth/SFH shape connection---high (low) $B/T$ implies SFHs that do (not) turn over---a robust conclusion of this work and an independent consequence of G13. Indeed, if \csf's $\tau$ error bar---and perhaps G13's large-$\tau$ tail---reflect the fact that they cannot formally be captured by a lognormal, this may suggest that galaxies with near-constant SFHs form a special class.}
\label{fig:btPredict}
\end{figure}

Figure \ref{fig:btPredict} presents a final test. Here, we plot the data against the $B/T$--$\tau$ prediction from \citet[][see their Figure 14]{Abramson16}. This is {\bfb calibrated} at $z=0$ by rank-ordering G13 $\tau$ and \citet{Gadotti09} $B/T$ measurements at fixed $\Mstel$ assuming the smallest $\tau$ {\bfb corresponds to} the largest $B/T$. Point colors reflect $z=1.27=\langle z_{\rm GLASS}\rangle$ model and observed $\ssfr$s. Since no information on the $y$-axis was used in its construction, the properties of this locus represent blind predictions of the G13 model. As such, this comparison is a {\bfb true} experiment.

Two points are salient. First, the general $B/T$--$\tau$ trends in the data {\bfb reflect those in} the G13 model: smaller {\it inferred} $\tau$ is associated with systems of higher {\it measured} $B/T$. Second, the sample's measured $\ssfr$s (point colors) are consistent with those of the G13 tracks at the appropriate $B/T$, though they are likely more massive than the typical model at that $y$ location, and \ssf\ and \psb\ have quantitatively lower $\tau$ than predicted. \csf---the lowest-$B/T$ sample galaxy---is especially well-matched in this regard. Considering the model locus was calibrated at $z=0$---some eight Gyr after the epoch probed here---we consider even this level of agreement somewhat striking. [Given their declining SFHs, the GLASS sample's $B/T$ should evolve little between $\tobs$ and today (Figure \ref{fig:evostuff}).]

If supported by larger future datasets, the implications of especially \csf's location in Figure \ref{fig:btPredict} amplify Section \ref{sec:peakSFR}'s discussion of the connection between $B/T$ and SFH shape as captured by $\tau$, the SF timescale: Apparently, G13 prefers galaxies with $\tau\gtrsim1.5$\,Gyr---i.e., the {\it flattest} SFHs---to be the most disk-dominated, even if they are relatively massive.\footnote{Note that constant SFHs cannot formally be captured in the lognormal framework. It is thus interesting to consider whether the large-$\tau$ tail in the G13 locus in Figure \ref{fig:btPredict}---or even \csf's $\tau$ error bar---fundamentally highlights a physical regime in which this paradigm breaks down (or something more abstract). If so, SFHs approaching this limit may form a special class. We encourage further investigation in this vein.} This {\bfb link between high (low) $B/T$ and SFHs that do (not) turn over} is perhaps the central {\bfb physical} inference of this work, and also an independent {\bfb feature} of G13.

\subsection{Implications for Methods and Theory}
\label{sec:discussionTheory}

\subsubsection{Future Applications}

{\bfb While other studies have used similar spectrophotometric data to reconstruct aspects of galaxies' longitudinal evolution at $z>1$ \citep{Whitaker13, Newman14, Nelson16, DominguezSanchez16, LeeBrown17, Toft17}, $S/N$ and spatial resolution limitations have forced these to rely on stacks, treat galaxies as monolithic objects, or quote SFH summary statistics (e.g., age). By exploiting gravitational lensing and GLASS' depth, this is, as far as we are aware, the first $z>1$ study to perform a full, spatio-temporally resolved, spectral continuum SFH reconstruction for individual galaxies to $r\sim2\,r_{e}$.

To {\bfb progress}, more such (or better) information} will be essential: Figures \ref{fig:massProf}, \ref{fig:sfhs}, \ref{fig:evostuff}, \ref{fig:sizeMassEvo} are (hopefully) crude representations of the kind of empirical inferences that will ultimately be available to constrain the galaxy evolution narrative.

Indeed, {\bfb next generation facilities will likely allow this type of analysis to become routine.} With its redder, higher spectral and spatial resolution slitless grisms, slitted spectrographs, IFU, and imagers, {\it JWST} could apply more sophisticated versions{\bfb---incorporating new insights into spatially variable emission line interpretations \citep{Sanders17}}---to practically any suitably bright $z\gtrsim0.7$ system. Further, it will extend samples vastly in redshift and provide critical spatially resolved $\Mstel$ constraints from rest-IR photometry that do not yet exist (and whose absence is a key weaknesses of this study).

{\it WFIRST} will more directly translate our methodology, but over huge areas of the sky, enabling tests of the critical galaxy--environment connection \citep[perhaps {\it the} key issue][]{Dressler80,Kelson16}. Combined with 30\,m class telescope observations of specific targets---which will replicate some of {\it JWST}'s capabilities but {\bfb at higher spatial resolution, given planned adaptive optics systems}---this may prove revolutionary. Lastly, as {\bfb envisioned}, spectrographs on all of the above platforms will measure spatially resolved stellar absorption lines, providing more robust SFH and metallicity constraints (e.g., Figure \ref{fig:highResCheck}; \citealt{Worthey92,Trager98}), another datum we could not incorporate here. To our knowledge, \citet{Newman15} and \citet{Toft17} present the only {\bfb such observations} at $z>1$ for two $\log\Mstel\sim11.1$--11.5 lensed systems, {\bfb with LEGA-C \citep{vanDerWel16} just now providing the first statistical database at $z\sim0.7$}. {\bfb Next generation facilities may make such data the norm.}

Other meaningful immediate advances include: {\bfb moving to true radial reconstructions using IFUs,} building a chordal-to-radial transformation calibration library using, e.g., GALFIT \citep{PengGALFIT}, and exploring the effects of moving from parametrized to free-form SFHs. The latter were adopted by the {\it Carnegie Spitzer IMACS Survey} (\citealt{Kelson14a, Dressler16}; Dressler et al., in preparation), and {\bfb this and similar approaches based on halo merger-tree motivated SFHs \citep{Pacifici13,Pacifici16} and fine landscapes of basis functions \citep{Iyer17} are proving powerful. Specifically, these have the advantage of} avoiding the assumption that SFHs rise and fall/each phase's relative mass contribution, and allowing each SFH component to have its own $A_{V}$ and metallicity{\bfb, enabling self-consistent chemical enrichment and mass-growth modeling \citep[][]{Munoz15}}. Both have potentially important ramifications we could not assess, and, though SFH discontinuities may complicate temporal, e.g., size reconstructions, further study will be fruitful.

\subsubsection{Irriducible Uncertainties}

Despite the above advances, a core issue will accompany all future work. {\bfb Assuming the above techniques yield consensus SFHs when applied to data of suitable quality, and irrespective of whether those data are resolved radially, in {\bfb sections}, or treated as independent spaxels,} the results will {\it never} correspond to the evolution of the {\it matter} in that aperture. Rather, they will reflect {\it the evolution of the aperture based on its contents at} $\tobs$. Because we only ever get one look at a galaxy, regardless of the fineness of the empirical mesh, inferences derived therefrom {\bfb must follow from each cell's state} at one and only one {\bfb instant. Yet, the galaxies themselves are collections of moving elements, so the stars/gas we aim to characterize may never occupy that cell at any other epoch. This fact---which also forces assumptions about the role of mergers, {\bfb which contribute to mass growth without leaving unique spectral signatures}---has been an ever-present if inevitable issue here, {\bfb whose effects only forward modeling can constrain}.

Of course, it may be that, once $t_{\rm Dyn}\ll t_{\rm Hubble}$ (which occurs quite early), material is moving so swiftly compared to the timescales of interest that the above {\bfb effects ``average-out,'' becoming} a non-issue. {\bfb Or, perhaps due to secular evolution, they may not.} Either way, observers and simulators should work together to understand and constrain the effects of this fundamental empirical limitation, thus clarifying when to conclude the project of understanding how galaxies' individual biographies led them to their observed states.

%



\section{Summary}
\label{sec:summary}

Based on deep, 17-band CLASH/HFF + G102/141 GLASS HST spectrophotometry, we reconstruct the integrated and spatially resolved star formation histories of four, $z\sim1.3$ cluster-lensed galaxies (Table \ref{tbl:sample}) using full-spectrum, rest-UV/-optical fitting at $|r|\lesssim2\,r_{e}$. The sample spans $\log\Mstel\simeq10.5$--11.2 and contains one anciently passive, one recently quenched, one continuously starforming, and one strongly starforming galaxy (Table \ref{tbl:derived}; Section \ref{sec:global}), enabling these case studies to probe spatio-temporal phenomena relevant to meaningful parts of parameter space (MW- to super-M31-mass progenitors; Figure \ref{fig:sfms}). Our work has three main implications:

{\bf Methodologically}, the best current data enable observers to resolve galaxies in both space and time, offering new ways to empirically connect their observed properties---here, $r_{e}$, $B/T$, $\Sigma_{\Mstel}(r)$, $\Sigma_{\rm \sfr}(r)$ (Section \ref{sec:resolved}, Figure \ref{fig:massProf}); perhaps one day $A_{V}(r)$, $Z(r)$---to aspects of their individual histories, moving data-driven inferences much closer to the domain of theoretical models. We demonstrate this using just three radial bins and the assumption of lognormal SFHs (Sections \ref{sec:data}, \ref{sec:models}), but future IFU studies or different SED fitting techniques could relax these constraints, enabling robust assessments of the dependence of astrophysical conclusions on such factors.

{\bf Astrophysically}, we find hints of a connection between $B/T$ and SFH shape/spatial uniformity (Sections \ref{sec:longitudinal}, \ref{sec:discussionPhysics}). The three sample galaxies with the highest $B/T$ are also those showing signs of inside-out growth---where the \inn\ regions move from leading to lagging those at larger-$r$ in $\sfr(t)$---and indeed those with obviously rising and falling SFHs (Figures \ref{fig:sizeMass}, \ref{fig:sfhs}, \ref{fig:evostuff}). Further, two of these are inferred to have had $\ssfr$s well above the SFMS at the time of peak \inn\ $\sfr$ (with the other marginally elevated). This is to be contrasted with the diskiest system ($B/T\sim0.3$), which had a spatially uniform, slowly declining SFH that never exceeded the mean $\ssfr$ of all starforming galaxies (Figure \ref{fig:feedback}). Both signals are consistent with recent theoretical suggestions positing a link between bulge growth and $\sfr$ declines \citep{Zolotov15,Tacchella16}.

Further, individual size- and $B/T$ evolutionary trajectories may diverge from cross-sectional inferences based on abundance matching \citep[][Figure \ref{fig:sizeMassEvo}]{Morishita15}, emphasizing the {\it temporal diversity} underlying single-epoch galaxy parameter projections. Half-mass radii are inferred to evolve negligibly across at least the 2\, Gyr bracketing $\tobs$, even as galaxies gain factors of $\sim$6 in $\Mstel$ over the same period. Similarly, $B/T$ evolution may be flat or decreasing in an absolute sense---suggestive of inside-out growth---but is more robustly negative relative to the mean at a given $\Mstel$; i.e., sample galaxies tend to start above the $B/T$--$\Mstel$ locus at low mass, but end-up at the appropriate value at their terminal masses, in closer agreement to M15's cross-sectional inferences.

While sample size forces these statements to remain tentative, {\it JWST}, {\it WFIRST}, and the 30\,m class telescopes will routinely produce data that can support this and more sophisticated analyses, and therefore contextualize or refute them. 

Finally, {\bf information-theoretically}, empirical spatio-temporal inferences will ultimately be limited in that they must forever be based on the matter in an aperture at one and only one epoch. As such, even given arbitrarily well resolved, high-$S/N$ IFU data and no systematic uncertainties (e.g., as to the IMF or chemical evolution), the fundamentally Eulerian nature of data-driven reconstructions will always place them in a slightly different domain (aperture evolution) than the Lagrangian theories they should be used to test (galaxy evolution). Understanding the relationship between these two entities will require the joint efforts of theorists and observers working towards a common interpretive framework. We are hopeful such partnerships will soon emerge, ushering in a qualitatively new relationship between data and theory.


\section*{}
LEA thanks Dan Kelson, Jordan Mirocha, and Simon Birrer for their toleration during many enlightening conversations. The same goes for Lindsay Young, who illuminated key longitudinal/cross-sectional distinctions underpinning this work, and Mike Gladders, whose ideas were central to it. {\bfb BV acknowledges support from an Australian Research Council Discovery Early Career Researcher Award (PD0028506).} GLASS (HST GO-13459) is supported by NASA through a grant from STScI operated by AURA under contract NAS 5-26555. This work uses gravitational lensing models by PIs Brada\v{c}, Natarajan \& Kneib (CATS), Merten \& Zitrin, Sharon, Williams, Keeton, Bernstein \& Diego, and the GLAFIC group. Lens modeling was partially funded by STScI's HST Frontier Fields program; models were obtained from the Mikulski Archive for Space Telescopes (MAST).\\

\noindent\facilities\ HST ACS/WFC3; Keck MOSFIRE.\\
\noindent\software\ IDL Coyote (D.\ Fanning; \url{http://www.idlcoyote.com}) and {\tt MPFIT} \citep{MPFIT} packages; SAO DS9.\\


\appendix

\section{A: Data Structures}
\label{sec:AA}

{\bf Spectra --} The GLASS pipeline outputs a FITS cube for each source in each grism at each orientation. The files contain the following objects/data types:

\bitem
	\item{\tt DSCI} -- Direct image of each source taken from a CLASH or HFF WFC3IR image stack to which the GLASS full field grism frames are aligned/referenced. (60$\times$60 pix image; ``{\tt DIRECT\_IMAGE}'' in text.)
	\item{\tt DWHT} -- {\tt DSCI}'s RMS error map. (60$\times$60 pix image.)
	\item{\tt SCI} -- Cutout of the full field GLASS grism image containing a target's 2D spectrum. (360$\times$60 pix image.) 
	\item{\tt MODEL} -- Model of the target's spectrum in {\tt SCI}; based on the CLASH/HFF alignment image SExtractor segmentation map. (360$\times$60 pix image.)
	\item{\tt CONTAM} -- Model of all {\it non-target} spectra in {\tt SCI}; based on the CLASH/HFF alignment image SExtractor segmentation map. (360$\times$60 pix image.)
	\item{\tt WHT} -- {\tt SCI}'s RMS error map. (360$\times$60 pix image.)
	\item{\tt YTRACE} -- Centerline of the target's 2D spectrum (spectra are not rectified). (360 element array.)
	\item{\tt SENS} -- 1D grism response curve along {\tt YTRACE} ($e^{-}/$s\,$\rightarrow10^{-17}$\,erg$/$s$/$cm$^{2}/$\AA). (360 element array.)
	\item{\tt WAVE} -- $\lambda(x)$ [\AA] along {\tt YTRACE}. (360 element array.)
\eitem

All 2D data derive from interlaced---not drizzled---WFC3IR reductions, maintaining pixel-to-pixel independence. 1D spectra derive from the {\tt SCI} minus {\tt CONTAM} image, masking ${\tt SCI}=0$ pix, and those with ${\tt CONTAM}\geq2\times{\tt WHT}$ \&\& ${\tt CONTAM}/{\tt MODEL}\geq0.01$. (I.e., pixels with significant contamination bright enough compared to the target to cause meaningful continuum biases over $\sim$hundred pixel extents.) Quality control (Section \ref{sec:selection}) ensures masking is minimal for the sources in this study, but it is visible, e.g., in the \out\ zone of \csf\ (Figure \ref{fig:spectra}).

Spectral extraction boxes are centered on and reflected over {\tt YTRACE}$(x)$ with widths set by $r_{e}$ as measured from {\tt DSCI} (Section \ref{sec:extraction}). Spectral uncertainties at each $\lambda$ are the quadrature sum of {\tt WHT}$(x)$ over all unmasked $y$ in the extraction zone.

For the optimal extractions, each column of a {\tt SCI}-sized image is filled with an object's normalized 1D spatial profile centered on {\tt YTRACE}$(x)$. This weights the $\tt SCI - CONTAM$ flux sum over all $N\leq60$ unmasked $y$ pixels at each $x$ \citep{Horne86}.

This process is repeated independently in G102 and G141, producing a pair of spectra and LSFs for each zone of each object. These are sent to {\tt PYSPECFIT} simultaneously, along with multiband, aperture-matched CLASH+HFF photometry.

\begin{figure}[h!]
\centering
\includegraphics[width = 0.9\linewidth, trim = 0.6cm 0cm -0.6cm 0cm]{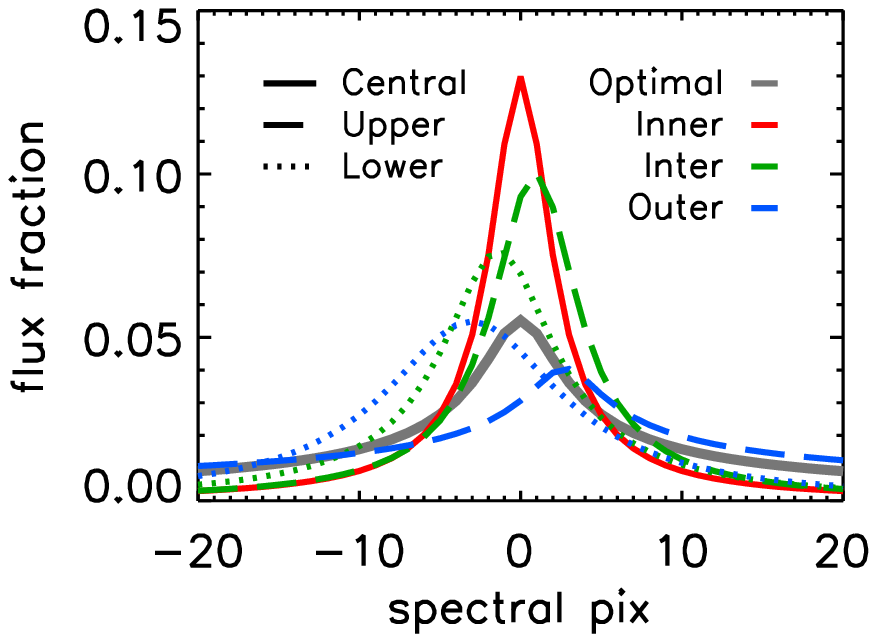}%
\caption{LSFs for \csf's unfolded radial and optimal extractions showing offsets and chord profile variations (Section \ref{sec:extraction}). These preclude {\it a priori} impositions of symmetry in objects with PA$\sim45\degree$ if studying line widths.}
\label{fig:lsfs}
\end{figure}

{\bf Photometry -- }In addition to the spectra, $N=17$ band image stacks are also cut for each source from PSF-matched (F160W resolution) CLASH/HFF imaging \citep{Morishita17}. These and their weight maps are rebinned from their native $0\farcs06$ to GLASS's $0\farcs065$ pixel scale and rotated to the GLASS {\tt DSCI} image orientation by finding the angle $\theta/{\rm deg}\in[0,359]$ that minimizes
\beq
	\chi^{2}_{\theta}=\sum_{\rm pix}\frac{(I^{\rm NIR}_{\theta}-{\tt DSCI})^{2}}{{\tt DWHT}^{2}},
\label{eq:rotate}
\eeq
where $I^{\rm NIR}_{\theta}$ is a stack of the F105--160W CLASH/HFF stamps (rotated by $\theta$). The best-fit $\theta$ is applied to each CLASH/HFF image individually. This angle will generally not correspond to the GLASS orientation---the HST roll angle when the spectra were taken---as {\tt DSCI} is from an independent alignment image. Pixel interpolation is cubic.

Once properly orientated, each band of CLASH/HFF imaging is photometered in the same zones as the spectral extractions. We assume {\tt YTRACE} follows the NIR image centroid, which we find using SExtractor on the rotated F105--160W stamp stack, $I^{\rm NIR}$. As with $\theta$, we fix this centroid to $r=0$ in {\it all} bands (in the blue, prominent starforming regions can and do shift independent SExtractor {\tt X\_} and {\tt Y\_IMAGE} estimates). The $r_{e}$-based extraction zone offsets derived from the {\tt DSCI} light profile (Section \ref{sec:extraction}) are then applied based on this centroid and all flux in the CLASH/HFF stamp summed within each zone (60$\times y_{\rm zone}$ pix aperture). Comparisons to synthetic photometry derived directly from the GLASS spectra suggest this process is absolutely accurate to within $\pm$$\sim$30\% (1.5\,$\sigma$). Regardless, the spectra are scaled to the photometry---which is corrected for galactic extinction---during SED fitting (Section \ref{sec:models}). Figures \ref{fig:mosaic} and \ref{fig:spectra} demonstrate that relative spectrophotometric accuracy is excellent.

\section{B: Further Details On Spectral Synthesis Modeling}
\label{sec:AB}

\subsection{Fitting Scheme}

The fitting scheme is as follows:
\benum
	\item Assume a delayed exponential SFH;
	\item Find the galaxy's global redshift and integrated properties using the optimal extraction;
	\item Find the redshift and properties of the \inn\ extraction using the redshift from Step 2 as strong prior;
	\item Lock the \mid\ and \out\ extractions to \inn's redshift, shifting the traces by the offsets identified during LSF measurement if these are larger than the formal $z$ uncertainty (Section \ref{sec:extraction}).
	\item Find the properties of the \mid\ and \out\ extractions given the central redshift solution;
	\item Repeat Steps 2--5 assuming a lognormal SFH fixed to the correct redshift.
\eenum

\subsection{Fitting Parameters}

Details differ slightly depending on the SFH, but the full parameter and prior set is as follows:
\bitem
	\item Redshift ($z$) in the case of the delayed exponential model and optimal/\inn\ extractions. Broad gaussian prior based on GLASS database redshifts.
	\item Stellar mass ($\Mstel$) and star formation rate ($\sfr$) in all cases. No prior.
	\item Equivalent widths and fluxes for H$\alpha$, H$\beta$, [\ion{O}{3}] $\lambda\lambda$4959,\,5007, [\ion{S}{2}] $\lambda\lambda$6718,\,6733 in all cases. Uniform priors from [0,\,500]\,\AA\ for H$\alpha$ and [\ion{O}{3}]; [0,\,50]\,\AA\ for [\ion{S}{2}]. Balmer ratios fixed to unextinguished values in the delayed exponential SFH; free for lognormals.
	\item Dust extinction to the stellar continuum, $A_{V}$. Uniform [0,\,3] mag prior; \citet{Calzetti00} reddening law.
	\item SFH parameters---$({\rm age}, \tau_{\rm exp})$ or $(T_{0},\tau)$ for delayed exponential and lognormal parameterizations, \resp. Priors are: age---log-uniform in $[10^{7},\,10^{10}]$ yr; $\tau_{\rm exp}$---log-uniform in $[10^{7},\,10^{10}]$ yr; $T_{0}$---log-uniform in $[10^{9},\,10^{10}]$ yr; $\tau$--- uniform in [0.1,\,3] $\ln$\,Gyr.
\eitem

{\bfb Emission lines are fitted jointly with the stellar continuum/corresponding absorption, but, beyond \csf\ AGN identification (Figure \ref{fig:AGN}) and gross EW(H$\alpha$+[\ion{N}{2}]) trends (Figure \ref{fig:ewHa}), we defer a discussion of these to a later paper.}

{\bfb Both passes use a $\chi^{2}$ likelihood function \citep[see Appendix B of ][]{Newman14}, but we quote the medians of the resulting posterior distributions as parameter estimates, which are more robust than those from the best-fit model.}

\section{C: SFH Reconstruction Self-Consistency}
\label{sec:resWhole}

A concern may be that inferences derived from summing the \inn, \mid, and \out\ results diverge from those based on the integrated optimal extractions. At $\lesssim$1.2\,$\sigma$, this is not the case at $\tobs$, and fortunately over much of cosmic time.

Figure \ref{fig:crossCheck} presents this cross-check, showing $\sfr(t)$ and $\Mstel(t)$ inferred from the integrated vs.\ summed resolved results. These are consistent, albeit within large uncertainties on the sums for the passive \pas\ and \psb\ systems. The fact that all four galaxies are past their peak SFRs (Figure \ref{fig:sfhs}) is brought to the fore here, {\bfb with no system likely to gain ``substantial'' mass in the future}. This highlights a meaningful distinction between starforming and rapidly growing galaxies---those with rising SFHs (Sections \ref{sec:spatResSFH}, \ref{sec:discussion})---but the agreement between integrated and summed inferences when \ssf\ and \psb\ {\it were} rapidly growing is nevertheless reassuring.

\begin{figure}[h!]
\centering
\includegraphics[width = \linewidth, trim = 0cm 0.5cm 0cm 0cm]{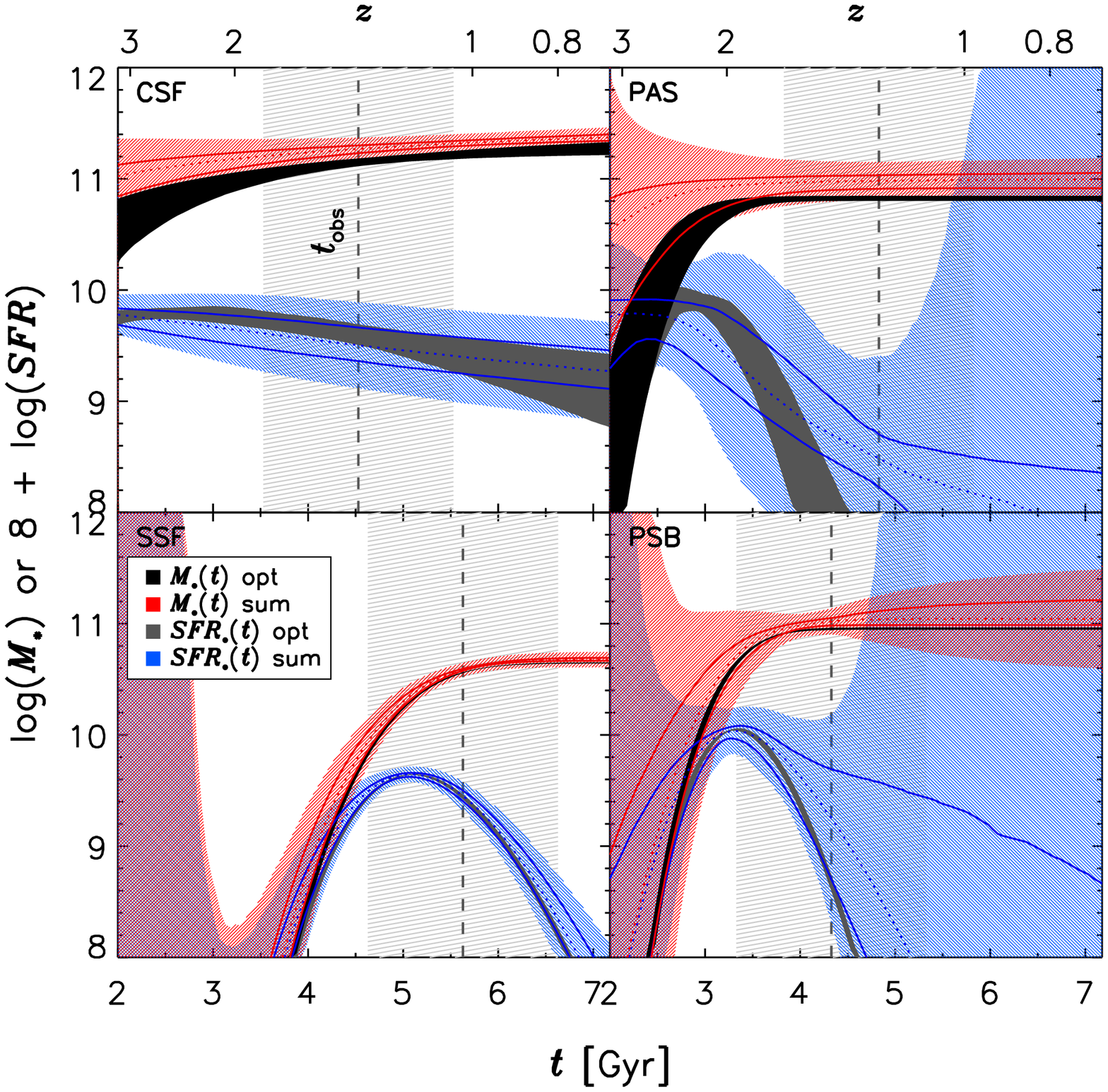}
\caption{SFHs and mass growth curves from the optimal integrated extractions (black/grey, \resp) vs.\ those summing \inn, 2$\times$\mid, and 2$\times$\out\ results (blue/red, \resp). Vertical dashes show $\tobs$ with the $\pm1$\,Gyr projection window shaded (Section \ref{sec:spatResSFH}). 1\,$\sigma$ summed uncertainties are calculated as a fraction of the median (shading), and as the log of the high/low linear $\sfr(t)$ or $\Mstel(t)$ estimates (solid lines). The former leads to mostly unconstrained (future) summed uncertainties for \pas\ and \psb\ due to their low $\sfr(\tobs)$. Even so, the largely independent sum of the parts and inferences from the whole broadly agree.}
\label{fig:crossCheck}
\end{figure}

\begin{figure*}[t!]
\centering
\includegraphics[width = 0.65\linewidth, trim = 0cm 0cm 0cm 0cm]{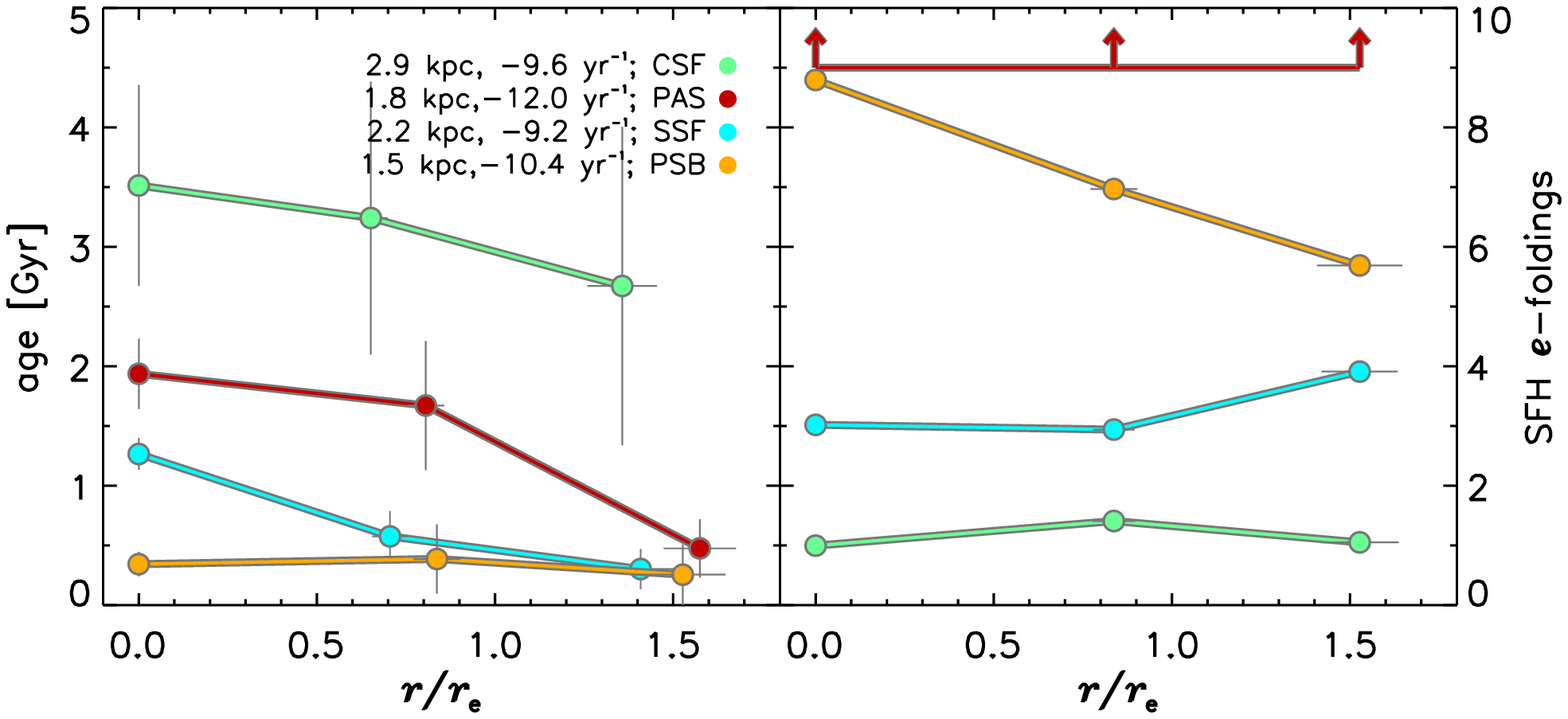}
\includegraphics[width = \linewidth, trim = 0cm 0cm 0cm 0cm]{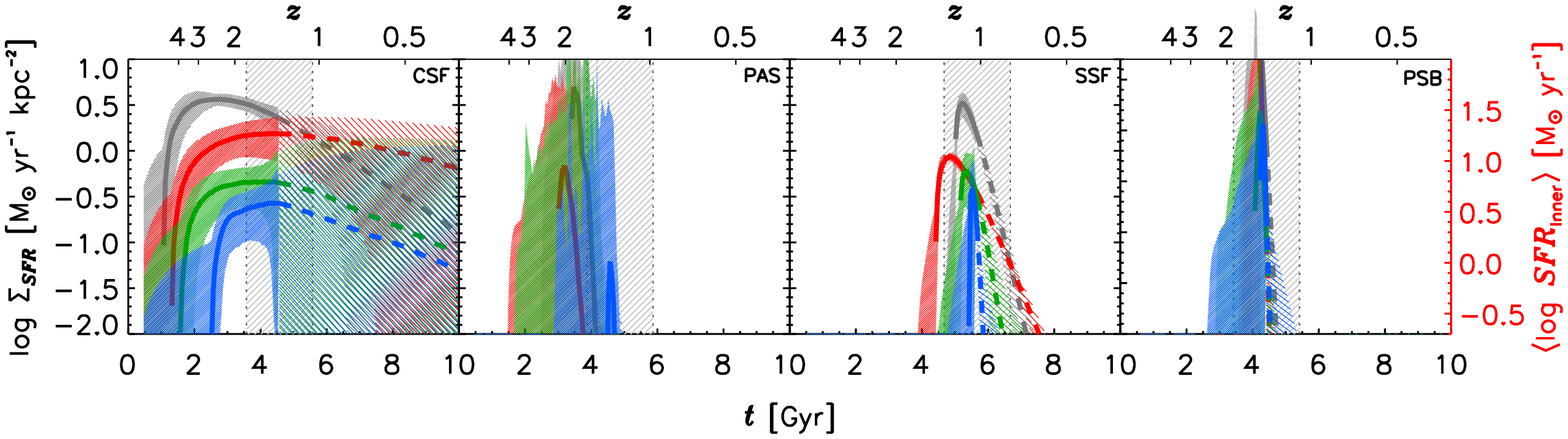}	
\caption{Summary plots based on the delayed exponential SFH fits (Section \ref{sec:models}). The top panels show radial parameter trends---${\rm age}\equiv\tobs-t_{\rm start}$, $e$-foldings$\,={\rm age}/\tau_{\rm exp}$---which reveal similar signatures of inside out growth to the main, lognormal analysis (\inn\ regions older/having undergone more $e$-foldings). The bottom panels reproduce Figure \ref{fig:sfhs}, showing the SFHs in full. These qualitatively reflect the lognormal results for \csf\ and \ssf's \inn\ region, but \psb\ and \pas\ become essentially a convolution of $\delta$-functions, implying unrealistic growth rates in \psb's case and forbidding time-domain reconstructions. Note that inferences such as the \inn/\out\ SFH crossing epoch (inside-out transition) in \ssf\ agree using either method (cf.\ Figure \ref{fig:evostuff}).}
\label{fig:expSfhs}
\end{figure*}

\begin{figure*}[t!]
\centering
\includegraphics[width = 0.4\linewidth, trim = 0cm 0cm 0cm 0cm]{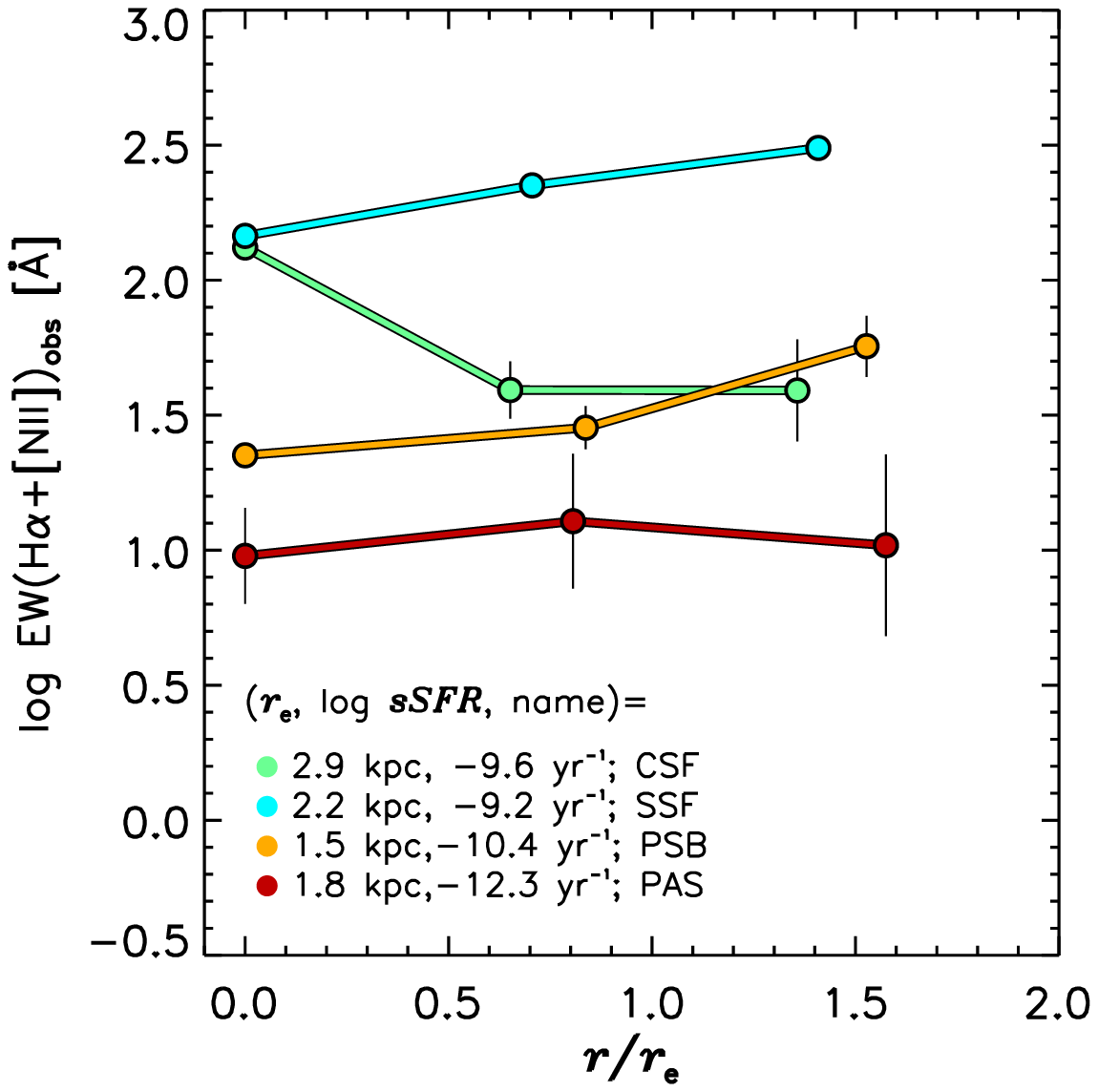}
\includegraphics[width = 0.4\linewidth, trim = 0cm 0cm 0cm 0cm]{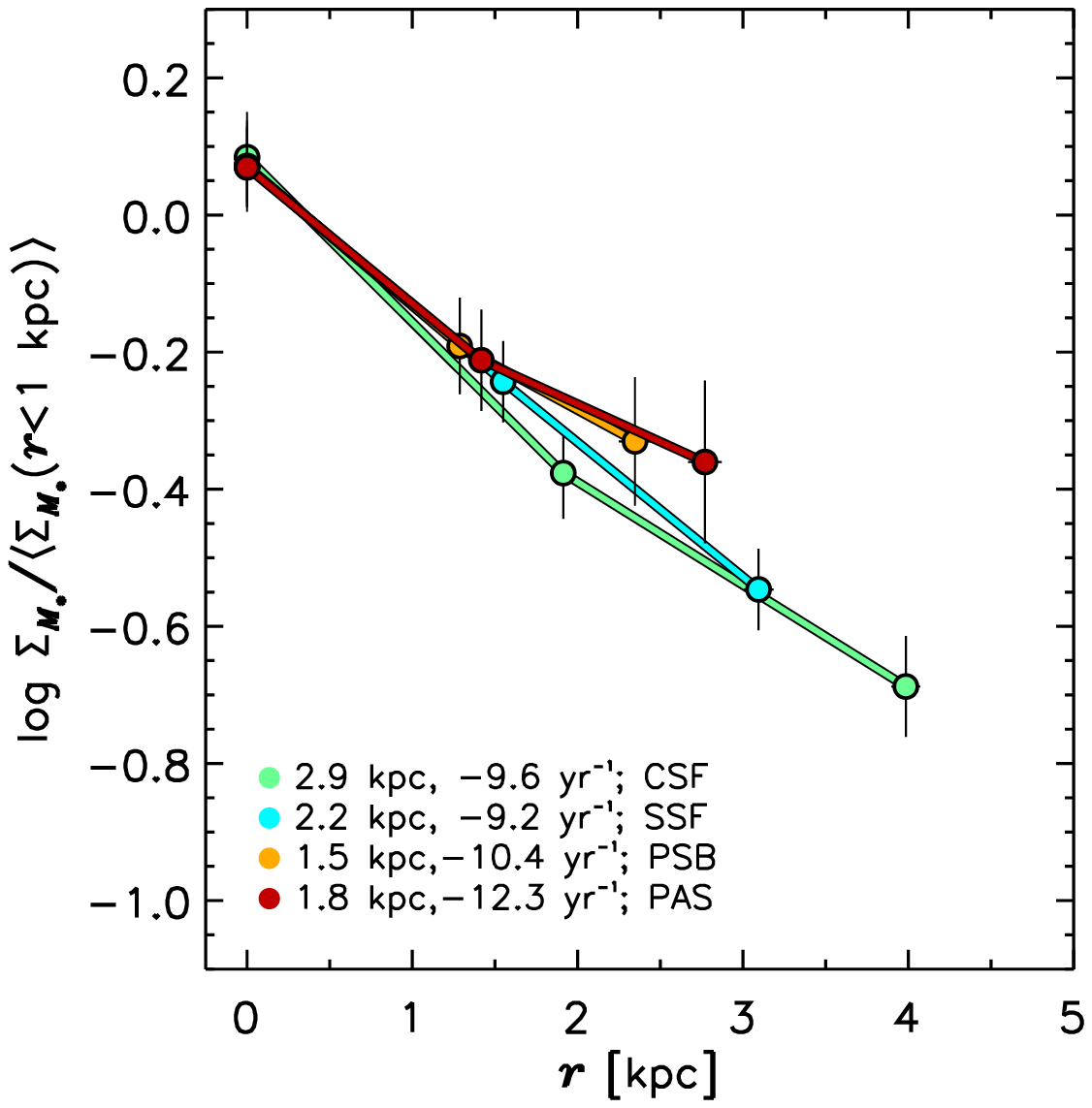}
\caption{Reinterpretations of Figure \ref{fig:massProf}. {\it Left}: Dust-uncorrected EW(H$\alpha$+[\ion{N}{2}]) yield similar trends to SED-fit $\ssfr$s. \csf\ is an exception: AGN activity elevates the \inn\ inference (Figures \ref{fig:spectra}, \ref{fig:AGN}), while those at larger-$r$ are $\sim$10$\times$ depressed relative to \ssf. IFU data will reveal if this is due to increased extinction towards \csf's \ion{H}{2} regions or a real suppression of $\ssfr$ beyond the factor of $\sim$3 inferred from the full SED. {\it Right}: Inner kpc-normalized $\Sigma_{\Mstel}$ profiles vs.\ physical radii (as opposed to $r/r_{e}$), highlighting the sample galaxies' sizes. Trends are the same as in the main-text figure, with the larger starforming systems having steeper mass profiles than the passive ones, perhaps suggesting the latter have already some dry merging (Section \ref{sec:resolved}; e.g., \citealt{Newman12a}).}
\label{fig:ewHa}
\end{figure*}

\section{D: Delayed Exponential Results Summary}
\label{sec:AC}

Figure \ref{fig:expSfhs} shows some relevant results based on the delayed exponential SFHs not used in the main analysis. The {\it top} panel shows spatially resolved age inferences ({\it left}; ${\rm age}\equiv\tobs-t_{\rm start}$; Equation \ref{eq:delayedExponential}), and $e$-foldings ({\it right}; ${\rm age}/\tau_{\rm exp}$). The {\it bottom} shows the resolved SFHs themselves (replicating Figure \ref{fig:sfhs}).

A similar picture emerges from these panels as with the lognormal SFHs, though its expressed slightly differently/less conveniently given our intent. First, from the top-left panel, inside-out age gradients characterize \pas\ and \ssf\ (\inn\ regions are oldest). While, unlike in the main analysis, \psb\ does not appear to have such a gradient, the top-{\it right} panel pushes back on this, showing an inside-out $e$-folding, i.e., {\it maturity} gradient (\inn\ regions most evolved).

Figure \ref{fig:expSfhs}, {\it bottom}, elaborates on this swap: \psb\ has $\delta$-function-like (zero-age) best-fit exponential SFHs. This follows from discussions in Sections \ref{sec:global}, \ref{sec:resolved} regarding how this form expresses what is likely an abruptly truncated (poststarburst) SFH. The other systems have $e$-folding trends that mimic the lognormal inferences, with the starforming systems having fewer (i.e., wider SFHs) and \pas\ having the most.

Indeed, the starforming delayed exponential SFHs grossly correspond to the lognormals (Figure \ref{fig:sfhs}), especially for \csf, where the same quasi-flat, rank-ordered SFHs emerge at all $r$. \ssf\ has similar \inn\ properties, but to capture this system's $\ssfr$ gradient (Figure \ref{fig:massProf}), the \mid\ and \out\ SFHs become dramatically compressed, illustrating the this form's expression of inside-out growth. (Note, however, that \ssf's \inn\ SFH still moves from a leading to lagging position roughly at $\tobs$ as in Figure \ref{fig:evostuff}.) The abruptness of these histories---and those for the the passive systems---contributes to our disfavoring them for the time-domain reconstructions that at the core of the main analysis.

\section{E: ${\rm EW(H\alpha}$+[\ion{N}{2}]$)$ Trends, $\Sigma_{\Mstel}(\lowercase{r})$ in Physical Radii, and the $\lowercase{z}\gtrsim1$ $\Sigma_{\Mstel}$--$\Sigma_{\sfr}$ Law}
\label{sec:AD}

Figure \ref{fig:ewHa} reinterprets Figure \ref{fig:massProf}'s {\it top-} and {\it middle-right} panels using dust-uncorrected EW(H$\alpha$+[\ion{N}{2}]) as an $\ssfr$ proxy ({\it left}), and physical---not $r_{e}$-normalized---radii ({\it right}). These yield very similar trends to those discussed in-text and preserve the systems' rank ordering. The exception is \csf's EW profile, where, compared to SED-derived $\ssfr(r)$, \inn\ levels are elevated and larger-$r$ values depressed relative to \ssf's. The former effect is largely if not entirely due to \csf's AGN activity (Figures \ref{fig:spectra}, \ref{fig:AGN}), and higher resolution IFU spectroscopy will reveal if the latter is due to increased extinction towards its \ion{H}{2} regions or a real suppression of $\ssfr$ beyond the factor of $\sim$3 inferred from the full SED.

\begin{deluxetable}{cccc}
	\tabletypesize{\footnotesize}
	\tablewidth{0pt}
	\tablecolumns{4}
	\tablecaption{Delayed Exponential-Derived Integrated Properties from Optimal Extractions}
	\tablehead{
	\colhead{Tag} & 
	\colhead{$\log \Mstel/\Msun$\tablenotemark{a,b}} & 
	\colhead{$\log \sfr/\Msun\,{\rm yr^{-1}}$\tablenotemark{a,b}} & 	
	\colhead{$A_{V}$ [mag]}
}
\startdata
\pas & 10.77$\pm$0.07 (0.10) & $<$-9.13\tablenotemark{c} & 0.25$\pm$0.13\\ 
\ssf & 10.54$\pm$0.06 (0.09) &  1.58$\pm$0.06 (0.07) & 1.31$\pm$0.03\\
\psb & 11.05$\pm$0.07 (0.10) &  1.16$\pm$0.15 (0.27) & 1.50$\pm$0.09\\
\csf & 11.17$\pm$0.07 (0.16) &  1.65$\pm$0.14 (0.20) & 1.28$\pm$0.14
\enddata
\tablecomments{$^{\rm a}$\,Quadrature sum of \inn\ + 2\,\mid\ + 2\,\out\ $\Mstel$ and $\sfr$ errors in parentheses; $\Mstel$ are consistent with lognormal estimates. $^{\rm b}$\,Magnification-corrected (Table \ref{tbl:sample}); uncertainties incorporate $\mu$ errors. $^{\rm c}$\,{\it Formal} 2\,$\sigma$ limit.}
\label{tbl:derivedExp}
\end{deluxetable}

\begin{figure}[t!]
\centering
\includegraphics[width = \linewidth, trim = 0.7cm 0cm 0.7cm 0cm]{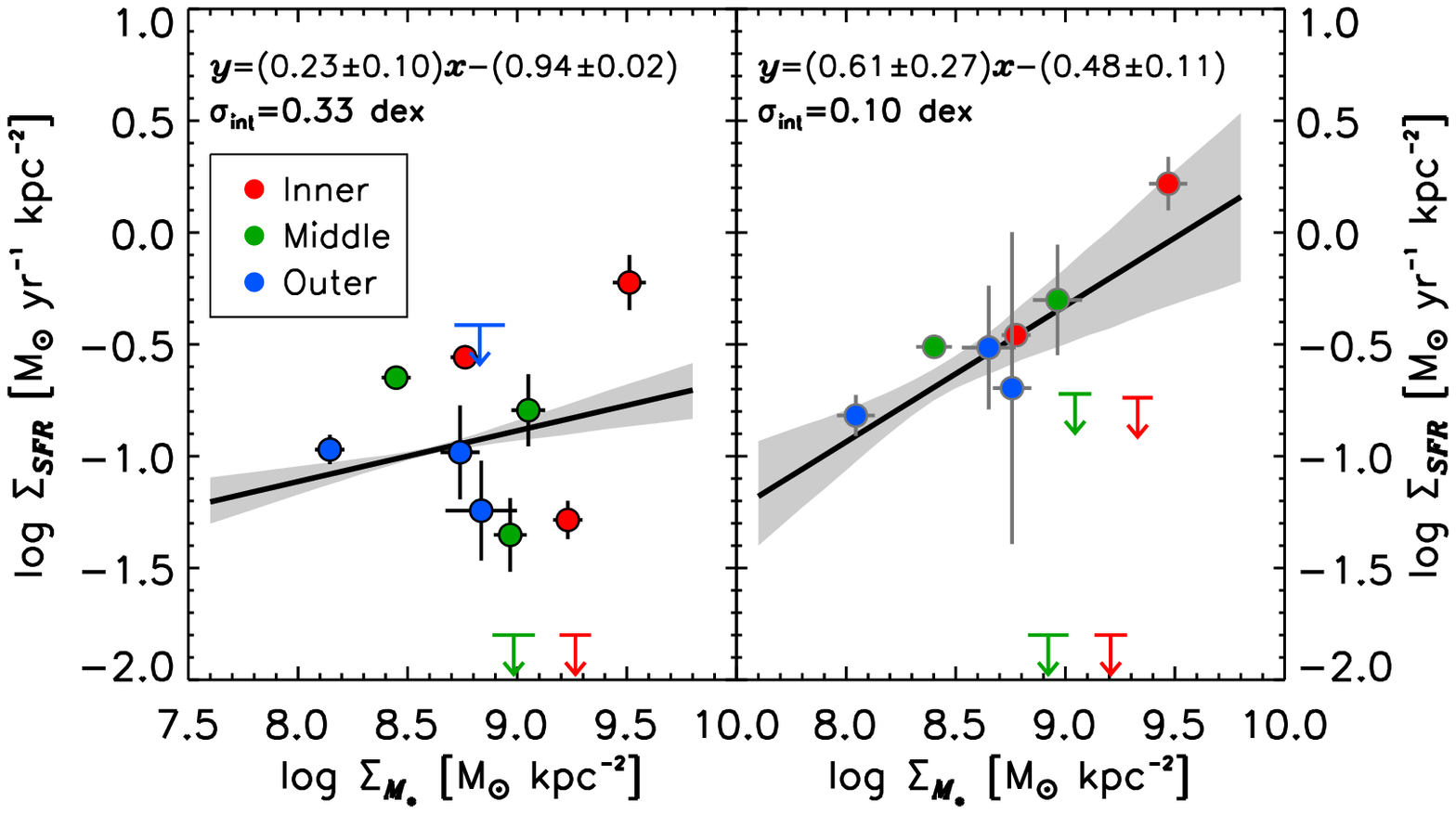}
\caption{Resolved SFR vs.\ $\Mstel$ surface densities (analogous to \citealt{Schmidt59}; \citealt{Kennicutt98}) assuming lognormal ({\it left}) and delayed exponential SFHs ({\it right}). Both have $x\equiv\log\Sigma_{\Mstel}-8.75$, with linear best-fits performed only to detections by bootstrapping and varying points by their $x$ and $y$ error-bars. While quite different from each other, both inferences are similar to $z=0$ IFU/resolved SED estimates over a similar mass (but not $\Sigma_{\Mstel}$) range: \citet{CanoDiaz16} find a slope/dispersion of $0.68\pm0.04$ and 0.23 dex, \resp, corresponding to our delayed exponential results; \citet{Abdurrouf17} find 0.33 and 0.7, \resp, closer to our lognormal results. This highlights the dependence of scaling relations on SFH choice.}
\label{fig:ksrel}
\end{figure}

Finally, Figure \ref{fig:ksrel} plots the full sample's 12 resolved $\Sigma_{\sfr}$ and $\Sigma_{\Mstel}$ values, a stellar mass reinterpretation of the Schmidt-Kennicutt Law \citep{Schmidt59,Kennicutt98}. The {\it left} panel shows lognormal inferences, with delayed exponentials at {\it right}. While these techniques yield quite different best-fit relations (Equations \ref{eq:lgnKSrel} and \ref{eq:expKSrel} below) both are nevertheless similar to previous $z=0$ IFU-/resolved SED based estimates: \citet{CanoDiaz16} find a slope and dispersion of $0.68\pm0.04$ and 0.23 dex, \resp, corresponding to our delayed exponential results, while \citet{Abdurrouf17}, find 0.33 and 0.7, \resp, closer to (though with larger scatter than) our lognormal results, highlighting the dependence of such scaling relations on SFH choice.

Using lognormal SFHs, we find:
\beq
	\begin{aligned}
		\log\left[\frac{\Sigma_{\sfr}}{\Msun\,\yr^{-1}\,\kpc^{-2}}\right] = &- (0.94\pm0.02)\\		
		&+(0.2\pm0.1)\log
			\left[\frac{\Sigma_{\Mstel}}{10^{8.75}\,\Msun\,\kpc^{-2}}\right],		
	\end{aligned}
\label{eq:lgnKSrel}
\eeq
with intrinsic scatter $\sigma_{\rm int}=0.33$\,dex. 

Using delayed exponentials yields:
\beq
	\begin{aligned}
		\log\left[\frac{\Sigma_{\sfr}}{\Msun\,\yr^{-1}\,\kpc^{-2}}\right] = &- (0.48\pm0.11)\\
		&+(0.6\pm0.3)\log
			\left[\frac{\Sigma_{\Mstel}}{10^{8.75}\,\Msun\,\kpc^{-2}}\right],
	\end{aligned}
\label{eq:expKSrel}
\eeq
with intrinsic scatter $\sigma_{\rm int}=0.10$\,dex.


\bibliographystyle{apj}
\bibliography{/Users/labramson/lit}


\end{document}